\documentclass[journal]{IEEEtran}
\ifCLASSINFOpdf
\else
\fi
\usepackage{xcolor}
\usepackage{graphicx}
\usepackage{subfigure}
\usepackage{caption}
\usepackage[textwidth=0.9]{todonotes}
\usepackage{makeidx}         
\usepackage{graphicx}
\usepackage{epstopdf}
\usepackage{color}
\usepackage{multicol}        
\usepackage[bottom]{footmisc}
\usepackage{url}
\usepackage{marvosym}
\usepackage{amssymb}
\usepackage{subfigure}
\usepackage{tabularx}
\usepackage{amsmath}
\usepackage{amsthm}
\usepackage{algorithm}
\usepackage{algorithmic}
\usepackage{setspace}
\usepackage{pifont}
\usepackage[T1]{fontenc}
\usepackage{color}
\usepackage{tabularx}

\usepackage{changepage}

\usepackage{booktabs}

\begin{document}
%
\title{MPTCP meets FEC: Supporting Latency-Sensitive Applications over Heterogeneous Networks }
%
%
%

\author{

    \IEEEauthorblockN{Simone Ferlin\IEEEauthorrefmark{1},
    Stepan Kucera\IEEEauthorrefmark{2}, Holger Claussen\IEEEauthorrefmark{2}, \"{O}zg\"{u} Alay\IEEEauthorrefmark{1}}
    
    \IEEEauthorblockA{\IEEEauthorrefmark{1}Simula Research Laboratory, Norway
    \\\{ferlin, ozgu\}@simula.no}
    
    \IEEEauthorblockA{\IEEEauthorrefmark{2}Nokia-Bell-Labs, Dublin, Ireland
    \\\{stepan.kucera, holger.claussen\}@nokia-bell-labs.com}
}

%
%

\markboth{IEEE/ACM TRANSACTIONS ON NETWORKING}%
{Shell \MakeLowercase{\textit{et al.}}: Bare Demo of IEEEtran.cls for IEEE Journals}
%



\maketitle

\begin{abstract}
Over the past years, TCP has gone through numerous updates to provide performance enhancement under diverse network conditions. However, with respect to losses, little can be achieved with legacy TCP detection and recovery mechanisms. Both \textit{fast retransmission} and \textit{retransmission timeout} take at least one extra round trip time to perform, and this might significantly impact performance of latency-sensitive applications, especially in lossy or high delay networks. While forward error correction (FEC) is not a new initiative in this direction, the majority of the approaches consider FEC inside the application. In this paper, we design and implement a framework, where FEC is integrated within TCP. Our main goal with this design choice is to enable latency sensitive applications over TCP in high delay and lossy networks, but remaining application agnostic. We further incorporate this design into multipath TCP (MPTCP), where  we focus particularly on heterogeneous settings, considering the fact that TCP recovery mechanisms further escalate head-of-line blocking in multipath. We evaluate the performance of the proposed framework and show that such a framework can bring significant benefits compared to legacy TCP and MPTCP for latency-sensitive real application traffic, such as video streaming and web services.
\end{abstract}


%
\IEEEpeerreviewmaketitle
\section{Introduction}

\IEEEPARstart{T}{he} enormous growth in mobile wireless devices and mobile traffic led to increased dependency on mobile infrastructures. Today, mobile operators are expected to deliver high capacity and reliable networks to meet the demand from many stakeholders. One approach to increase both reliability and capacity is to better foster network resources. For example, smartphones can leverage both cellular and WLAN connections, or air-to-ground communications can utilise both mobile satellite terminals and cellular networks. While the number of use-cases and applications vary and steadily grow, the choices of transport protocols to address these demands do not evolve at the same pace, with UDP and TCP being the main options in the Internet, and, in their original designs, unable to explore multiple networks simultaneously.

\begin{figure}[t]
  \vspace{-3mm}
\begin{center}
\includegraphics[width=0.95\columnwidth]{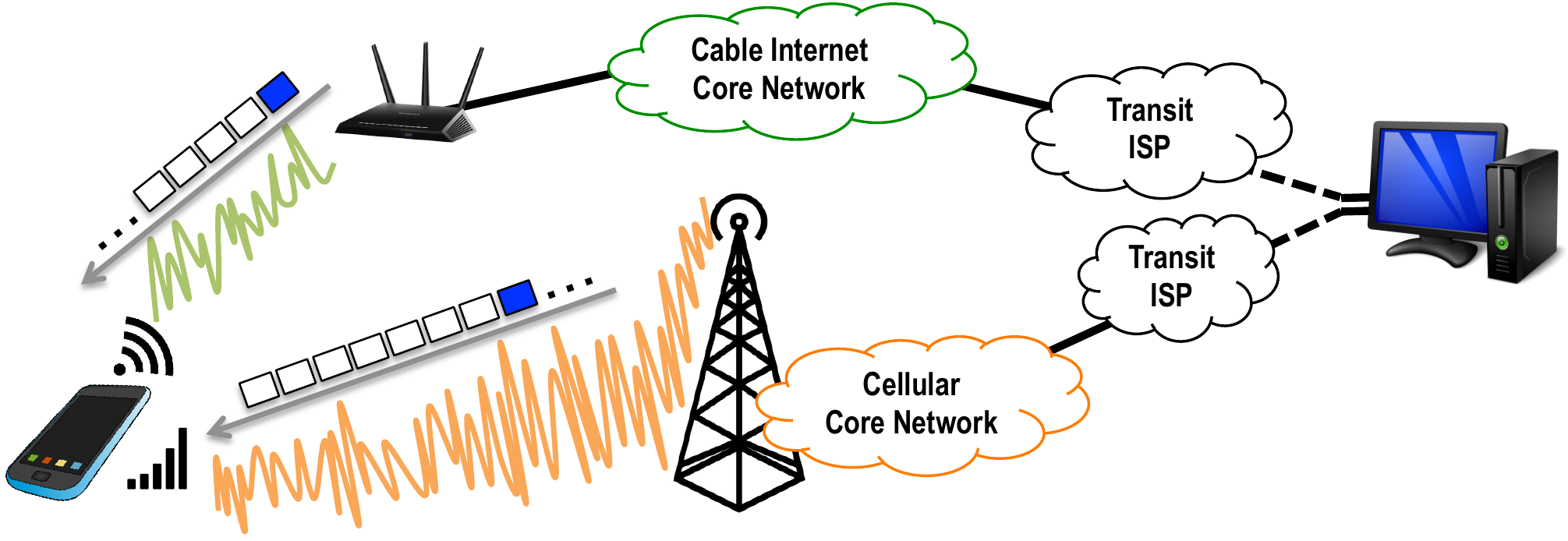} 
\end{center}
  \vspace{-3mm}
\caption{System Building Blocks: Overview}\label{fig:overview}
  \vspace{-3mm}
\end{figure}

In light of this diversity for network access technologies, we find multipath TCP (MPTCP). MPTCP maintains applications unchanged, where \textit{subflows} running on different networks are unnoticed by the application. In Figure~\ref{fig:overview}, we illustrate the smartphone scenario with cellular and WLAN links.
Two main advantages with MPTCP are envisioned: Capacity aggregation across multiple links and the ability to maintain the connection, if at least one path remains active. Capacity aggregation is, however, quite challenging with heterogeneous paths, in particular due to delay and loss heterogeneity. This
heterogeneity results in packet reordering, leading to head-of-line (HoL) blocking, increased out-of-order (OFO) buffer at the receiver and, ultimately, reduced goodput.


Furthermore, since MPTCP is closely tied to TCP, TCP's time-dependent loss recovery can, in turn, also be a bottleneck for high delay and lossy networks. In TCP, both fast recovery (FR) and retransmission timeout (RTO), are strictly tied to round trip time (RTT). Hence, regardless of how the network capacities grow, the required RTTs recovery remain the same. For MPTCP, while the scheduler is commonly the place to improve performance, tackling network heterogeneity, little can be achieved with legacy TCP loss recovery if, in addition to delay, the~\textit{subflows} have heterogeneous loss characteristics\footnote{MPTCP's default recovery mechanism for FR resends a packet on the same~\textit{subflows}, whereas, for an RTO, it reschedules a packet on the~\textit{subflows} with space in its congestion window (CWND) and the next lowest RTT.}.


In this paper, we address TCP's loss recovery mechanism in order to improve TCP's performance in high delay and lossy networks as well as improve MPTCP's performance in heterogeneous settings. The main contributions of this paper can be summarised as:

\begin{enumerate}
\item We integrate forward error correction (FEC) in TCP in order to provide \emph{zero-RTT} loss recovery for latency-sensitive applications. To achieve this, we propose TCP with dynamic FEC (TCP-dFEC) building on TCP instant recovery (TCP-IR)~\cite{flach00,flach13a} that uses XOR-based FEC within TCP. TCP-dFEC extends TCP-IR in two major ways: \emph{(i) making it fair to regular TCP} and \emph{(ii) designing a dynamic FEC mechanism to better cope with changing channel conditions}. 

\item We further extend this framework and propose MPTCP with dynamic FEC (MPTCP-dFEC) where each TCP~\textit{subflow} runs TCP-dFEC. We follow an intra-subflow FEC approach, in order to better understand the interaction of FEC within MPTCP, without considering its interaction with the scheduler and congestion control algorithms.
The proposed MPTCP-dFEC works seamlessly with MPTCP's connection-level management signalling without sacrificing resources of good subflow(s) with FEC for other(s).

\item The proposed TCP-dFEC and MPTCP-dFEC algorithms are implemented into linux kernel. This enables the proposed framework to be application agnostic and as well as deployable.
Our evaluations show that the proposed TCP-dFEC and MPTCP-dFEC significantly improve the completion times for HTTP/2 web traffic and the frame rate for video streaming with H.264.
\end{enumerate}

The reminder of this paper is organised as follows: Section~\ref{sec:motivation} motivates our work putting the features provided by TCP and MPTCP protocols, FEC mechanisms and latency-sensitive application requirements into perspective. Section~\ref{sec:approach} explains our dynamic FEC (dFEC) algorithm design as well as the necessary system building blocks to integrate it into TCP and MPTCP. Section~\ref{sec:measurement:setup} explains our measurement setup with different applications, network settings and the end-host configuration. Section~\ref{sec:evaluation} presents the results of proposed algorithms compared to regular TCP and MPTCP. Section~\ref{sec:related_work} puts our work into perspective with other proposals to integrate FEC into either applications or into the transport layer. Finally Section~\ref{sec:conclusion} concludes our work, hinting to the future directions with dFEC design and evaluation setups.

\section{Motivation and Background}\label{sec:motivation}
The performance of the Transport Control Protocol (TCP) over wireless high delay and lossy networks is known to be suboptimal~\cite{Martin2002,Johansson16}, with one of the main limiting factors being TCP's loss recovery time. In such scenarios, it is often an option to replace TCP by UDP, at the expense of compromising benefits such as flow and congestion controls. When MPTCP~\cite{RFC6182} emerged, enabling simultaneous use of multiple network paths by a single data-stream, it had to take operability and deployment in the Internet into account, hence, making MPTCP look like regular TCP from the network's perspective. Although this integration brings many benefits, it also comes with challenges that hinder MPTCP. Particularly, when the underlying network paths are heterogeneous, MPTCP often underperforms TCP especially for certain latency-sensitive applications~\cite{Kiran2016,ferlin16_IFIP}. 

Next, we will first put MPTCP in perspective with TCP, also mentioning its performance challenges under certain network environments. We will then discuss our motivation for designing a dynamic XOR-based Forward Error Correction (dFEC) inside TCP, and how this can aid multipath transport for heterogeneous paths. Finally, we will summarise the applications chosen for the evaluations and their requirements.

\subsection{Transport Protocols}\label{subsection:transport:protocol}
When TCP and the Internet Protocol (IP) were specified more than 30 years ago, end-hosts were typically connected to the Internet via a single network interface, and TCP was built around the notion of a single connection between them. Nowadays, the picture has been changing with end-hosts commonly accommodating multiple interfaces, e.g., smartphones with cellular and WLAN interfaces. Standard TCP is not able to efficiently explore the multi-connected infrastructure as it ties applications to source and destination IP addresses and ports. MPTCP emerged to close this gap, by allowing the use of multiple network paths for a single data-stream simultaneously, providing a great potential for higher application throughput and resilience against network path failure~\cite{RHW09}.

Although MPTCP enables better utilisation of network resources, scenarios with heterogeneity remain a challenge: Delay and loss heterogeneity result in packet reordering, which lead to Head-Of-Line (HoL) blocking, increased out-of-order (OFO) buffer and, ultimately, reduced overall throughput, causing MPTCP at times perform worse than TCP~\cite{Kiran2016,ferlin16_IFIP}.


Not only MPTCP-specific elements, e.g., the scheduler, are critical to enhance multipath performance with heterogeneity~\cite{ferlin16_IFIP}, but also TCP specific elements should be addressed. For example, TCP's performance in certain scenarios, e.g., high delay or lossy networks, is suboptimal~\cite{Martin2002,Johansson16}. Focusing on TCP first, one of the limiting factors is the loss recovery time, with its legacy loss detection and recovery mechanisms being strictly tied to time: A Retransmission Timeout (RTO) after a timer expires, or Fast Retransmission (FR) after three duplicated acknowledgements (\texttt{DupACK}) arrive from the receiver\footnote{The standard value in Linux TCP stacks.} to detect a loss, and at least one extra RTT for a retransmission to perform. For multipath, MPTCP~\textit{subflows} belonging to different technologies with distinct delay and loss profiles, can have the situation with one of the~\textit{subflows} stalling the multipath connection~\cite{Kiran2016}.

\subsection{Forward Error Correction (FEC)}\label{subsection:fec}
TCP loss recovery mechanisms, such as Fast Recovery (FR) and Retransmission Timeout (RTO), are strictly tied to Round Trip Time (RTT). Thus, regardless of how the network capacity evolves, time to recover from losses remains the same. One approach addressing this challenge is proposed in TCP-Tail Loss Probe (TLP)~\cite{dukkipati01}, focusing on reducing web latency, duplicating packets at the flow's tail to avoid RTOs\footnote{In addition to TCP-TLP, TCP-RACK~\cite{Cheng2016} is under study to change TCP's hardcoded~\texttt{DupACK} loss detection threshold.}. 
 
\begin{figure}[h!]
   \vspace{-4mm}
  \centering
   \includegraphics[width=.35\textwidth]{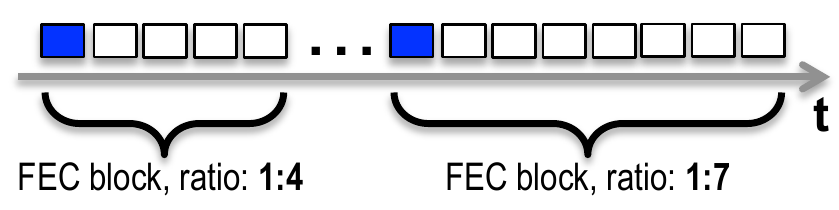}
   \vspace{-2mm}
  \caption{XOR-based FEC block size: View from the wire.}
  \label{fig:fec:wire}
   \vspace{-4mm}
\end{figure}

One common approach to improve reliability in wireless networks has been the use of FEC, where block erasure codes are used to correct transmission errors with redundant information added to the data stream. For example, in an $(n, k)$ block erasure code, there are a total of $n$ packets, with $k$ source packets and $(n-k)$ redundant parity packets. The parity packets are generated in such a way that any $k$ of the $n$ encoded packets are sufficient to reconstruct the $k$ source packets, resulting in a overhead of $m/(k+m)$. For many applications of block erasure codes,
encoding and decoding complexity is the key concern behind the choice of the codes. XOR-based codes are very beneficial with pure XOR operation, efficient in both hardware and software implementation, although limited to a single loss within a block of packets for the XOR-FEC packet, see Figure~\ref{fig:fec:wire}.

Taking the fact that the average link loss rate is generally unknown, and loss can be manifested through actual network congestion or, specially in wireless networks, as random channel effects or medium access control schedulers; there has been interest to integrate FEC-like approaches into transport protocols. In particular, we observe initiatives for TCP~\cite{dukkipati01,flach00}, and, recently, for QUIC~\cite{hamilton16}. The reasons for that are multifold: First, improve performance by decoupling loss detection from recovery to better scale bandwidth with latency. Second, improve performance over wireless networks when recovered loss by link layer mechanisms can arrive too late at the transport layer, being discarded and retransmitted. 

Along this line of thoughts, TCP-Instant Recovery (TCP-IR)~\cite{flach00,flach13a} aims to reduce TCP's loss recovery to \textit{zero-RTT}, by applying XOR-based FEC injecting encoded packets within TCP to provide N+1 redundancy. However, XOR-based FEC can be disadvantageous if more than one packet per FEC~\textit{block} is lost and a fixed-rate FEC that always reserve Congestion Window (CWND) for FEC, as it is the case in TCP-IR, wastes link capacity if FEC is not used for recovery. 
However in QUIC, details about its earlier experiments with FEC are not publicly available, although unofficial reports state that applications such as YouTube performed worse with FEC, which may have been the reason to its deprecation in QUIC's current development. Later, in Section~\ref{section:tcpir}, we show some shortcomings of TCP-IR, which at this stage it can be only source of speculation, whether QUIC's experiments with FEC could not have suffered from similar issues.

\textbf{Summary: }FEC within the transport layer has been proposed, but it has been prohibitively complex inside TCP, with most of the proposals focusing on application layer FEC or simulations. The goals are multifold to improve TCP's performance over wireless networks and to decouple loss detection and recovery mechanisms. TCP-TLP and TCP-IR initiated the work to reduce TCP's loss detection and recovery to~\textit{zero-RTT}, however, both approaches do not respect TCP's CWND nor specify adaptiveness for FEC at run-time. In order to support latency sensitive applications, with the benefits that TCP provides, there is a strong need for dynamic FEC (dFEC) adaptation that also respects TCP's CWND. This should be further propagated to MPTCP, where link heterogeneity can amplify the problem, especially with links that are heterogeneous in terms of delay and loss.

\subsection{Latency-Sensitive Application Traffic}\label{subsection:applications}
Although web traffic still constitutes a large fraction of today's Internet~\cite{Labovitz-IOR-2009}, video is becoming the most dominant and bandwidth intensive application. Recent reports~\cite{sandvine-2013} show that more than 53\% of North America's downstream traffic is already video streaming. Forecasts~\cite{cisco-white-paper-2013} also point that Internet video will continue to grow. Even though web and video differ in many ways, they are both sensitive to latency in a way that users have a better experience when web pages are loaded faster and when video has a more fluid delivery. In this paper we use video and web traffic to assess whether MPTCP with FEC can be suitable for latency-sensitive applications. The remainder of this section describes the main characteristics of the applications and discusses their requirements.

\subsubsection{HTTP}\label{subsection:http}
HTTP/1.1 has now served the Internet for more than 15 years, being the dominant application protocol for web requests. However, loading web pages efficiently nowadays is more resource intensive, with HTTP/1.1 allowing only one outstanding request per TCP connection, hence, leaving data splitting to applications themselves. This has shown very quickly to lead to self-inflicted congestion, hurting performance. For this reason, HTTP/2 if becoming to be de-facto substitute, addressing such shortcomings, e.g. HTTP/2 is fully multiplexed, allowing multiple requests within a single TCP connection and using a single connection for parallelism. 


\textbf{Requirements:} The quality of user experience when accessing a website is highly linked to the download completion time. For example,~\cite{why-latency-matters-2013} reports that ``an additional $500$~ms to compute (a web search) resulted in a 25\%~drop in the number of searches done by users.''. Although the download completion time may not be the most relevant metric for modern browsers, as they often start rendering pages before completion, it is the most suitable metric to use when evaluating transport protocols as it is browser agnostic. 

\subsubsection{Video}\label{subsection:video}

We consider non-adaptive live video streaming with H.264 in our experiments, with frames that are not delivered on time being dropped by the receiving application. In such applications, users' good quality of experience watching a live video delivered over networks that have high base delay induced or not by~\textit{bufferbloat}~\cite{jiang12}, e.g. cellular or satellite air-to-ground networks, is the ability to receive data as early and as complete as possible. Here, retransmissions caused by full or partial frame loss are hardly affordable, resulting in frames being dropped at the receiving application. Similarly, video delivery over TCP, e.g. Skype uses TCP as a backup transport protocol, would be penalised in such scenarios.

\textbf{Requirements:} The quality of user experience for video when consider in our experiments is related to latency, however, we quantify it as the average fully received frames for non-adaptive H.264.

Finally, in addition to HTTP and non-adaptive video, we also include bulk in the evaluation, the most common application with MPTCP.

\section{System Design and Implementation}\label{sec:approach}

\begin{figure*}
   \vspace{-3mm}
  \centering
   \includegraphics[width=.905\textwidth]{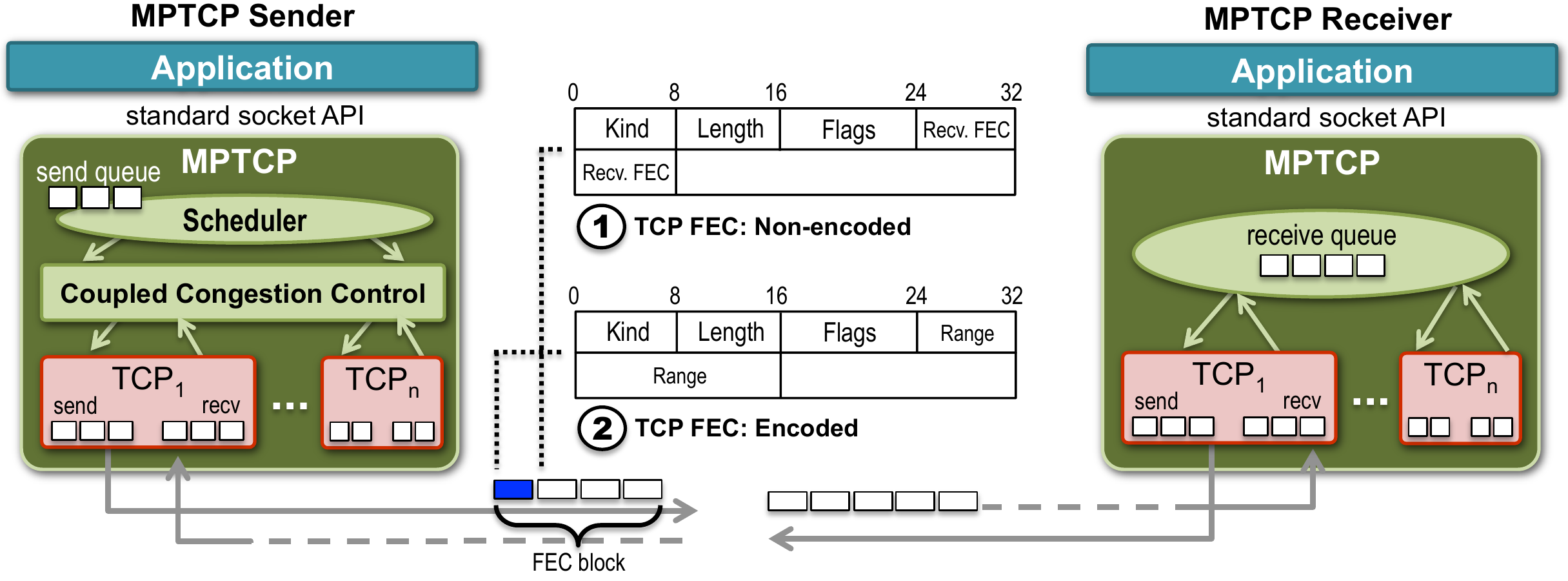}
   \vspace{-2mm}
  \caption{\textbf{System Overview:} During TCP's three-way handshake, the FEC option is negotiated between sender and receiver. Right after it, shown by \ding{192}, the sender includes the FEC option in~\textit{all} subsequent packets of the same connection to be able to discern encoded from non-encoded packets. In \ding{193}, one FEC packet that encodes a certain number of preceding unencoded packets is sent from the sender to the receiver, on the link it is the last packet in what we call a \textit{FEC block}, see Figure~\ref{fig:fec:wire}.}
  \label{fig:system}
   \vspace{-3mm}
\end{figure*}

We illustrate different proposals to integrate FEC within transport protocols in Table~\ref{tab:fec:approaches}. We observe that the majority of these proposals opt for an application layer approach, simplifying deployability at the expense of implementation complexity and maintenance. However, by doing so, they compromise a generic application agnostic scheme as well as sacrifice the benefits of kernel space operations such as high granularity about the connection state, e.g., RTT, flow and congestion controls. Also, some of the proposals use MPTCP solely as a multipath protocol, not taking the underlying subflows' characteristics directly into account inside FEC. 

In this paper, we opt for a pure transport layer XOR-based FEC within TCP to aid MPTCP with heterogeneous networks. Our goal with this design choice is to provide a clearer interface to MPTCP to manage FEC on each of its subflows independently. This is particularly relevant in the presence of heterogeneity, where MPTCP subflows have different delay and loss rates. In other words, we aim at not sacrificing capacity with FEC on low loss subflow, while avoiding HoL-blocking, due to FEC sent on a path with higher delay. 

We illustrate the system building blocks in Figure~\ref{fig:system}.
During TCP's three-way handshake, the FEC option is negotiated between both end-hosts. Afterwards, as depicted in both~\ding{192} and~\ding{193}, the sender include the FEC option in all subsequents packets, marked inside the~\textit{Flags} field, allowing the receiver to distinguish between encoded and non-encoded packets. Likewise, the receiver keeps the same format, signalling inside the~\textit{Flags} field, whether FEC failed to recover or not. As one can see, FEC signalling takes place entirely in the TCP-level at this stage, which raises questions related to deployment, e.g., if FEC options are removed or not successfully negotiated. In this case, the connection is terminated as stated in~\cite{flach00}.

In the remainder of this section, we explain in detail how we departed from TCP-IR towards dynamic FEC (dFEC) and how it is finally integrated into MPTCP.

\subsection{FEC within TCP}\label{subsection:fec:implementation}
Both TCP-TLP and TCP-IR approaches duplicate data at a fixed rate, not respecting TCP's CWND, even though TCP-IR integrates FEC into the congestion control~\cite{flach00}. Also, both focus on web latency, although other latency-sensitive applications can profit from such a mechanism, e.g., video streaming. However, these applications have a different behaviour, e.g., application-limited, bursty or greedy traffic, and must be also taken into evaluation for a generic FEC scheme. Hence, our proposed TCP dynamic FEC (dFEC), similar to TCP-IR, chooses a XOR-based FEC scheme due to its low computational overhead and implementation simplicity. However, TCP-dFEC extends TCP-IR framework in two major ways: \emph{(i) respect CWND} and \emph{(ii) Dynamic FEC adaptation}. TCP-dFEC aims at being fully compatible to TCP's congestion control, application agnostic, adjusting FEC dynamically at run-time.
By adopting a XOR-based FEC, we send FEC systematically every $\textrm{X}$ TCP segments, see Figure~\ref{fig:fec:wire}.
We argue that a pure transport layer FEC implementation is necessary mainly due to two factors: First, due to TCP's small CWNDs in very lossy environments, FEC segments may not be guaranteed every RTTs at times, depending mostly on the FEC adaption rate. Hence, a dynamic FEC is strictly necessary, see Part~\ref{section:dynamic:fec}.
Second, and a particular corner-case: When FEC ratio=10 and packet number \#7 within this block is lost, the receiver will send \texttt{DupACKs} back, as it is expected with TCP, telling the sender that \#7 is missing, triggering a Fast-Retransmission (FR) after three \texttt{DupACKS}\footnote{This TCP's default behaviour to recover before the timer expires. There is however controversy, whether the \texttt{DupACK} threshold should be hard-coded as it is set to 3 in Linux-TCP stacks~\cite{rfc4653} and TCP-RACK~\cite{Cheng2016}.}, whereas, meanwhile, the FEC belonging to this block could have arrived and recovered \#7. As a rule of thumb, in our implementation, the \texttt{DupACK} threshold should be changed to the current $\texttt{FEC block size}$ to avoid early retransmissions\footnote{Note that when TCP triggers a FR, the CWND is reduced, e.g., CWND/2 with a Reno-based congestion control over one RTT, according to TCP's Proportional Rate Reduction (PRR).}. We would like to point out our implementation, with the exception of adjusting the \texttt{DupACK} threshold, does not change how much data is sent into the network, which is originally TCP's congestion control task. The XOR-FEC implementation changes, however, only what is sent in terms of the ratio between data and FEC packets covering the data. Hence, there is no impact on congestion control rather than adapting the \texttt{DupACK} threshold, which has shown to be a source of concern in~\cite{rfc4653,Dukkipati11}.

\subsection{The dynamic FEC (TCP-\MakeLowercase{d}FEC) algorithm}\label{section:dynamic:fec}

\begin{figure*}
  \vspace{-3mm}
  \vspace{-3mm}
  \centering
  \subfigure [\textbf{25 ms}\label{fig:}]{
   \includegraphics[width=.29\textwidth]{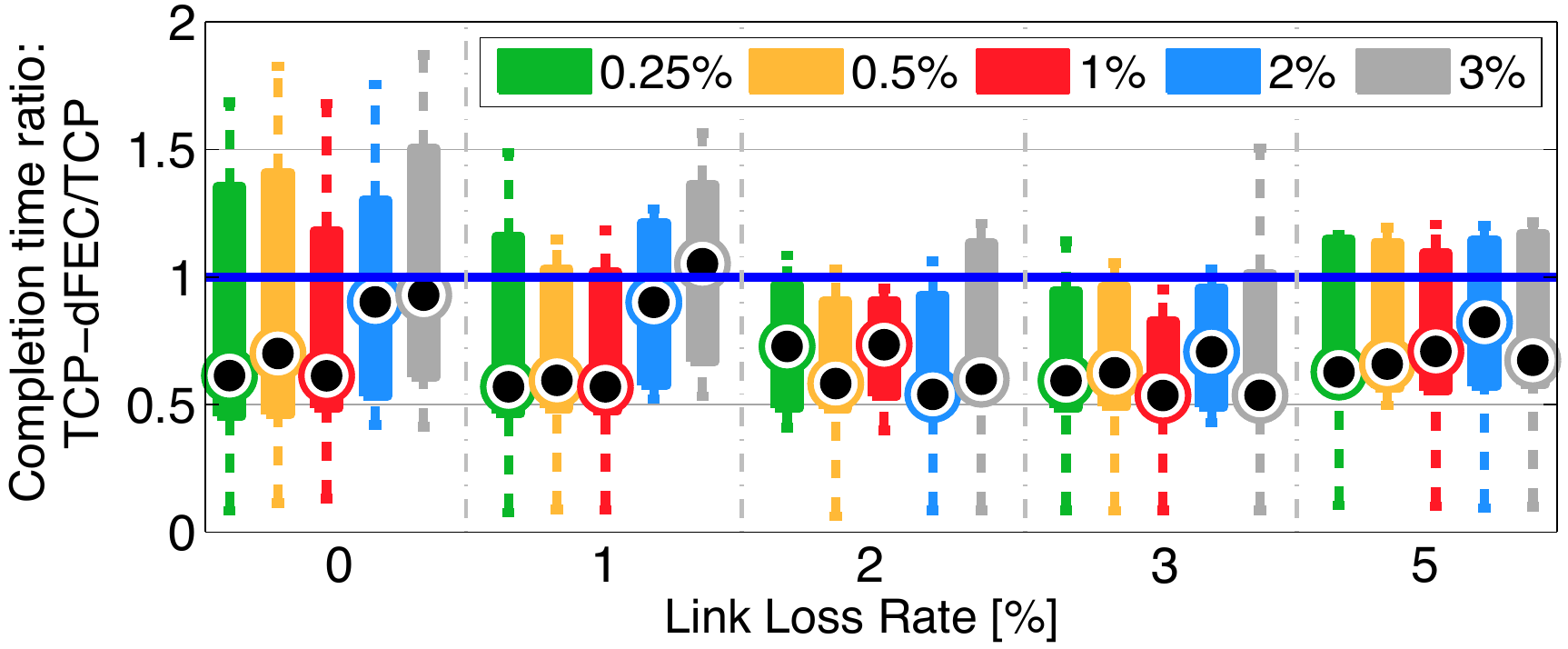}}
  \hspace{3mm}
  \subfigure[\textbf{100 ms}\label{fig:}]
  {\includegraphics[width=.29\textwidth]{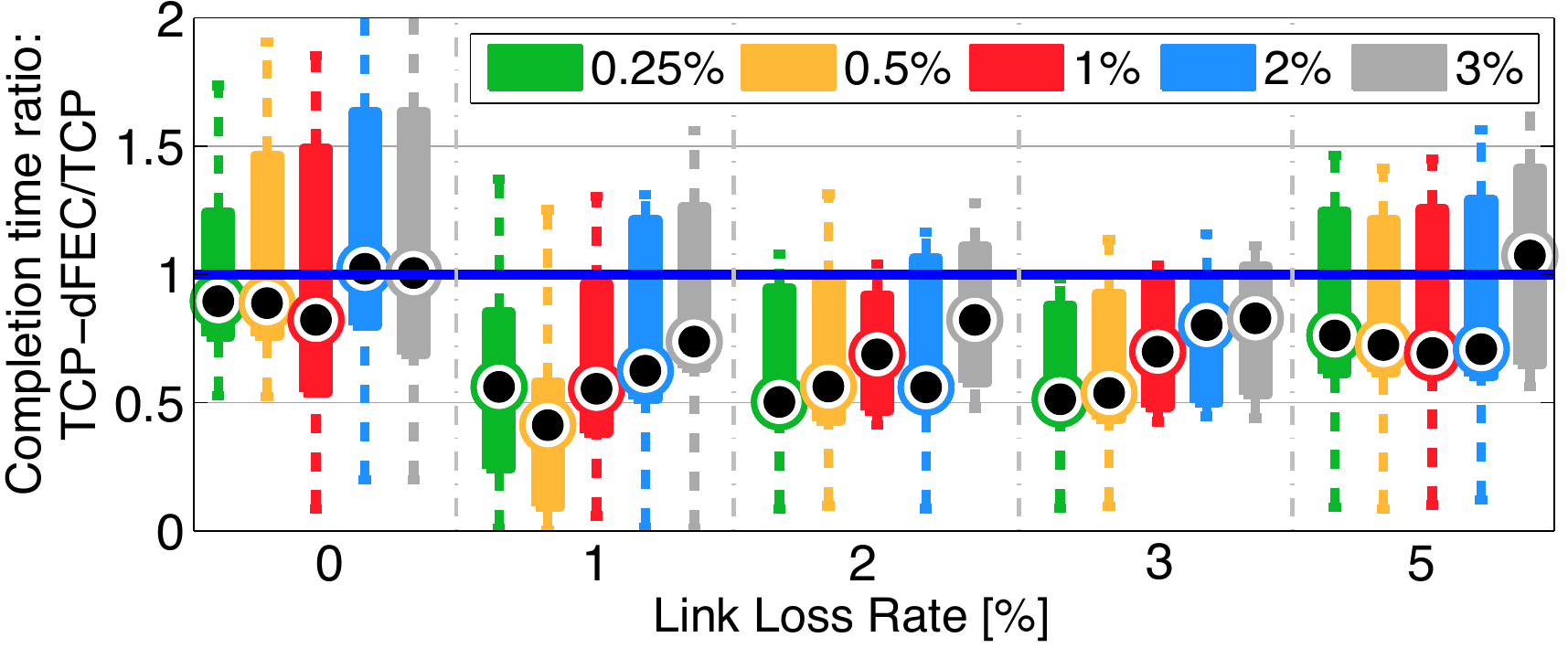}}
  \hspace{3mm}
  \subfigure[\textbf{400 ms}\label{fig:}]
  {\includegraphics[width=.29\textwidth]{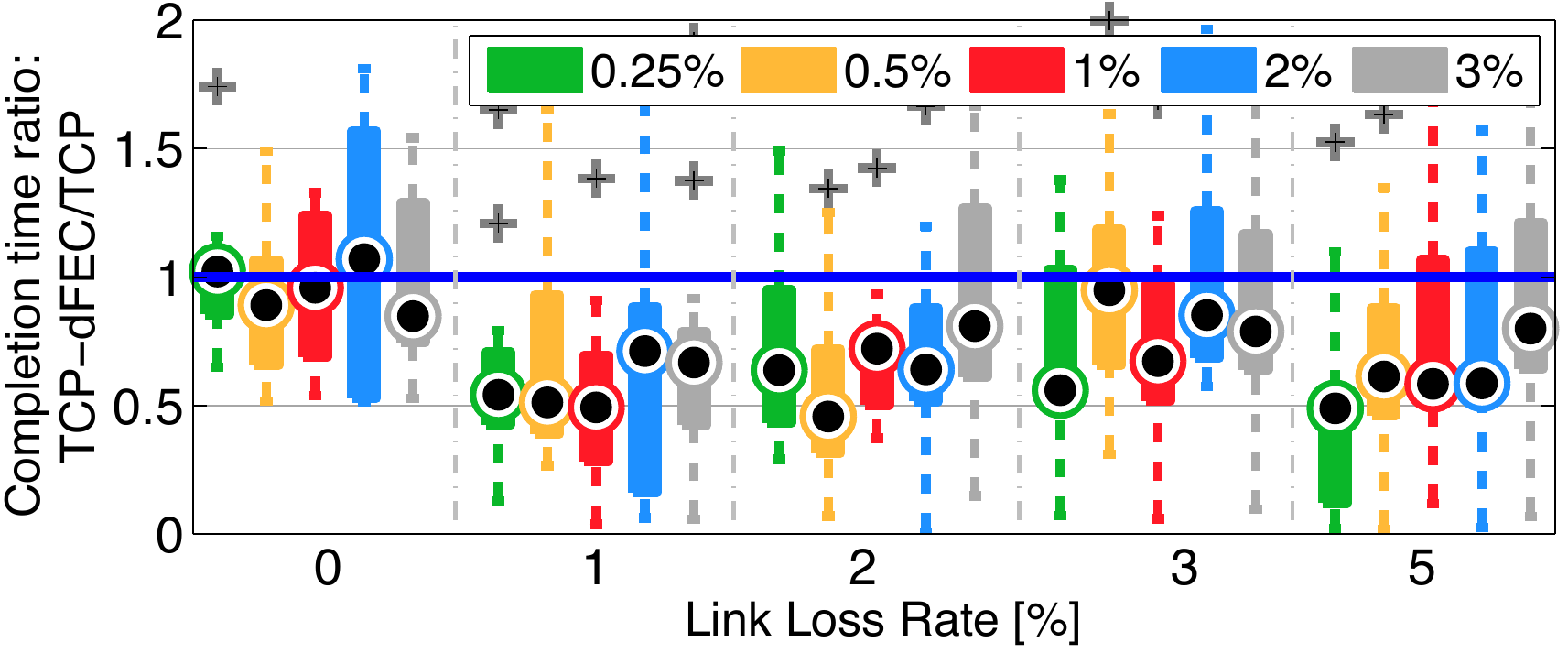}}
  \vspace{-3mm}
  \caption{\textbf{TCP-dFEC vs TCP:} Varying the tolerance between 0.25, 0.5, 1, 2 and 3\% with fixed correction rate $\delta$ = 0.33.}
  \label{fig:fec:dynamic:intermediate:tolerance}
  \vspace{-2mm}
\end{figure*}

\begin{figure*}
  \vspace{-3mm}
  \centering
  \subfigure [\textbf{25 ms}\label{fig:}]{
   \includegraphics[width=.29\textwidth]{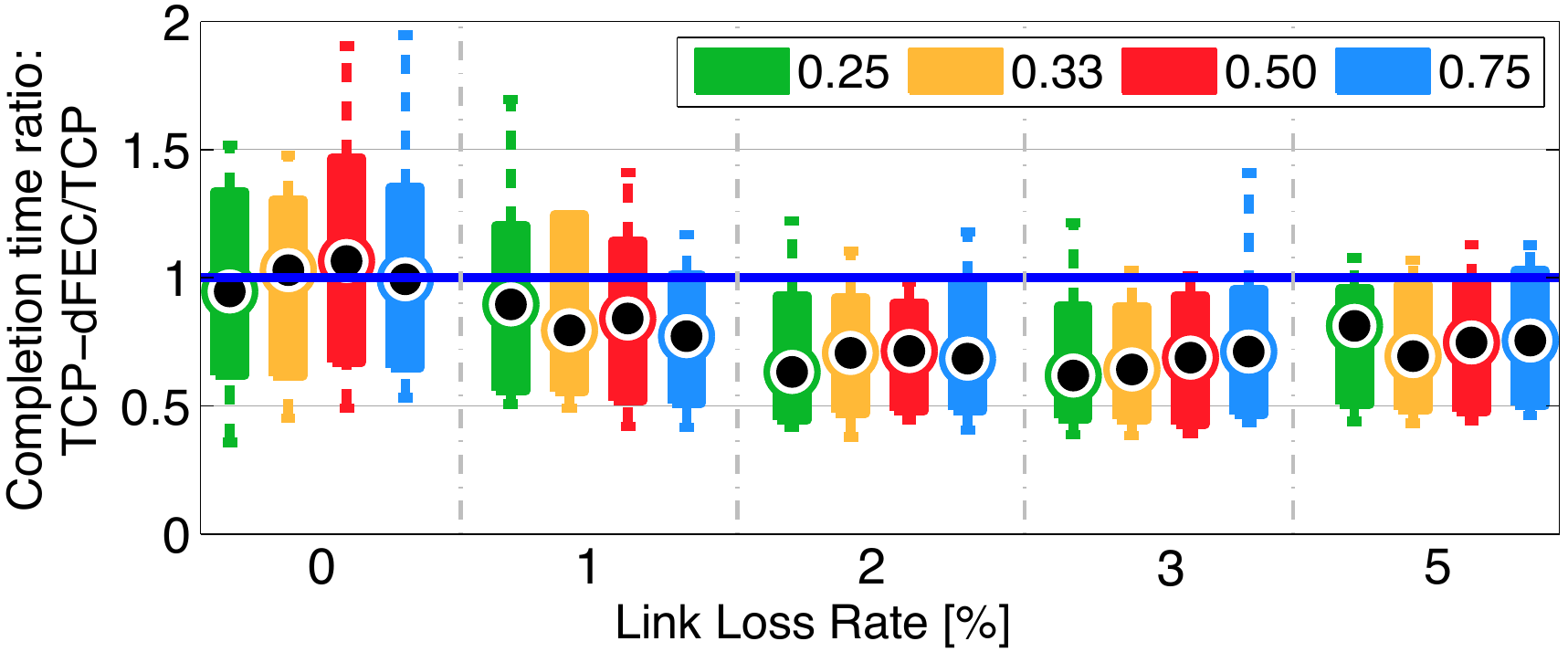}}
  \hspace{3mm}
  \subfigure[\textbf{100 ms}\label{fig:}]
  {\includegraphics[width=.29\textwidth]{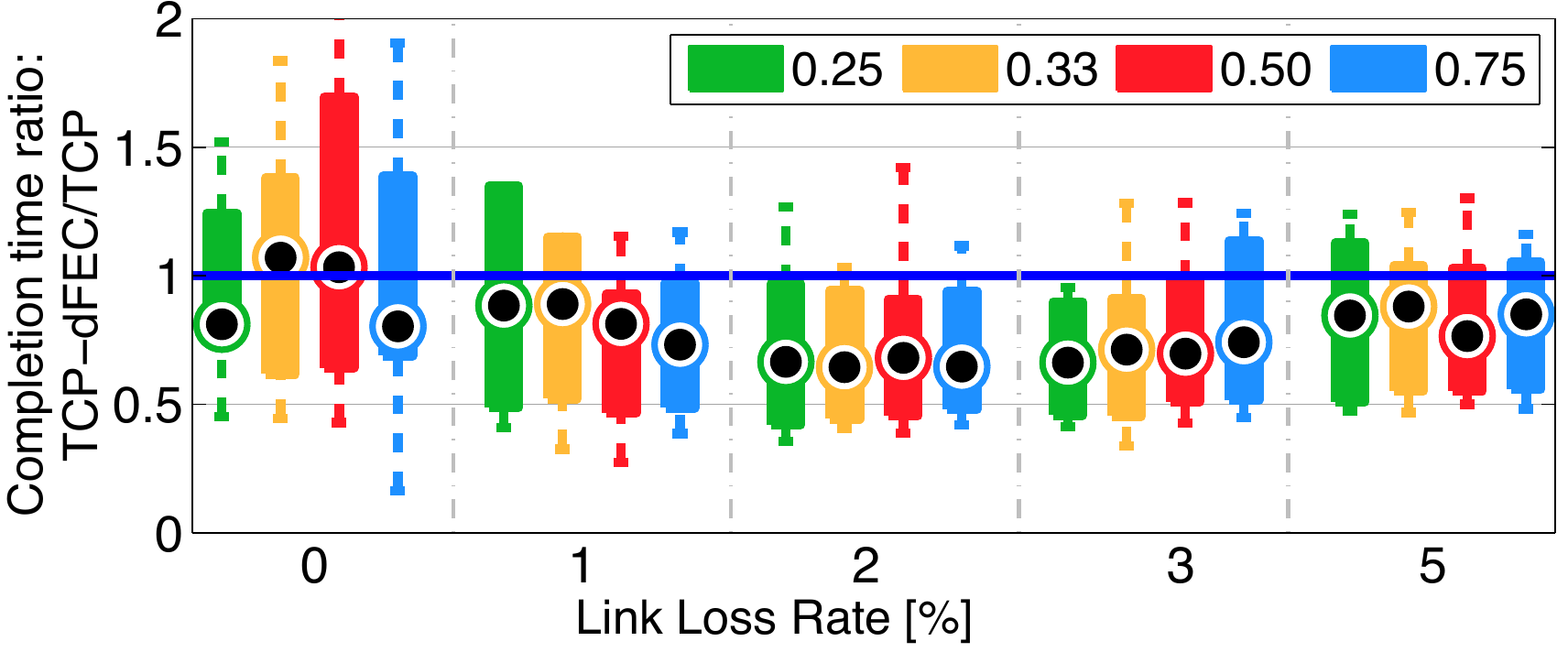}}
  \hspace{3mm}
  \subfigure[\textbf{400 ms}\label{fig:}]
  {\includegraphics[width=.29\textwidth]{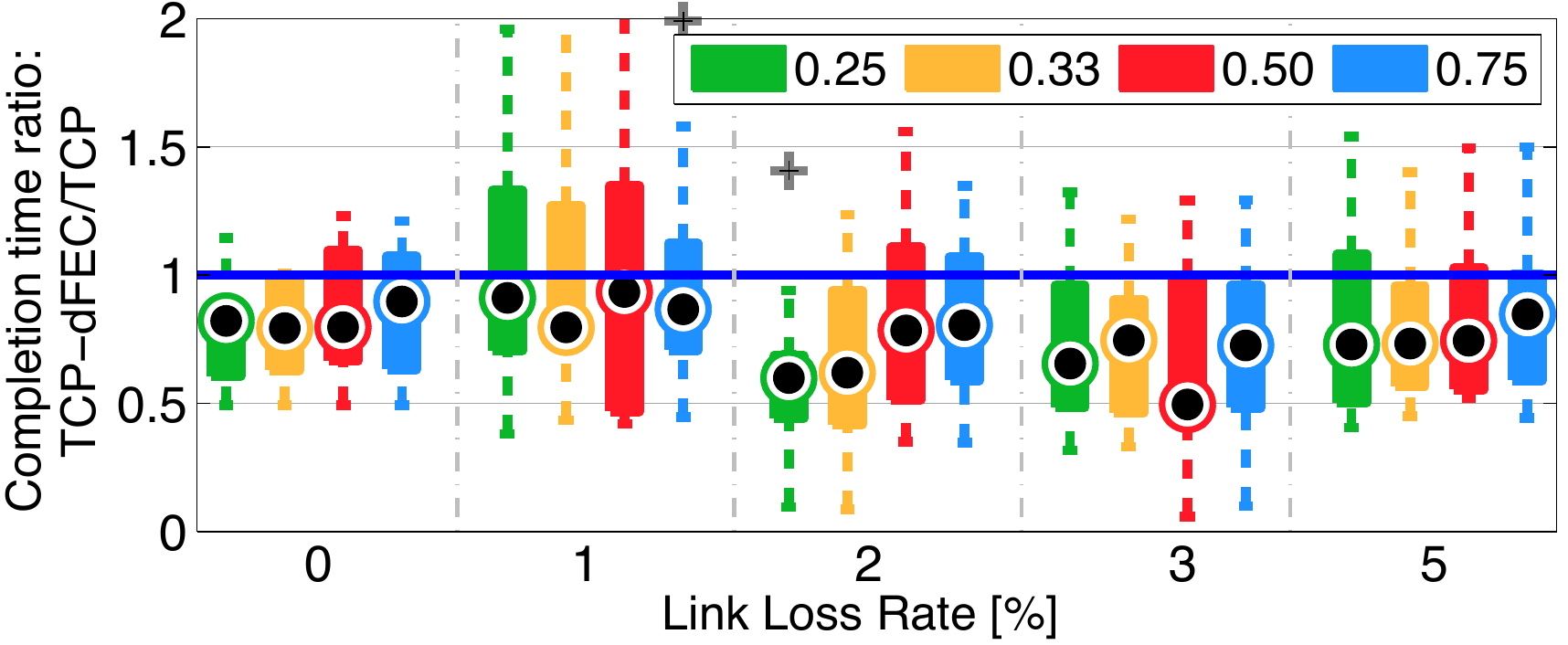}}
  \vspace{-3mm}
  \caption{\textbf{TCP-dFEC vs TCP:} Varying the correction rate $\delta$ between 0.25, 0.33, 0.50 and 0.75 with fixed tolerance = 1\%.}
  \label{fig:fec:dynamic:intermediate:delta}
  \vspace{-5mm}
\end{figure*}

The sender requires three pieces of feedback information to account FEC in TCP's congestion control: \emph{(1) FEC that successfully recovered data}, \emph{(2) FEC that failed recovering data} and \emph{(3) FEC acknowledgements}. The proposed TCP-dFEC, depicted in Figure~\ref{fig:system}, although using the same FEC scheme as TCP-IR due to its low overhead and simplicity, it extends it in two major ways:

\textbf{\emph{(i) Respect CWND:}} The sender tracks FEC packets in order to determine whether they successfully recovered data at the receiver or not. If more than one packet is lost within a block, FEC will then fail. If FEC is successful, then XOR-based FEC provides single loss recovery avoiding retransmissions. This, in turn, provides zero-RTT loss recovery. 

In TCP-IR, both successful and failed FECs are used in the feedback, and integrated into the congestion control. However, it is not clear whether single losses in TCP should be treated in the same way by the congestion control~\cite{Martin2002} in all settings. Hence, in our design, the sender accounts for FEC in the congestion control, i.e., reducing the CWND\footnote{With a FR, the CWND is reduced, e.g., CWND/2 with a Reno-based congestion control over one RTT, according to TCP's Proportional Rate Reduction (PRR).}, only if FEC is lost or fails. Similarly, acknowledged FEC triggers a CWND increase in congestion avoidance.
Furthermore, TCP-dFEC only sends FEC if the CWND has space, remaining compliant to TCP's congestion control. There is, however, an obvious trade-off between the FEC block size, probability of multiple losses within a block and, hence, the chance of FEC to fail.

\textbf{\emph{(ii) Adjust FEC ratio: TCP Dynamic FEC (TCP-dFEC): }}
TCP-IR addresses many beneficial aspects of FEC within TCP, however, they do not specify a ratio between TCP and FEC segments, but rather use a hard-coded approach by sending a FEC packet every 0.25 RTT. We introduce the ability to set after how many TCP segments FEC should be sent, e.g., ratio=4 means that after 4 TCP segments 1 FEC is generated. However, FEC should adapt to link changes and be application agnostic. TCP-dFEC's adaptivity is based on steering~\textit{residual losses}, with residual loss being packets that need retransmission due to FEC failing to recover, hence, triggering TCP's default loss detection and recovery behaviour. Over a period  $\textrm{T}$, as the fraction of retransmitted to first-time transmitted packets, the average~\textit{residual losses} is taken ($\text{Residual}_i$). Then, $\textrm{N}$ of $\textrm{T}$ periods is taken ($\overline{\text{Residual}}$) and compared against a~\textit{target} residual loss rate: If the average link loss rate is higher than the target, the FEC ratio is reduced, otherwise, increased. Then, the algorithm can update the FEC ratio, following the \textit{target}, with a correction rate $\delta$. The residual loss is computed as:

%
%
\addtolength{\jot}{-0.75ex}
\begin{align}
  \text{Residual}_i &= \frac{Retransmit}{Total - Retransmit}
    \intertext{where $i$ identifies a particular Residual Loss measurement over interval $T$, taken from Total and Retransmitted packets. The average residual loss is then computed as:}
  \overline{\textbf{\text{Residual}}} &= \frac{\sum_{n=1}^N \text{Residual}_n}{N}
  \intertext{where $N$ is the average Residual Loss period, where \texttt{target} and $\delta$ are configurable, and determine the tolerance to FEC recovery fail and correction rate, respectively:}
    \begin{split}
    \textbf{if}~~~\overline{\text{\textbf{Residual}}} > \texttt{target}~~\textbf{then} ~~~~~~~~~~ \\
    \text{ratio'} = \text{ratio}  \times (1 - \delta) ~~~~~~~~~~\\
    \textbf{else}~~~~~~~~~~~~~~~~~~~~~~~~~~~~~~~~~~~~~~~~~~~ \\
    \text{ratio'} = \text{ratio}  \times (1 + \delta) ~~~~~~~~~~\\
    \textbf{end if}~~~~~~~~~~~~~~~~~~~~~~~~~~~~~~~~~~~~~~~~ \\
  \end{split}
\end{align}
We choose $\textrm{T}$ = 3 RTTs as a minimal period during which we can capture how TCP recovers with loss: If during one RTT a loss occurs, retransmissions will be performed during the second RTT and, possibly, concluded during the third. 

With a start FEC ratio=9, $\textrm{N}$ = 2 and $\delta$ = 0.33, the algorithm includes one FEC in TCP's Initial Window (IW), updating FEC in short $\textrm{N}$ intervals at $\delta$ rate. On low-loss links, we expect FEC block to grow quickly reducing overhead, while on high-loss links FEC block will oscillate between low values\footnote{We evaluated the algorithm with several $\textrm{N}$ and $\delta$, e.g., $\delta$ = 0.25 and 0.50, where $\delta$ = 0.33 yield best results.}. Note that we restrict the ratio to be not smaller than 4, which corresponds to a maximum overhead of 20\%. We also enforce an upper bound of 256, hence, limiting the amount of buffering at the receiver. Both values can be, however, set by the user.

\begin{figure}[h!]
   \vspace{-3mm}
  \centering
   \includegraphics[width=.415\textwidth]{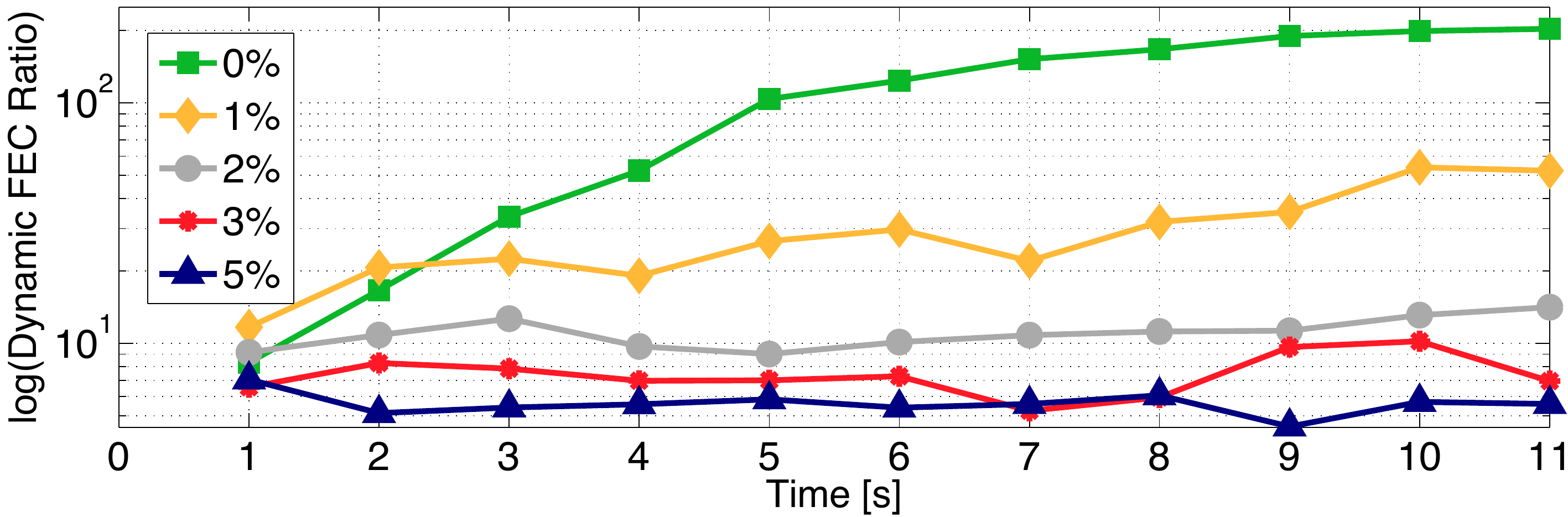}
   \vspace{-2mm}
  \caption{\textbf{TCP-dFEC: }10~s of a TCP bulk with RTT~25ms and loss between 0, 1, 2, 3 or 5\%, see~\ding{174} in Section~\ref{subsection:fec:implementation}.}
  \label{fig:fec:dynamic}
   \vspace{-3mm}
\end{figure}

The TCP-dFEC's adaptation rate depends on RTT, i.e., the adaptation rate is slower in connections with higher RTT. For short flows, this might be suboptimal, and, as a remedy, end-hosts could cache the FEC ratio per connection or per interface, just like TCP does with \texttt{ssthresh}\footnote{As a rule of thumb: FEC should be approximately twice the amount of the average link loss: 2\% FEC for 1 to 2\% random loss.}. In Figures~\ref{fig:fec:dynamic:intermediate:tolerance}, ~\ref{fig:fec:dynamic:intermediate:delta} and~\ref{fig:fec:dynamic} we show preliminary TCP-dFEC results with NewReno, and due to its simplicity to understand, run in the measurement setup described in Section~\ref{sec:measurement:setup}, including background traffic. The same way we have later run our emulation experiments.

Figure~\ref{fig:fec:dynamic} shows the first 10~s of a TCP bulk transfer with 25~ms RTT and how the FEC ratio changes over time for different link losses. Figure~\ref{fig:fec:dynamic:intermediate:tolerance} shows TCP-dFEC with different tolerance values, which indicate the FEC fail occurrences (\%) before the FEC ratio is changed. There, we show tolerance values between 0.2, 0.5, 1, 2 and 3\% with a fixed $\delta$, with $\delta$ being the correction rate for the FEC ratio in the next RTT, i.e., the lower the value the milder the correction. One can observe that \textit{tighter} tolerance values up to 1\%, although mild, yield in general better results, regardless of the RTT or link loss rates. Figure~\ref{fig:fec:dynamic:intermediate:delta} shows preliminary results keeping tolerance fixed at 1\%, but varying $\delta$ between 0.25, 0.33, 0.50 and 0.75.

\begin{figure}
  \vspace{-6mm}
  \centering
  
  \subfigure[\textbf{Random loss 3\%}\label{fig:tcp:dynamic:3random}]{
   \includegraphics[width=0.34\textwidth]{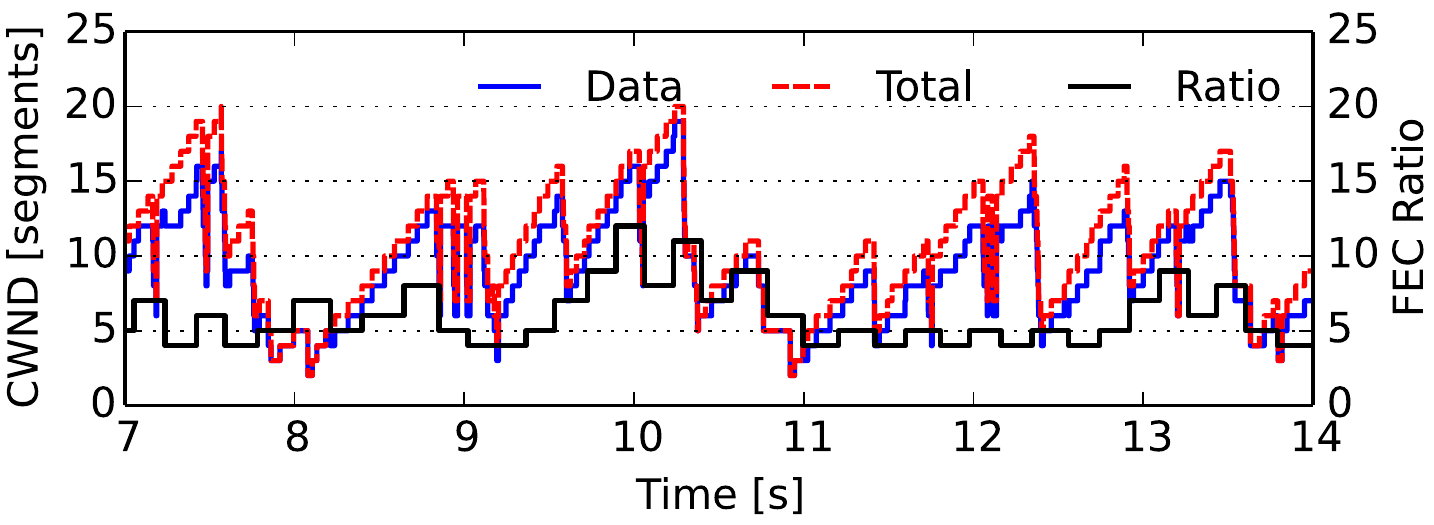}}
  
    \vspace{-3mm}

  \subfigure[\textbf{Gilbert-Elliot 3\%, average burst size: 2 packets}\label{fig:tcp:dynamic:3random:burst2}]{
   \includegraphics[width=0.34\textwidth]{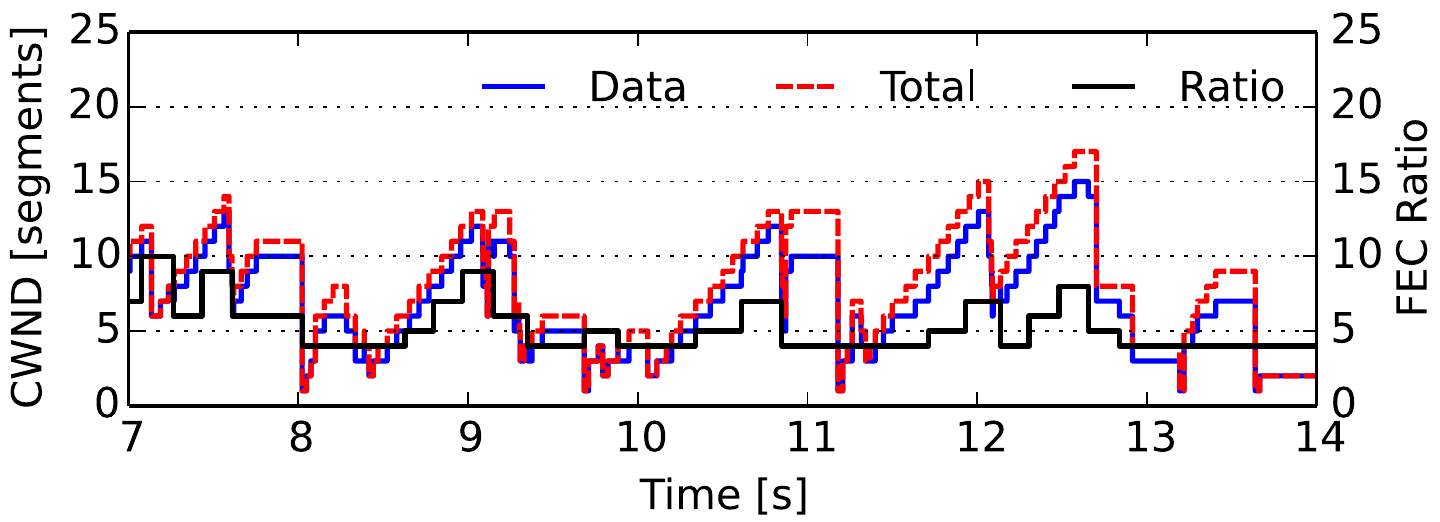}}
   
      \vspace{-2mm}

    \caption{\textbf{TCP-dFEC: }CWND evolution and its respective FEC ratio with 25 ms RTT, 3\% random loss and Gilbert-Elliot loss model~\cite{hasslinger08} with 3\% loss rate and average burst size of 2.\label{fig:fec:dynamic:cwnd:ratio}}
    \vspace{-3mm}
\end{figure}

\subsubsection{\textbf{TCP-\MakeLowercase{d}FEC adaptation}}\label{}Figure~\ref{fig:fec:dynamic:cwnd:ratio} shows a snapshot over 15s of the CWND evolution accompanied by the respective FEC ratio stepwise adaptation for 3\% loss rate and we included some preliminary results with Gilbert-Elliot burst loss model~\cite{hasslinger08} with 3\% loss rate and average burst size of 2 packets measured in the scenario described in Section~\ref{sec:measurement:setup}. Herewith, we would like to illustrate how the FEC ratio stably adapts over time with different injected loss rates and pattern, i.e., random and burst. We use these results to demonstrate the stability of the FEC adaptation algorithm, also because it is more controlled compared to real experiments. We would like to point out in Figure~\ref{fig:fec:dynamic:cwnd:ratio} that the CWND increase is not altered with TCP-dFEC, rather the amount of FEC over a new set of new packets to be sent, forming a FEC block size, is changed according to the calculate FEC ratio. For the burst loss Figure~\ref{fig:tcp:dynamic:3random:burst2} the CWNDs are slightly smaller compared to Figure~\ref{fig:tcp:dynamic:3random}, where we also observe lower FEC ratios, hovering above its minimum of 1:4.

\subsubsection{\textbf{TCP-\MakeLowercase{d}FEC fairness}}\label{}
\begin{figure}[h!]
  \vspace{-4mm}
  \centering
  
  \subfigure[25 ms]{
       \includegraphics[height=.165\textwidth]{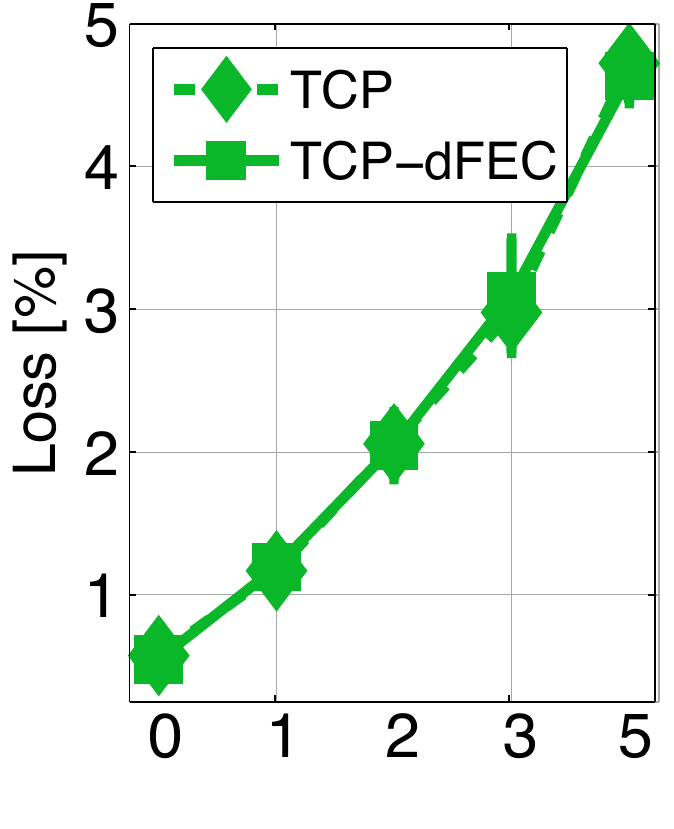}\label{fig:fec:tcp:fairness:25ms}}
  \hspace{2mm}
  \subfigure[100 ms]{
       \includegraphics[height=.165\textwidth]{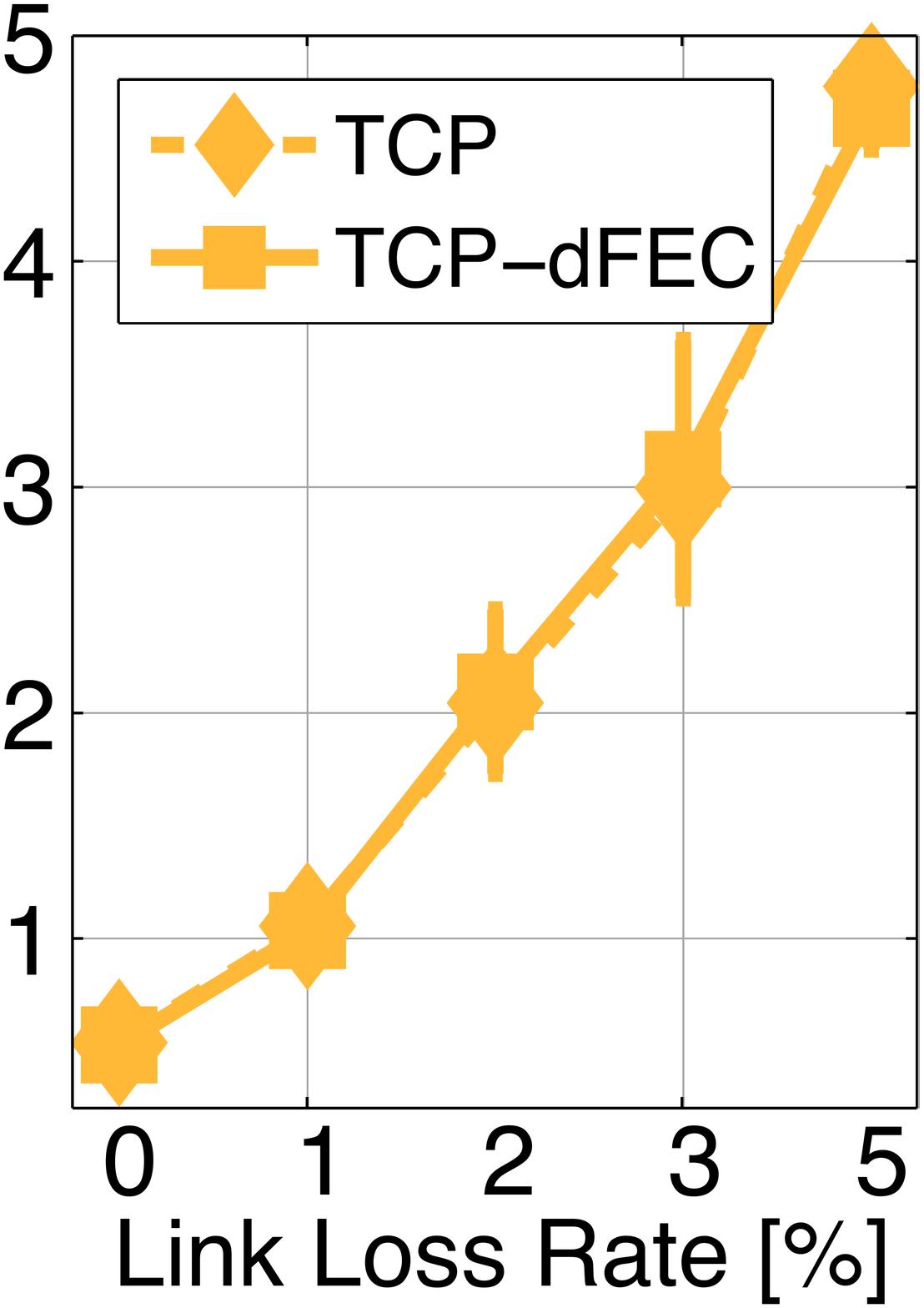}\label{fig:fec:tcp:fairness:100ms}}
  \hspace{2mm}
  \subfigure[400 ms]{
       \includegraphics[height=.165\textwidth]{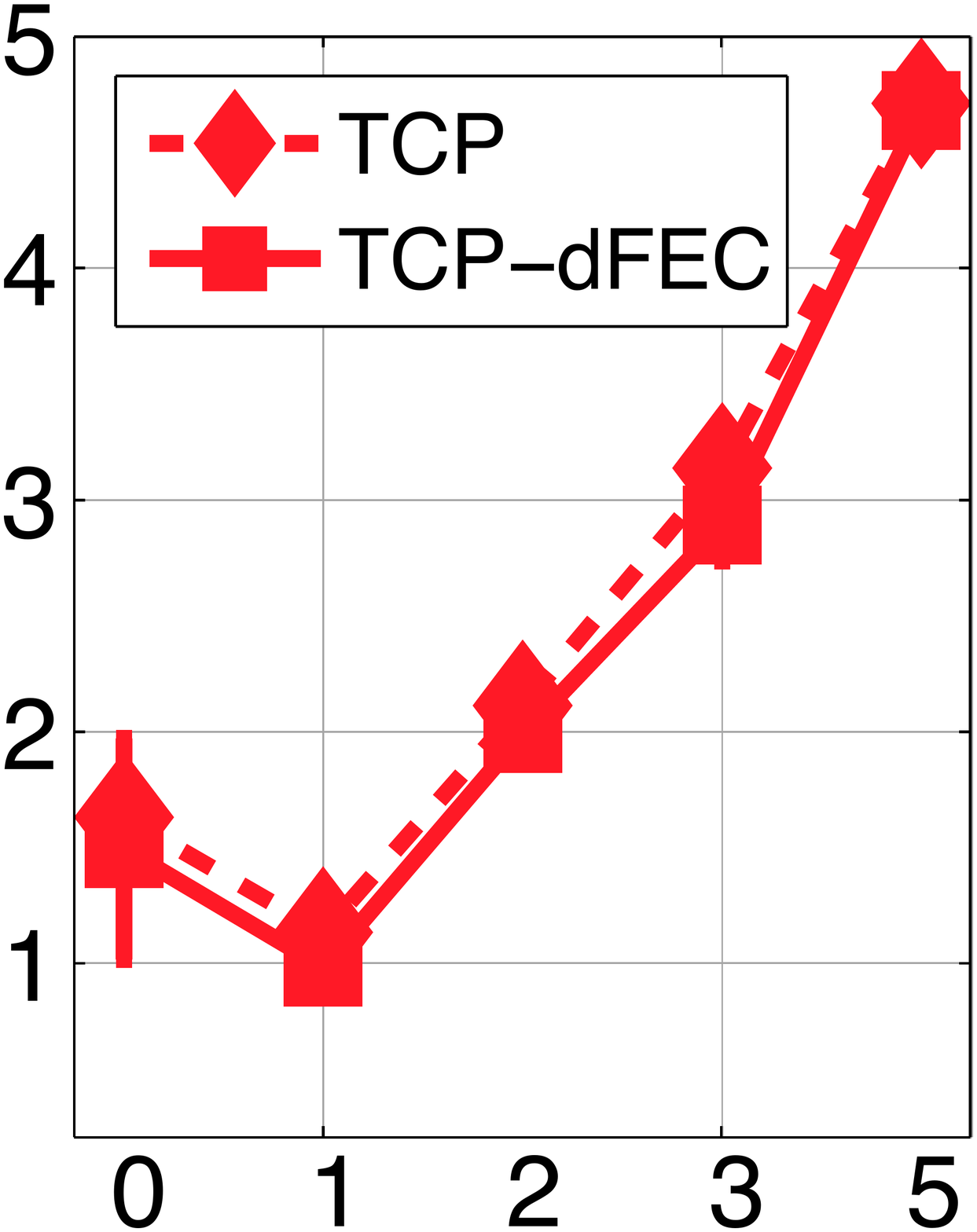}\label{fig:fec:tcp:fairness:400ms}}

    \vspace{-3mm}   

      \subfigure[25 ms]{
       \includegraphics[height=.165\textwidth]{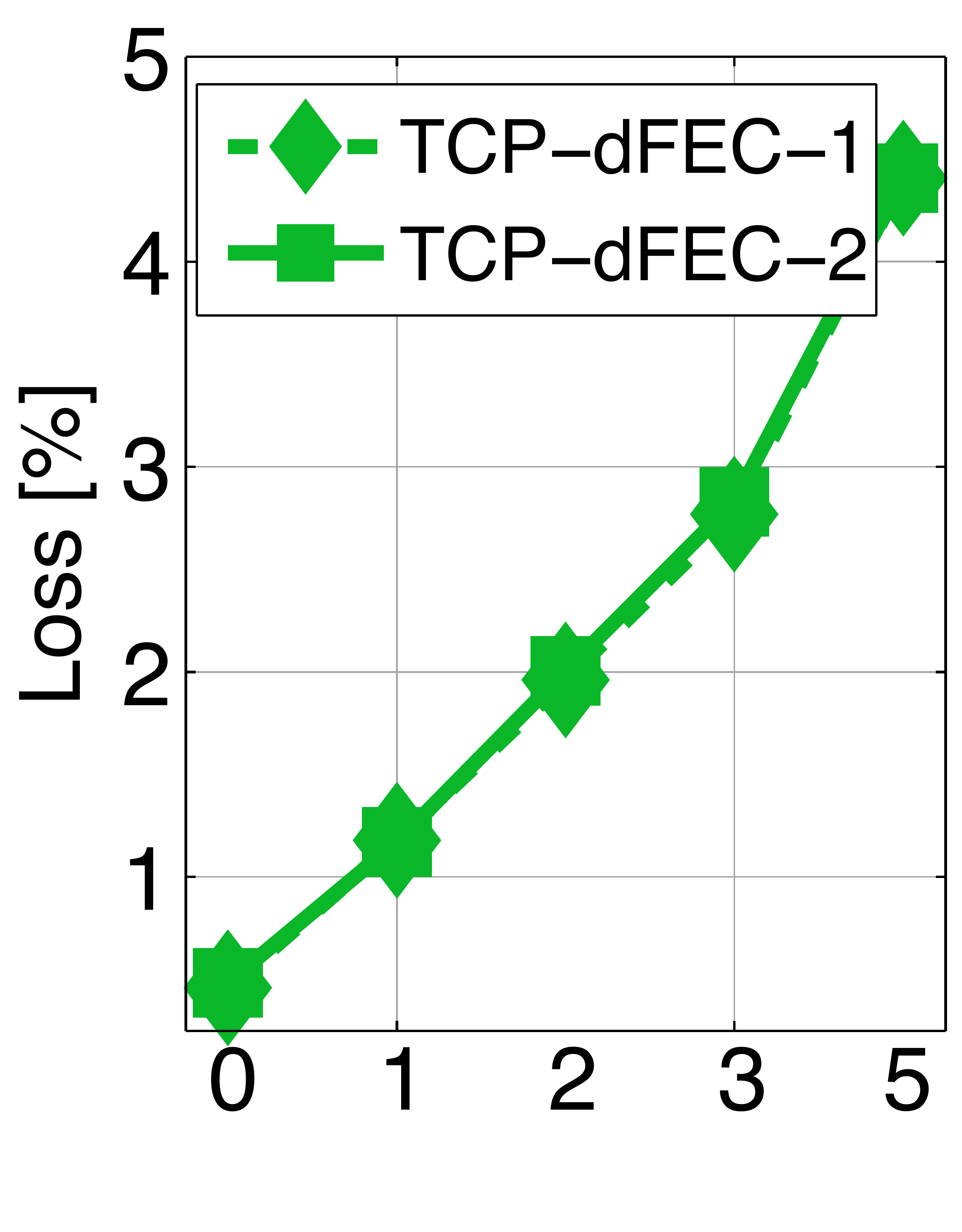}\label{fig:fec:tcpdfec:fairness:25ms}}
  \hspace{2mm}
  \subfigure[100 ms]{
       \includegraphics[height=.165\textwidth]{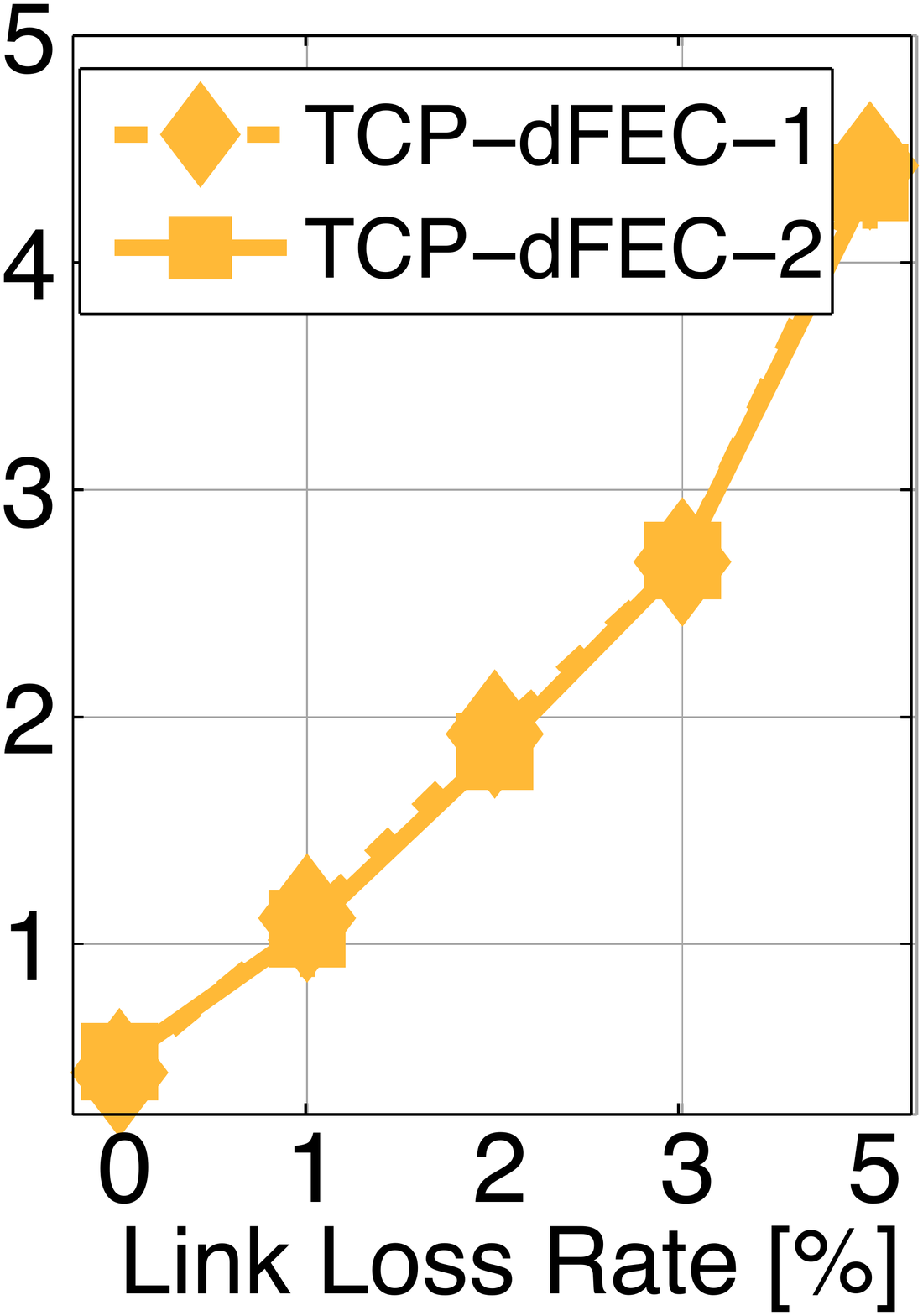}\label{fig:fec:tcpdfec:fairness:100ms}}
  \hspace{2mm}
  \subfigure[400 ms]{
       \includegraphics[height=.165\textwidth]{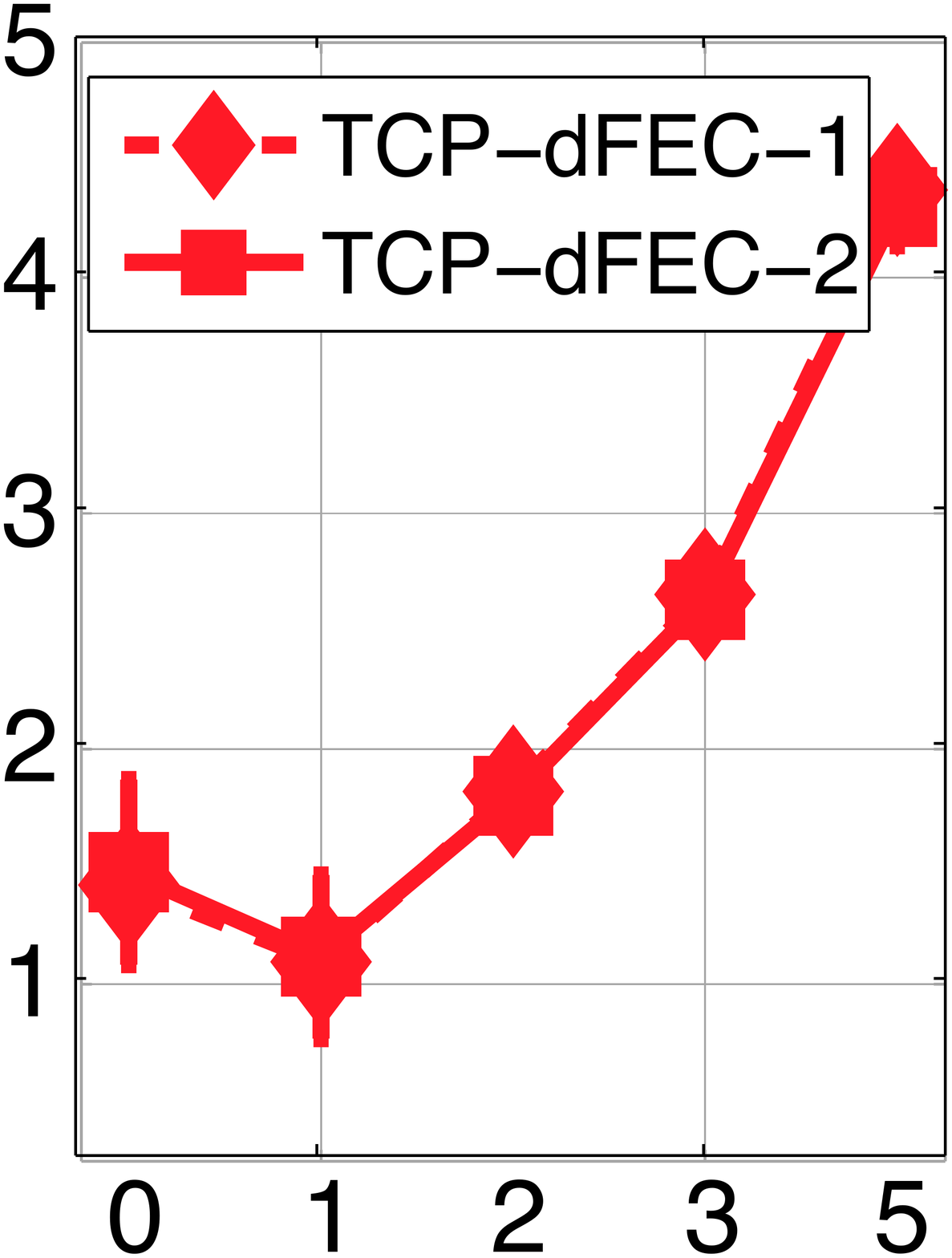}\label{fig:fec:tcpdfec:fairness:400ms}}
    \vspace{-3mm}   
  \caption{\textbf{TCP-dFEC fairness to TCP: }Figures~\ref{fig:fec:tcp:fairness:25ms},~\ref{fig:fec:tcp:fairness:100ms} and~\ref{fig:fec:tcp:fairness:400ms} and \textbf{TCP-dFEC fairness to TCP-dFEC: }Figures~\ref{fig:fec:tcpdfec:fairness:25ms},~\ref{fig:fec:tcpdfec:fairness:100ms} and~\ref{fig:fec:tcpdfec:fairness:400ms} with 0, 1, 2, 3 or 5\% injected losses and RTTs between 25, 100 and 400 ms.}
  \label{fig:tcpdfec:fairness}
    \vspace{-3mm}
\end{figure}

Finally, we also consider fairness in the bottleneck against regular TCP and against TCP-dFEC itself, because of how dFEC is implemented into TCP's congestion control might be interpreted as "loss masking" by QUIC~\cite{hamilton16}. However, Figure~\ref{fig:tcpdfec:fairness} shows that TCP-dFEC does not introduce losses on a concurrent TCP and TCP-dFEC, running against further TCP bulk flows in the bottleneck. 

The loss amount ratio in both experiments in Figure~\ref{fig:tcpdfec:fairness} is bound, regardless of the injected loss \% and the FEC block sizes. Remember that values less than 1 are in favour of TCP-dFEC. We relate these results to the FEC algorithm congestion control management in Section~\ref{subsection:fec:implementation}, part~\ding{172}.
 
\subsection{Dynamic FEC and MPTCP}\label{section:dynamic:fec:mptcp}
To finally achieve our goal to integrate dynamic FEC into MPTCP, the XORed packets on the TCP level, i.e., subflow level, need to be mapped onto MPTCP's connection level signalling and management. In MPTCP, data is multiplexed on all subflows belonging to the same MPTCP connection according to mainly the scheduler, but also the couple congestion control, e.g., via load balancing, however, at the receiver, data on the different subflows need to be reconstructed in the MPTCP level, before the application can read it. This is achieved through MPTCP's Data Sequence Signal (DSS), which in a non-FEC connection, normally maps subflow data directly to connection-level window. Therefore, in a FEC connection, relevant parts of the DSS option had to be also XORed so that the receiver can reconstruct the data and packet are not dropped on the MPTCP level for this reason.

The mapping choice to apply FEC on the subflow level rather than direct on the MPTCP (connection-level) is dictate by some reasons: Firstly, we wanted to guarantee compatibility and seamlessly operation to TCP, regardless of a single path or a multipath connection with or without FEC. Secondly, adding FEC directly to MPTCP would require major integration with MPTCP's scheduler and congestion control. Finally, and perhaps the strongest argument at this stage, we were focusing on equilibrating heterogeneity from the subflows for the multipath connection, and carrying FEC on a well-performing subflow, e.g. lower latency and loss rates, to compensate for other subflows' performace, e.g. with higher loss rates, could dismiss the advantage of FEC.

\textbf{Summary: }This section presented dFEC, a \textit{dynamic} FEC for TCP, which is application agnostic and adapts FEC dynamically according to the network condition. We also described how the algorithm extends TCP-IR and integrates into MPTCP to aid multipath transport with heterogeneous networks.

\section{Measurement Setup}\label{sec:measurement:setup}
Throughout this section, we explain our experiment setup, with the respective network settings and the applications.

\subsection{Experiment Setup}
\begin{figure}[h!]
  \vspace{-3mm}
\centering
    \includegraphics[width=0.65\columnwidth]{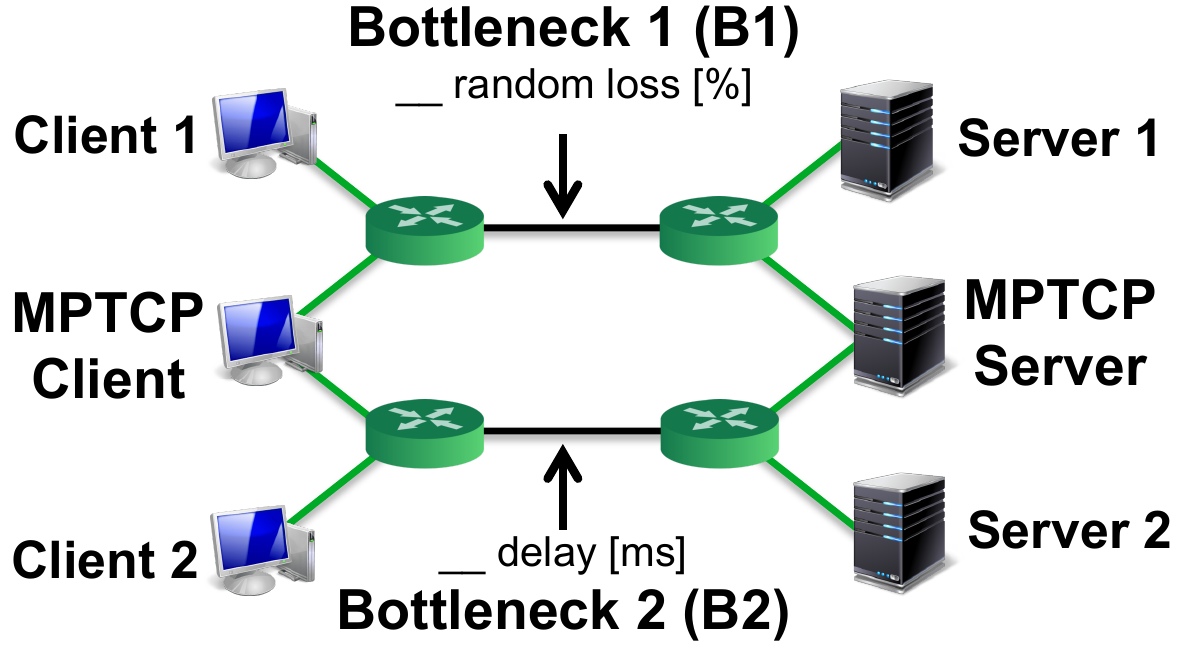}
    \label{fig:nsb}
\caption{Non-shared bottleneck emulation scenario}
\label{fig:emulation:nsb}
  \vspace{-3mm}
\end{figure}

We use CORE network emulator~\cite{Ahrenholz2008}, which enables the use of real protocols and applications with emulated network links, making the evaluation easy to control and replicate. We use the MPTCP v0.90 Linux kernel implementation, so this setup also allowed us to use most of the features\footnote{Linux MPTCP: \url{http://www.multipath-tcp.org}}. We use the default options of MPTCP, including e.g.\ receive buffer optimisation, and the socket buffer size adjustment to improve MPTCP's aggregation as suggested in~\cite{paasch14}. To guarantee independence between experiments, we flushed all TCP-related cached metrics after each run. The network characteristics are shown in Table~\ref{tab:exp:parameters} and the topology is illustrated in Figure~\ref{fig:emulation:nsb}.

\begin{table}[h!]
 \vspace{-2mm}
  \centering
  \caption{Emulation Network Characteristics.}
  \label{tab:exp:parameters}
    \begin{tabular}{lccc}
      \toprule
      \multicolumn{1}{c}{\textbf{}} & \textbf{WLAN} & \textbf{3G/4G}  & \textbf{Satellite} \\
      \midrule
      Capacity [Mbps]          & 20          & 5 -- 10   & 0.5 -- 1.5 \\
      End-to-end delay [ms]    & 20 -- 30    & 50 -- 85  & 250 -- 500 \\
      Loss [\%]                & 0 -- 5      & 0  & 5 -- 10 \\
      \bottomrule
    \end{tabular}
   \vspace{-2mm}
\end{table}

To create a more realistic emulation environment, the experiments are run with background traffic modelled as a synthetic mix of TCP and UDP generated with D-ITG~\cite{BottaDP12}.The TCP traffic is composed of greedy and rate-limited TCP flows with exponential distributed mean rates of 150\,pps. The UDP traffic was composed of flows with exponentially distributed mean rates varying between 395 and 995\,pps and Pareto distributed on and exponentially distributed off times with on/off intervals between 1\,s and 5\,s. The UDP and TCP generated flows have packet sizes with a mean of 1000\,Bytes and RTT between 25 and 1000\,ms. Note that, Bottlenecks $1$ and $2$ have different capacities and the UDP background load was adjusted accordingly. Also, for the video experiments with H.264, since both video files are encoded at 3.4 Mibps and maximum of 4 Mibps, we also increased the UDP background traffic load to keep the congestion levels comparable to the other experiments. We would like to motivate the configuration of the background traffic to control the load as well as the~\textit{burstiness} in the bottleneck. The choice of the traffic distribution is based on earlier studies we have made and followed~\cite{hayes14, hayes14_LCN}. In general terms, the bottleneck is loaded with bulk TCP flows and oscillations are caused with few bursty UDP flows with different average rates and distribution, which also represent popular Internet applications.

\subsection{Applications}
We perform experiments in a static scenario evaluating bulk, non-adaptive video with H.264 and~\texttt{ffmpeg} and web traffic via HTTP/2 with real application traffic, see Table~\ref{tab:web-profile}. 

For the non-adaptive H.264 streaming, we use \texttt{ffmpeg}\footnote{https://ffmpeg.org/ffplay.htm} with one minute of the~\textit{Big Buck Bunny} video H.264 encoded at 3.4 MiBps with 25 Frames Per Second (FPS). For the HTTP/2 experiments, Table~\ref{tab:web-profile} shows the websites with their corresponding number of objects and total transferred size in KiByte, where these are downloaded using a combination of different tools, such as \texttt{nghttp2}.

\begin{table}[h!]
 \vspace{-2mm}
  \centering
  \caption{Web Traffic Generation.}
  \label{tab:web-profile}
    \begin{tabular}{lcc}
      \toprule
      \multicolumn{1}{c}{\textbf{Domain name}} & \textbf{Objects} & \textbf{Total Transfer Size} \\
      \midrule
      \url{http://www.google.com}              & 6      & 1,080 KiB \\ 
      \url{http://www.youtube.com}             & 26     & 3,204 KiB  \\
      \url{http://www.espn.go.com}             & 111    & 6,072 KiB \\
      \bottomrule
    \end{tabular}
   \vspace{-2mm}
\end{table}

\subsection{Experiment Configuration}
To emulate a multipath scenario in the topology shown in Figure~\ref{fig:emulation:nsb}, we select a list of different path Bandwidth-Delay Products (BDP) and loss rates, mimicking different networks, such as cellular, WLAN and satellite. Following the settings from Table~\ref{tab:emulation:parameters}, we keep $B1$'s RTT fixed at 25~ms, varying the loss rate between 0, 1, 2, 3 and 5\%, while, in $B2$, only the RTT is changed to 25, 100 and 400~ms\footnote{Although there is evidence that LTE networks maintain buffer sizes larger than the path's BDP, associated to the effect known as~\textit{bufferbloat}~\cite{jiang12}, we adjusted the buffers to be the value of one BDP in our experiments.}. Hence, the scenarios under evaluation can be read as: 1) Loss heterogeneity, e.g., $B1$'s RTT is 25~ms but loss rate >~0\% relative to $B2$ and 2) loss and RTT heterogeneity when $B1$'s loss rate~> 0\% and $B2$'s RTT~> 25~ms, e.g., $B1$'s RTT is 25~ms and loss rate 1, 2, 3 or 5\% and $B2$'s RTT is 100 or 400~ms.
A more comprehensive summary is shown in Table~\ref{tab:emulation:parameters}.

\begin{table}[b]
    \vspace{-2mm}
  \centering
	\caption{Bottlenecks ($B1$ and $B2$) link capacity, RTT and average link loss (\%), see Figure~\ref{fig:emulation:nsb}.}
	\label{tab:emulation:parameters}
    \resizebox{\columnwidth}{!}{%
    \begin{tabular}{ccc}
      \toprule
 \textbf{\shortstack{Capacity \\ B1 and B2 [Mibps]}} & \textbf{\shortstack{RTT \\ B1 and B2 [ms]}} & \textbf{\shortstack{Loss \\ B1 and B2 [\%]}} \\
      \midrule
                                   & 25 and 25    &             \\
                           20 and 10 & 25 and 100   &    0, 1, 2, 3, 5 and 0     \\
                                   & 25 and 400   &             \\

      \bottomrule
    \end{tabular}
    }
        \vspace{-2mm}
\end{table}



\section{Evaluation and Discussion} \label{sec:evaluation}
This section is dedicated to present our results from the implementation described in Section~\ref{sec:approach} with all settings from Section~\ref{sec:measurement:setup}, starting with TCP and following with MPTCP.

\subsection{Dynamic FEC and TCP}
In Section~\ref{subsubsec:dfec:ir}, we show some of the results from~\cite{ferlin16_conext} of the original TCP-IR algorithm compared to regular TCP with bulk and web traffic, and then we compare TCP-dFEC to regular TCP with bulk, H.264 and HTTP/2 in Section~\ref{subsubsec:dfec:tcp}. Note that the following results evaluate TCP-dFEC and TCP-IR under the same conditions, using the same measurement setup from Section~\ref{sec:measurement:setup} and parameters from Table~\ref{tab:exp:parameters}. For more discussion and evaluation results about TCP-IR with web transfer of different sizes can be seen in~\cite{ferlin16_conext}.

\subsubsection{TCP-IR vs TCP}\label{subsubsec:dfec:ir}
\begin{figure}[h!]
  \vspace{-4mm}
  \centering
  \subfigure[\textbf{TCP-IR: }Google]{
       \includegraphics[height=.153\textwidth]{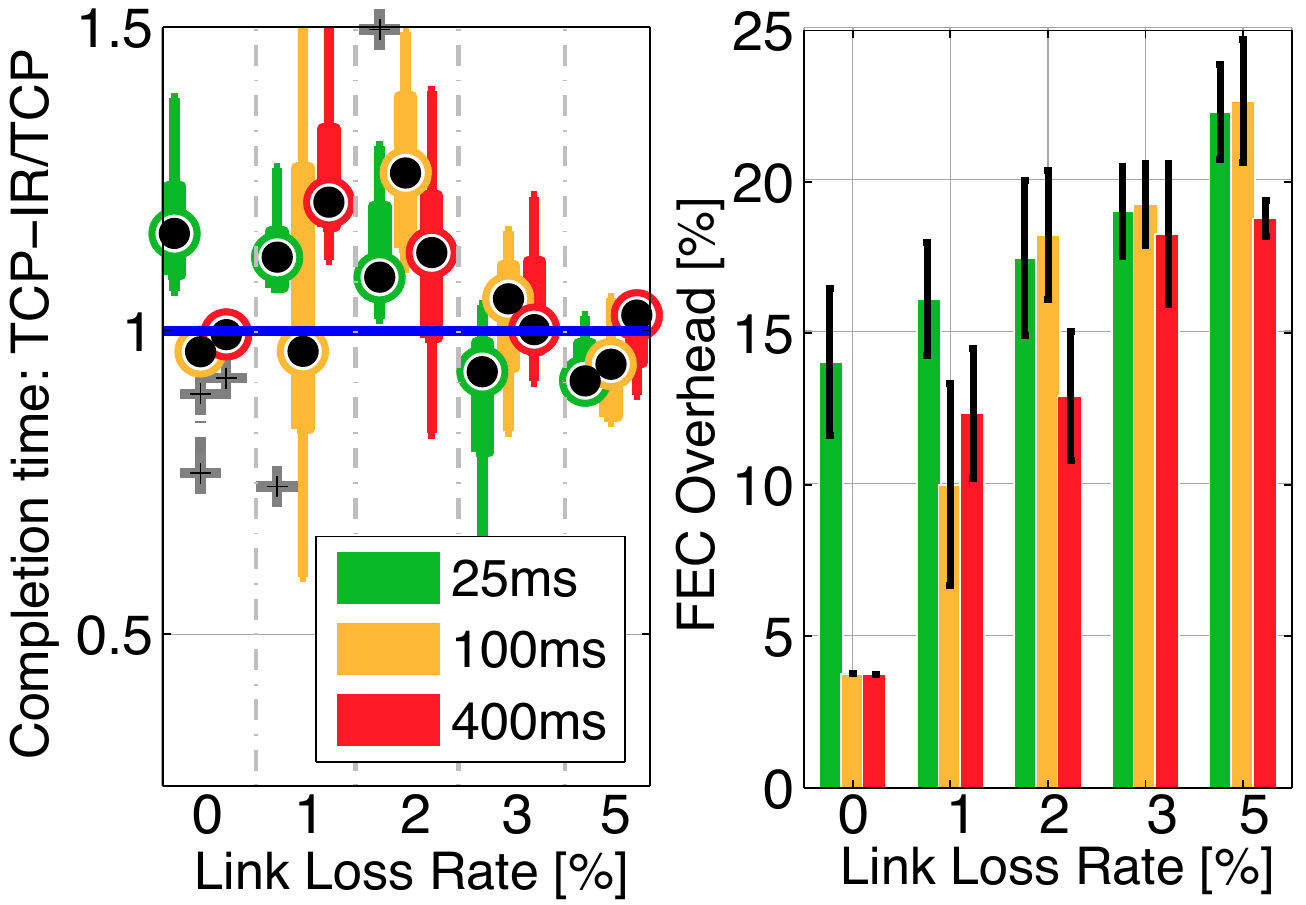}\label{fig:fec:tcpir:google}}
  \hspace{-2mm}
  \subfigure[\textbf{TCP-IR:} Bulk]{
       \includegraphics[height=.152\textwidth]{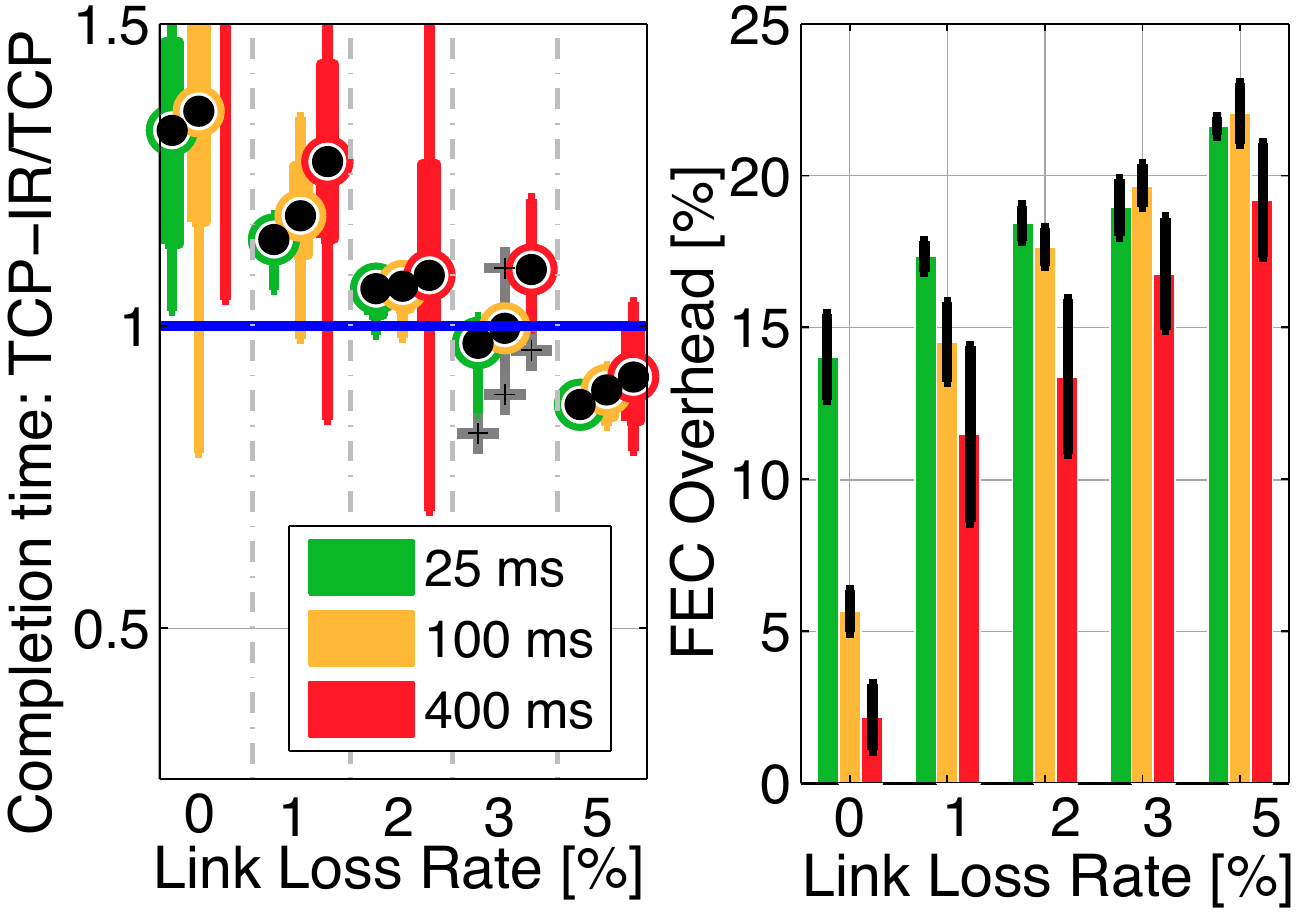}\label{fig:fec:tcpir:bulk}}
    \vspace{-3mm}
  \caption{\textbf{TCP-IR: }Google web traffic and bulk transfers. Completion time ratio (TCP-dFEC/TCP) and FEC Overhead.}\label{fig:fec:tcp:web}
    \vspace{-3mm}
\end{figure}

In Figure~\ref{fig:fec:tcpir:google} we show the completion time and FEC overhead for TCP-IR compared against regular TCP with Google web traffic. TCP-IR provides no benefit in terms of completion times regardless of RTTs or loss rates also with a relatively high FEC overhead. 

In Figure~\ref{fig:fec:tcpir:bulk}, with 5\% injected loss, the CWND becomes small and hence the FEC overhead increases to approximately 20\%, still providing no gains in terms of completion time for bulk transfers. Also, for low link loss rate, TCP-IR increases the completion time compared to regular TCP. The results from~\cite{ferlin16_conext} with TCP-IR provided in this section illustrate how a FEC algorithm can turn out to negatively impact applications. We emphasize the point here that a stronger integration into TCP and an adaptation mechanism to the link conditions at run time are strictly necessary. 

\subsubsection{TCP-dFEC vs TCP}\label{subsubsec:dfec:tcp}
\begin{figure}[h!]
  \vspace{-3mm}
  \centering
  \includegraphics[width=.28\textwidth]{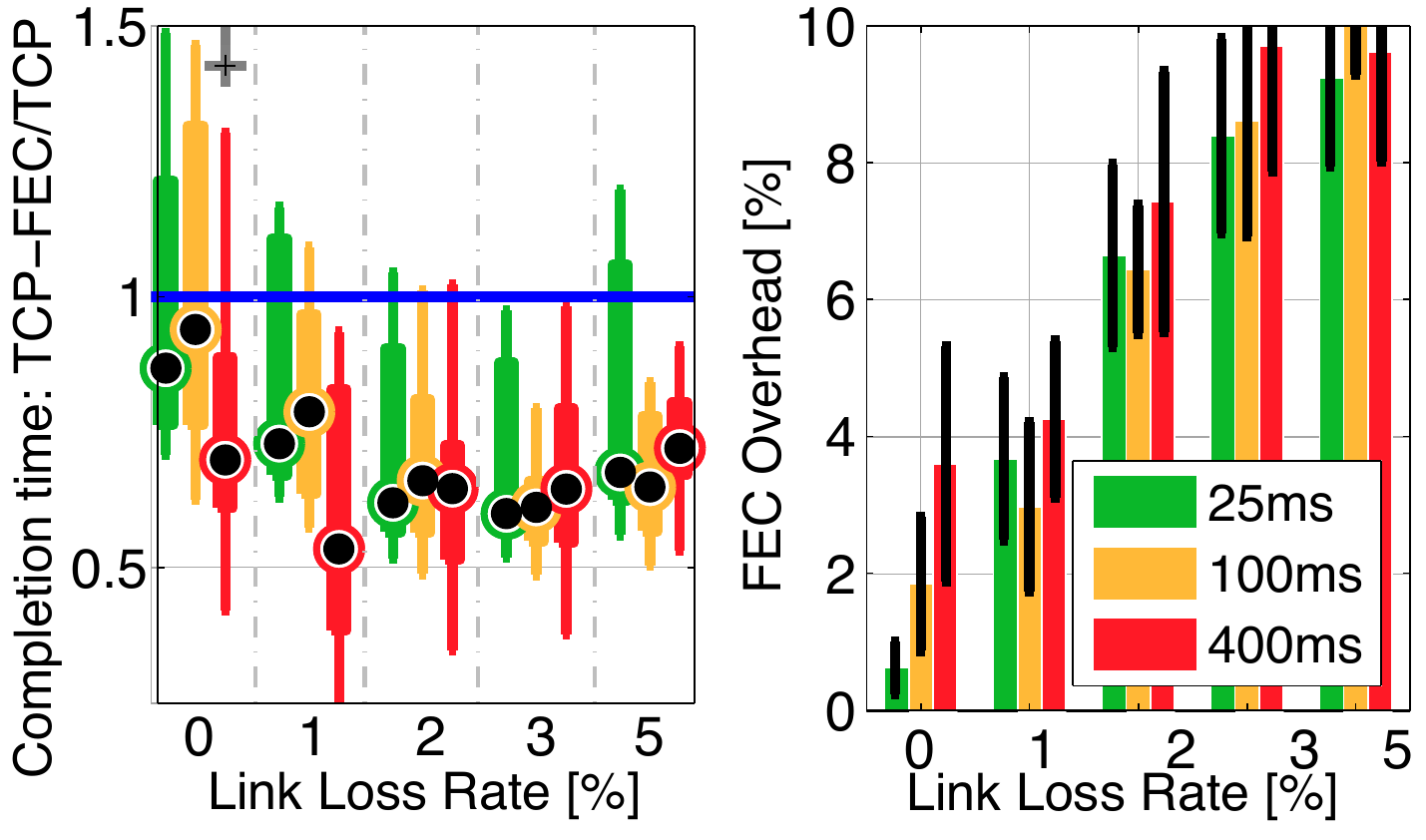}
    \vspace{-1mm}
    \caption{\textbf{TCP-dFEC for bulk: }Completion time ratio (TCP-dFEC/TCP) and FEC Overhead. \label{fig:bulk:tcp:dynamic}}
    \vspace{-3mm}
\end{figure}

\textbf{Bulk: }Figure~\ref{fig:bulk:tcp:dynamic} shows the completion time and FEC overhead of TCP-dFEC with bulk transfer against regular TCP. The left-hand side figure shows the completion times for 25, 100 and 400~ms, while the right-hand side figure shows the FEC overhead (\%). One can see that regardless of the RTTs, adjusting FEC dynamically to the link characteristics brings a clear benefit of up to 40\%. The benefit is lower with 25~ms, because the~\textit{price}, i.e., the time, for a retransmission to perform is lower compared to 400~ms RTT. We observe a small FEC overhead (\%) with low link losses, increasing with the average link losses. With 0\% injected loss, the FEC overhead is about 1\% for 25 and 100~ms and 3.5\% for 400~ms, whereas it reaches up to 9\% with 5\% link loss. Since TCP-dFEC adapts the FEC-ratio with the feedback from the receiver, the FEC overhead for 25, 100 and 400~ms scenarios is, hence, also distinct with same average link loss rates. To compare TCP-dFEC performance with the original TCP-IR directly under the same conditions, see Figure~\ref{fig:fec:tcpir:bulk}.

\begin{figure}[h!]
  \vspace{-4mm}
  \centering
  \includegraphics[width=.28\textwidth]{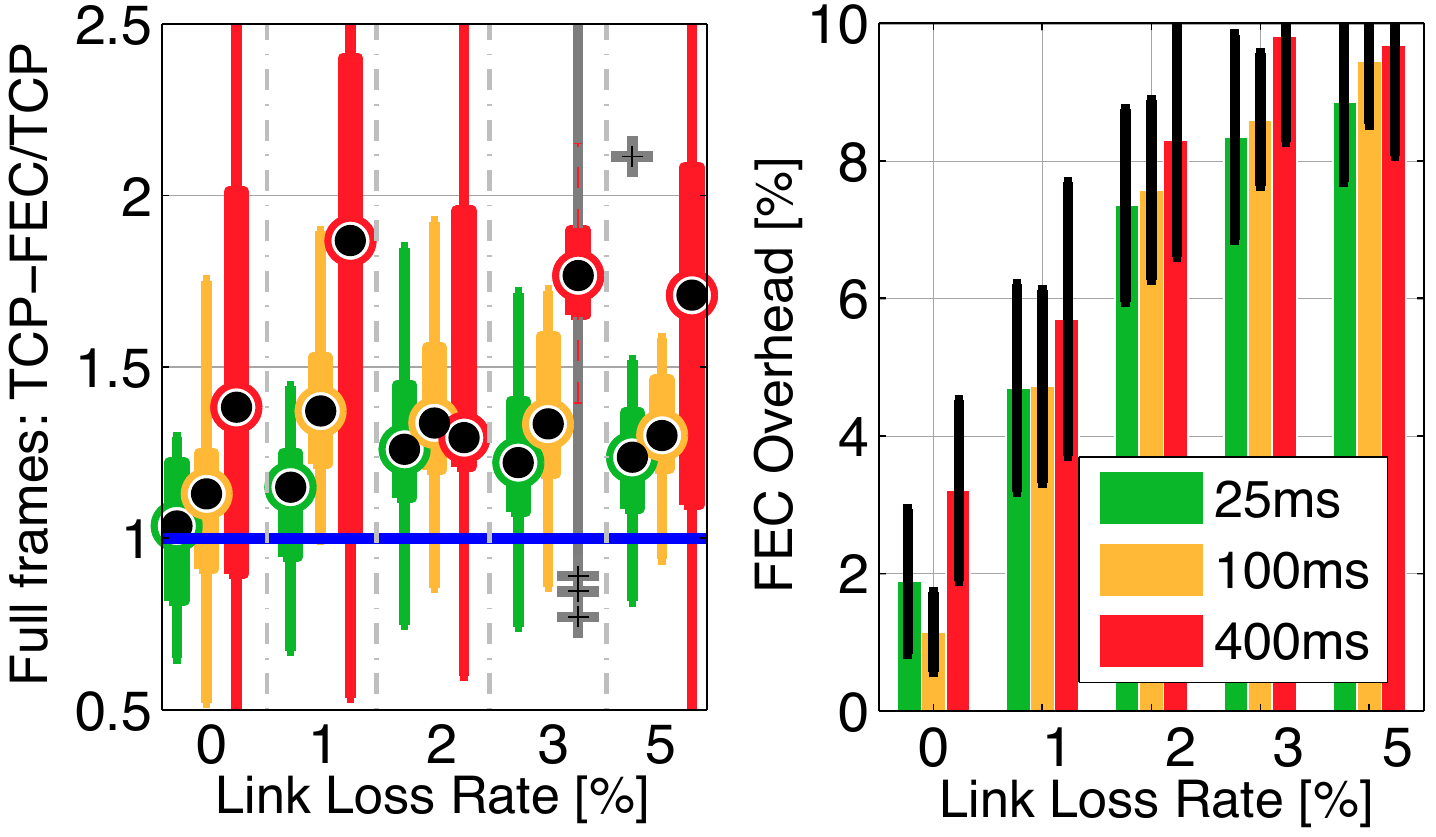}
    \vspace{-1mm}
    \caption{\textbf{TCP-dFEC for H.264: 1min. of \textit{Big Buck Bunny} encoded at 3,4 Mibps: } Average dropped frame (TCP-dFEC/TCP) and FEC Overhead.\label{fig:h264:tcp:dynamic}}
    \vspace{-3mm}
\end{figure}

\textbf{H.264 }Figures~\ref{fig:h264:tcp:dynamic} shows TCP-dFEC with H.264 with the left-hand side figure presenting fully received frames ratio compared against regular TCP for 25, 100 and 400~ms and the right-hand side figure showing the FEC overhead (\%). TCP-dFEC brings a steady benefit, although more variable compared to bulk: A constant benefit of ca. 20\% with 25~ms, and a larger benefit of up to 30 to 40\% with 100 and 400~ms.

\begin{figure*}[!]
  \vspace{-3mm}
  
  \centering
  \subfigure [\textbf{Google}
  \label{fig:http2:tcp:dynamic:google}]{
   \includegraphics[width=.29\textwidth]{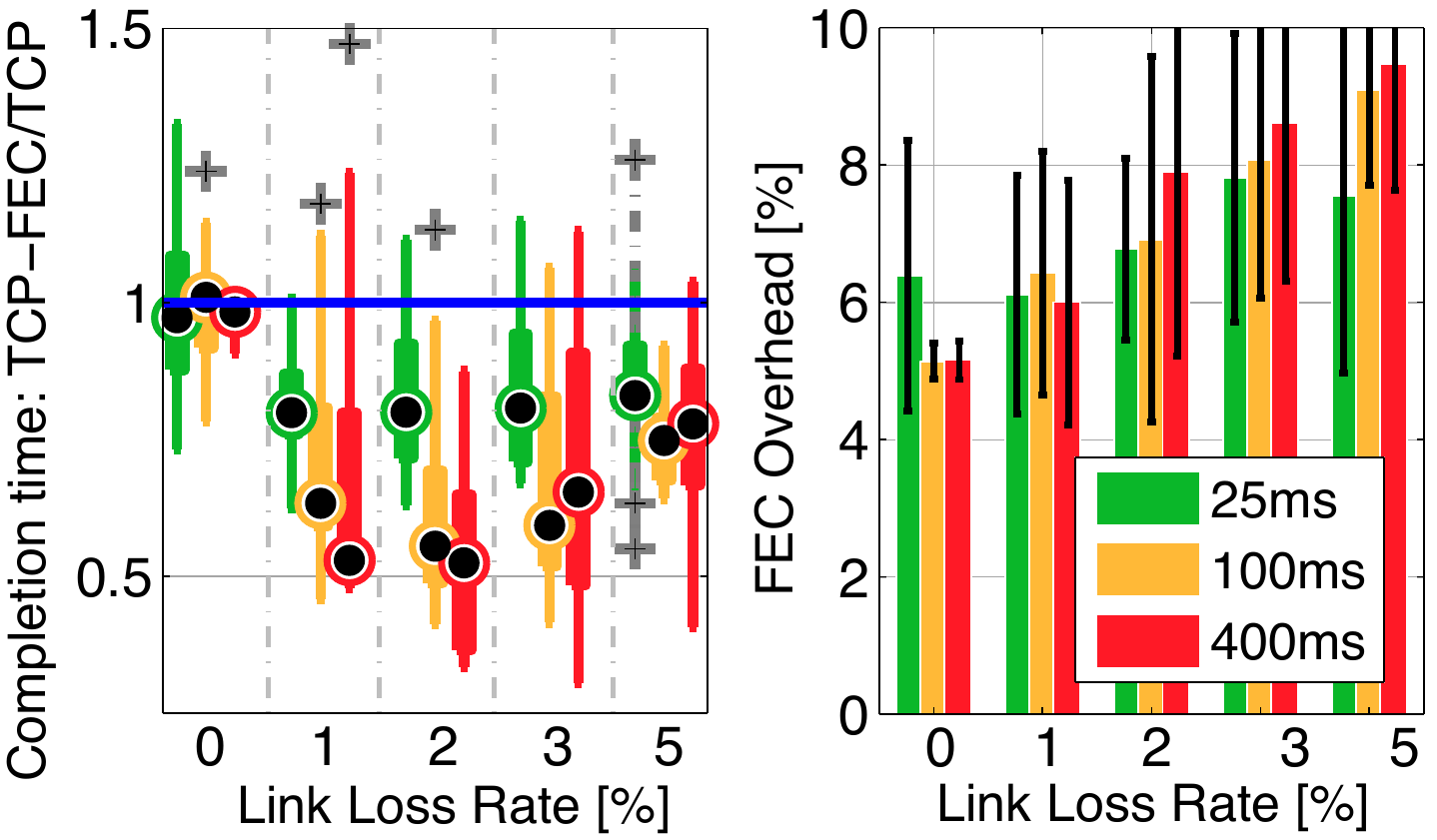}}
  \hspace{-1mm}
  \subfigure[\textbf{YouTube}\label{fig:http2:tcp:dynamic:youtube}]
  {\includegraphics[width=.29\textwidth]{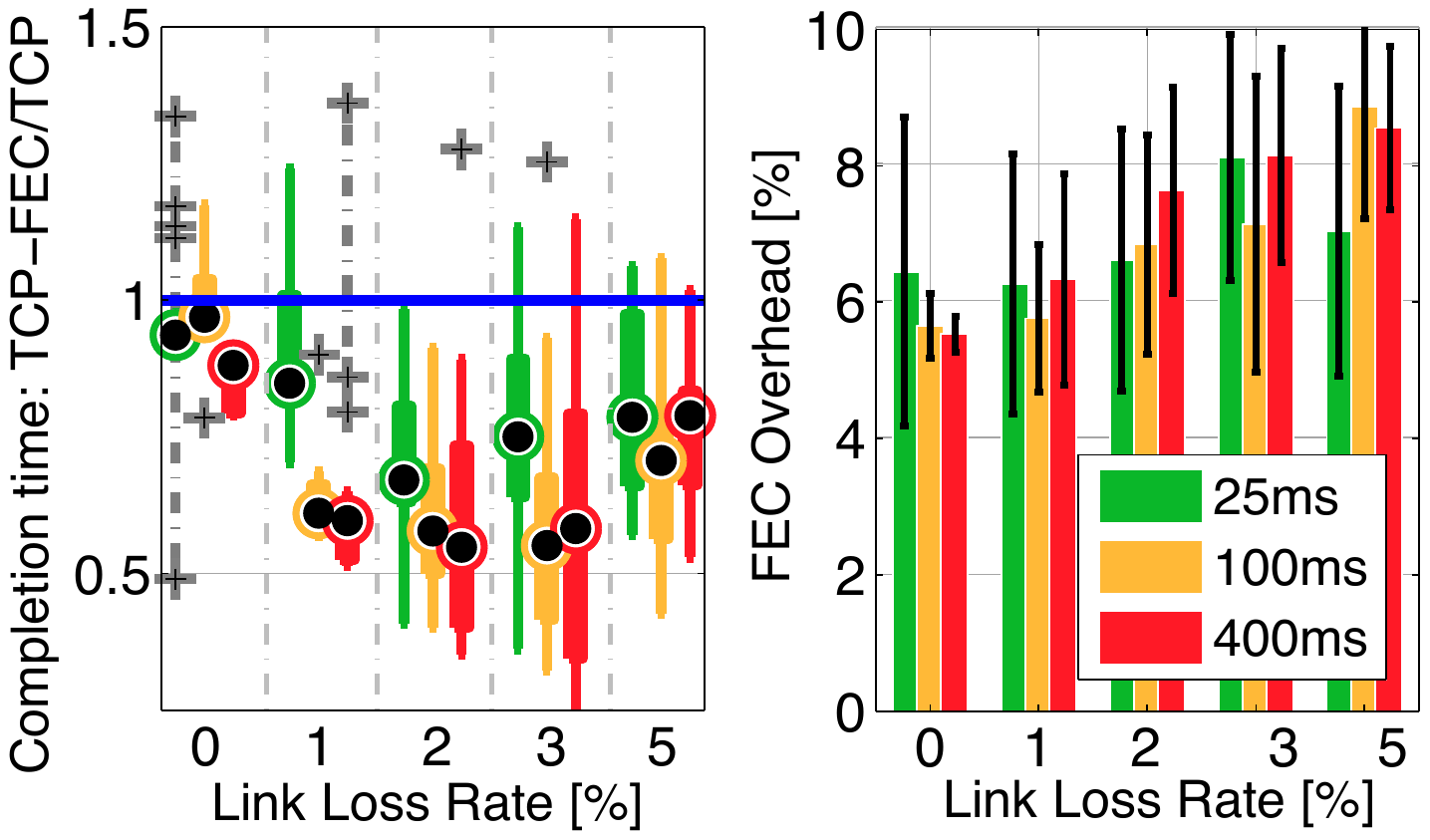}}
  \hspace{-1mm}
  \subfigure[\textbf{ESPN}
 \label{fig:http2:tcp:dynamic:espn}]
  {\includegraphics[width=.29\textwidth]{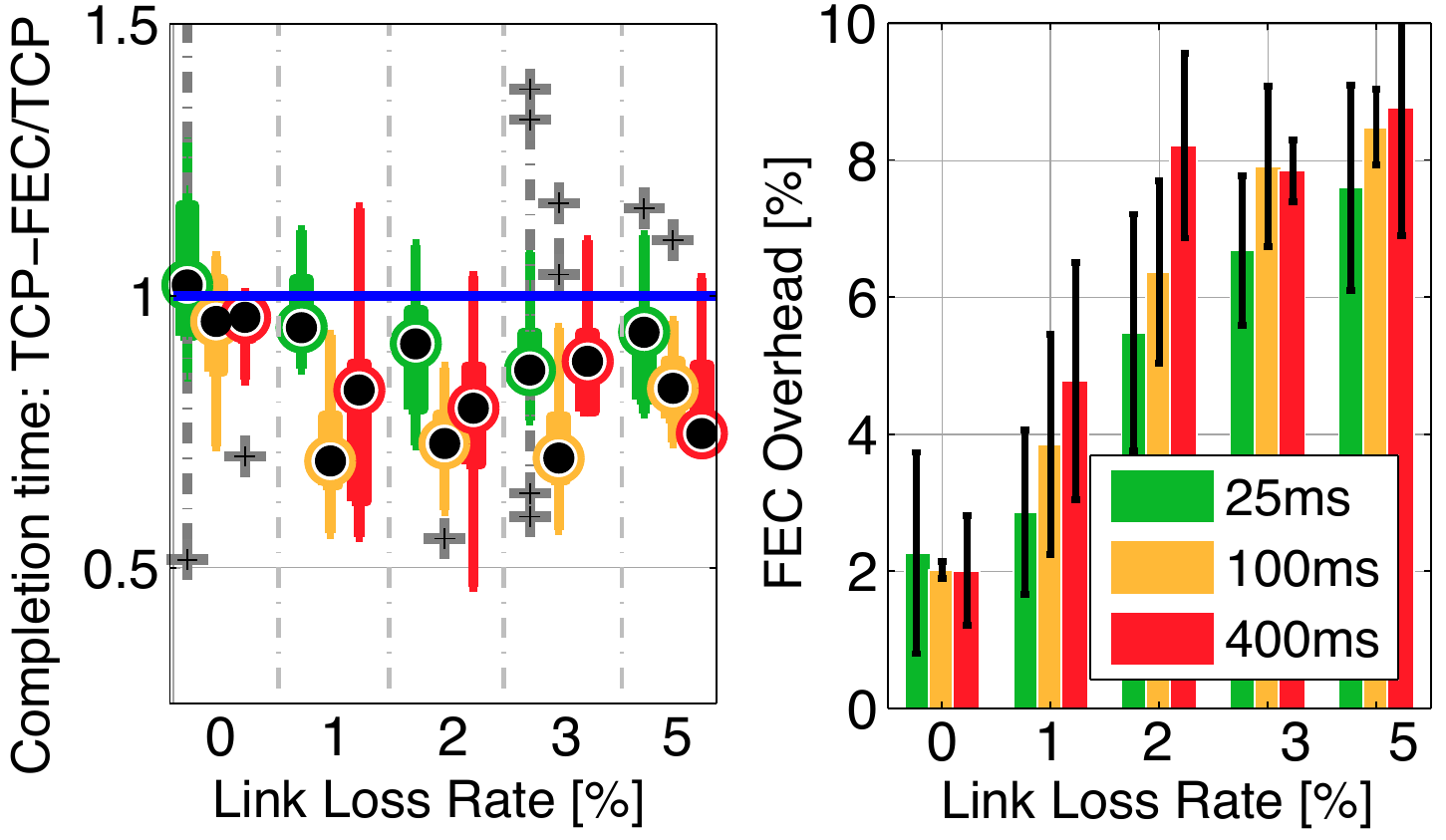}}
  
  \vspace{-3mm}
  
  \caption{\textbf{TCP-dFEC for HTTP/2: }Completion time ratio (TCP-dFEC/TCP) and FEC Overhead with background traffic, losses between 0, 1, 2, 3 or 5\% and RTT between 25, 100 and 400 ms, see Table~\ref{tab:emulation:parameters}.}
  \label{fig:http2:tcp:dynamic}
    \vspace{-3mm}
\end{figure*}

\textbf{HTTP/2: }Figure~\ref{fig:http2:tcp:dynamic} shows TCP-dFEC with HTTP/2 and different websites sizes, see Table~\ref{tab:web-profile}, with the left-hand side figures showing the completion time ratio to regular TCP and the right-hand side figures depicting the FEC overhead (\%). One can see that TCP-dFEC brings a clear benefit as the link gets lossier for Google and YouTube website sizes. The experiments with injected loss of 0\% show little benefit and a relatively high FEC overhead with ca. 5 to 6\% for 25~ms with Google. This is due to the FEC dynamic ratio starting with 1 FEC each 9 non-FEC packets, i.e., we can send 1 FEC packet within TCP's IW. Since the Google's website is relatively small, finishing within few RTTs, dFEC does not have the time to substantially reduce the overhead. However, with all other RTTs, loss rates and website sizes, the dynamic FEC ratio reduces the completion time by more than 30\% with YouTube and 20\% with ESPN. To compare TCP-dFEC from Figure~\ref{fig:http2:tcp:dynamic:google} under the same conditions directly against TCP-IR, see Figure~\ref{fig:fec:tcpir:bulk}.
\subsection{Dynamic FEC and MPTCP}
In this section we first explain how dynamic FEC can be beneficial in a multipath scenarios with heterogeneous links for bulk transfers. To emulate these links we use the topology shown in Figure~\ref{fig:emulation:nsb}, applying the configuration for $B1$ and $B2$ as shown in Table~\ref{tab:emulation:parameters}. Next we present the results for MPTCP with bulk, H.264 and HTTP/2.

\textbf{Bulk: }Figure~\ref{fig:bulk:mptcp:dynamic:goodput} shows MPTCP-dFEC's completion time compared to MPTCP, with $B2$'s RTT varying between 25, 100 and 400~ms and $B1$'s loss rate between 0, 1, 2, 3 and 5\%. One can see that regardless of the RTTs and loss rates, MPTCP-dFEC brings a benefit of more than 20\% for 100~ms and 400~ms as the link gets lossier. Figures~\ref{fig:bulk:mptcp:dynamic:overhead:s1} and~\ref{fig:bulk:mptcp:dynamic:overhead:s2} show per subflow FEC overhead for $B1$ and $B2$, respectively. One can see that there is a consistent higher utilisation of the $B1$ subflow compared to non-FEC subflows, the gains are up to 40\% across all RTTs and loss rates. Note that $B1$'s settings emulate the WLAN path, meaning that, MPTCP-dFEC better utilises the lossy WLAN subflow compared to regular MPTCP, shifting traffic away from the higher delay and commonly paid cellular $B2$ subflow. Finally, Figures~\ref{fig:bulk:mptcp:dynamic:overhead:s1} and~\ref{fig:bulk:mptcp:dynamic:overhead:s2} depict the FEC overhead for $B1$ and $B2$. One can see that the FEC overhead in Figure~\ref{fig:bulk:mptcp:dynamic:overhead:s2} for 400~ms is high due to dynamic FEC adjustment based on the response from the receiver. Although this is an optimisation aspect of dFEC, one can see from Figure~\ref{fig:bulk:mptcp:dynamic:utilisation:s2} that FEC was not sent in vain, improving $B2$'s utilisation up to 20\% with 100~ms RTT, reduced to ca. 10\% with 400~ms. Following the same setup for the experiments, we now comment on H.264 and HTTP/2:
\begin{figure*}[!]
  \vspace{-2mm}
  \centering
  \subfigure [\textbf{Completion time}\label{fig:bulk:mptcp:dynamic:goodput}]{
   \includegraphics[width=.2\textwidth]{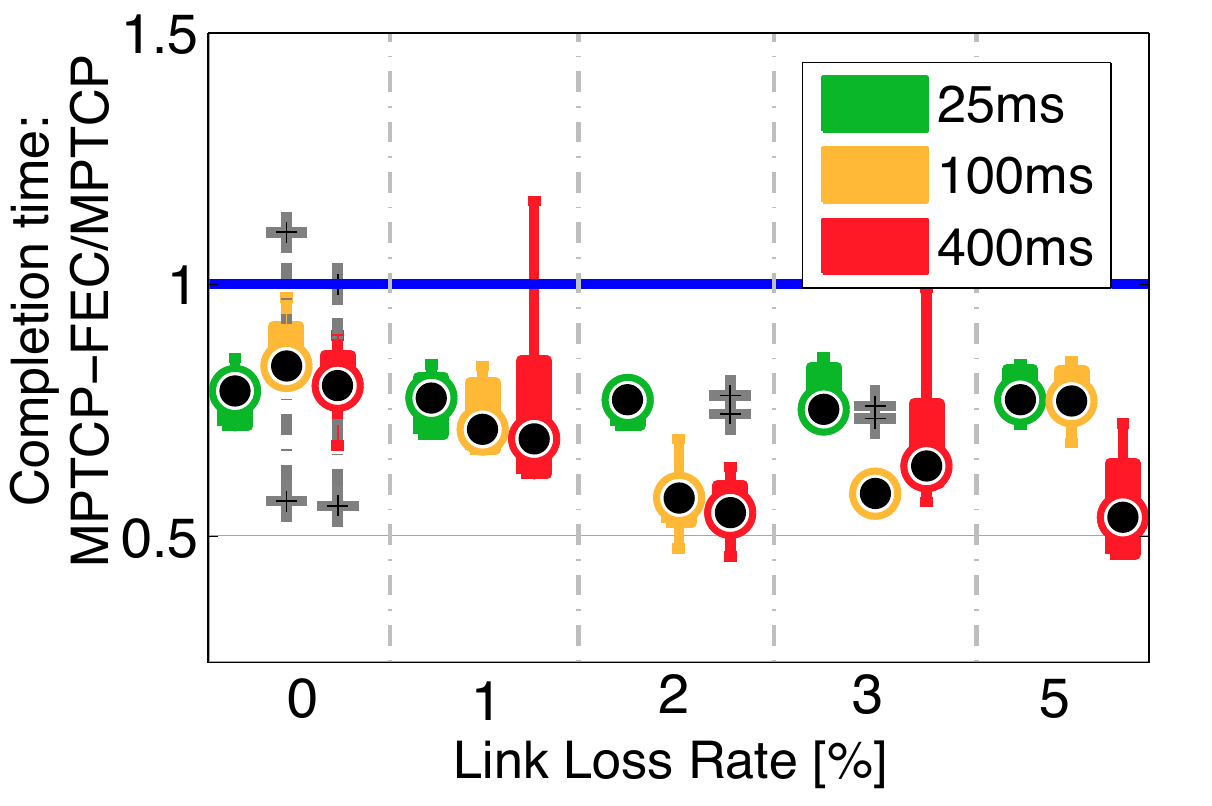}}
  \hspace{-2mm}
  \subfigure[\textbf{$B1$ utilisation}\label{fig:bulk:mptcp:dynamic:utilisation:s1}]
  {\includegraphics[width=.19\textwidth]{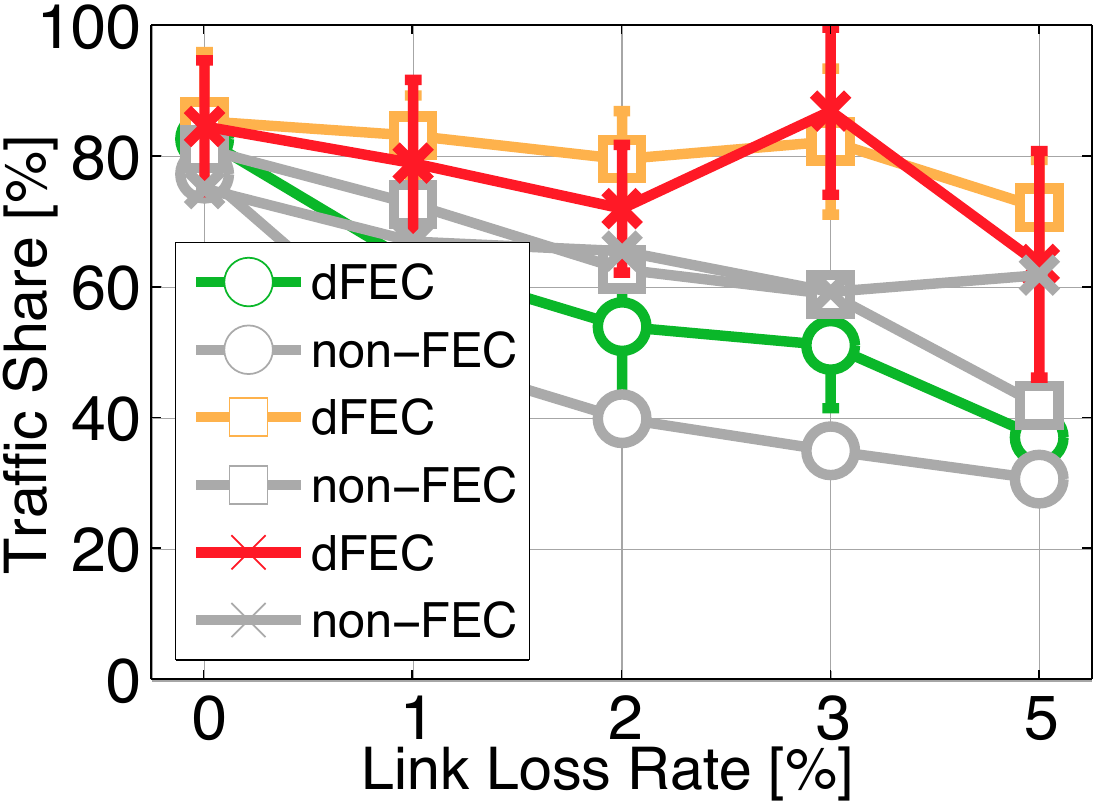}}
  \hspace{-2mm}
  \subfigure[\textbf{$B1$ FEC overhead}\label{fig:bulk:mptcp:dynamic:overhead:s1}]
  {\includegraphics[width=.185\textwidth]{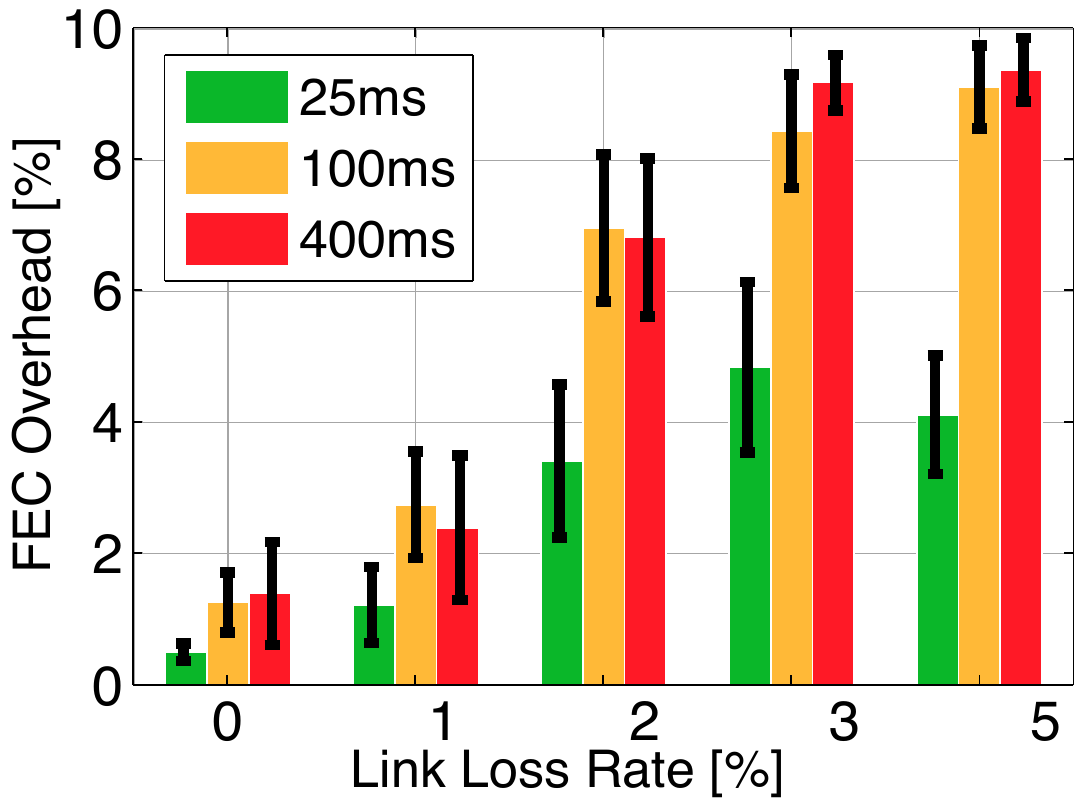}}
  \hspace{-2mm}
  \subfigure[\textbf{$B2$ utilisation}\label{fig:bulk:mptcp:dynamic:utilisation:s2}]
  {\includegraphics[width=.19\textwidth]{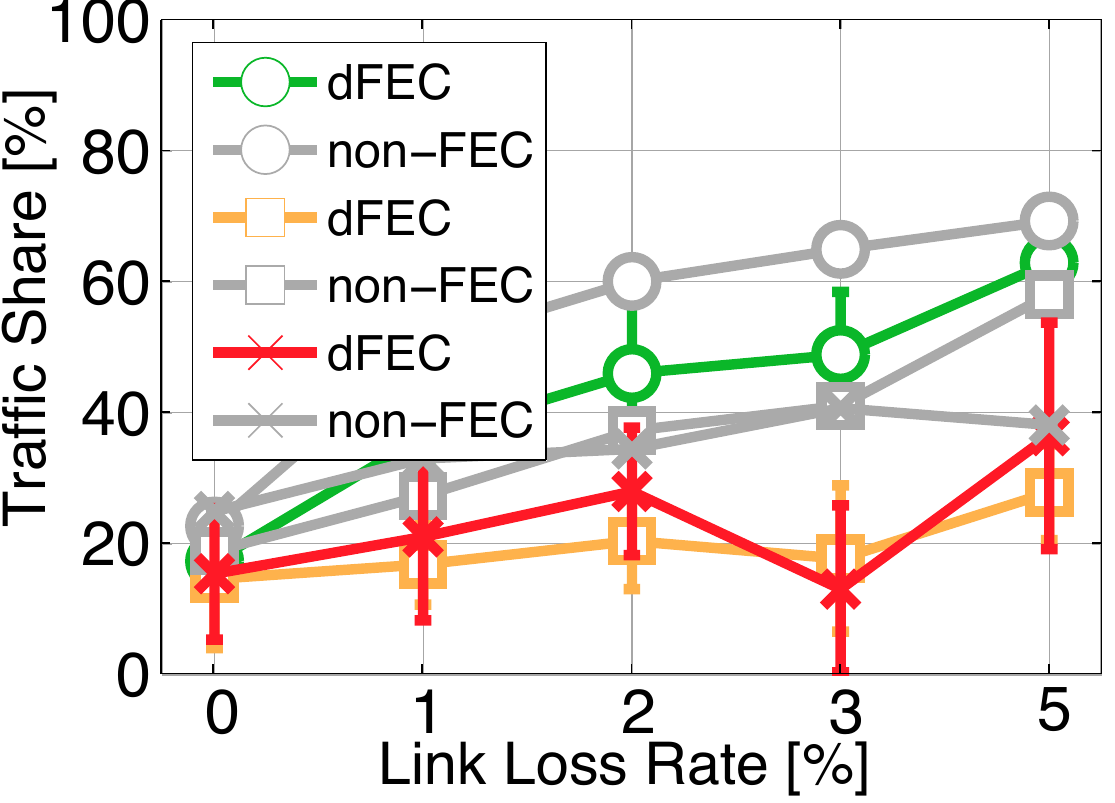}}
  \hspace{-2mm}
  \subfigure[\textbf{$B2$ FEC overhead}\label{fig:bulk:mptcp:dynamic:overhead:s2}]
  {\includegraphics[width=.185\textwidth]{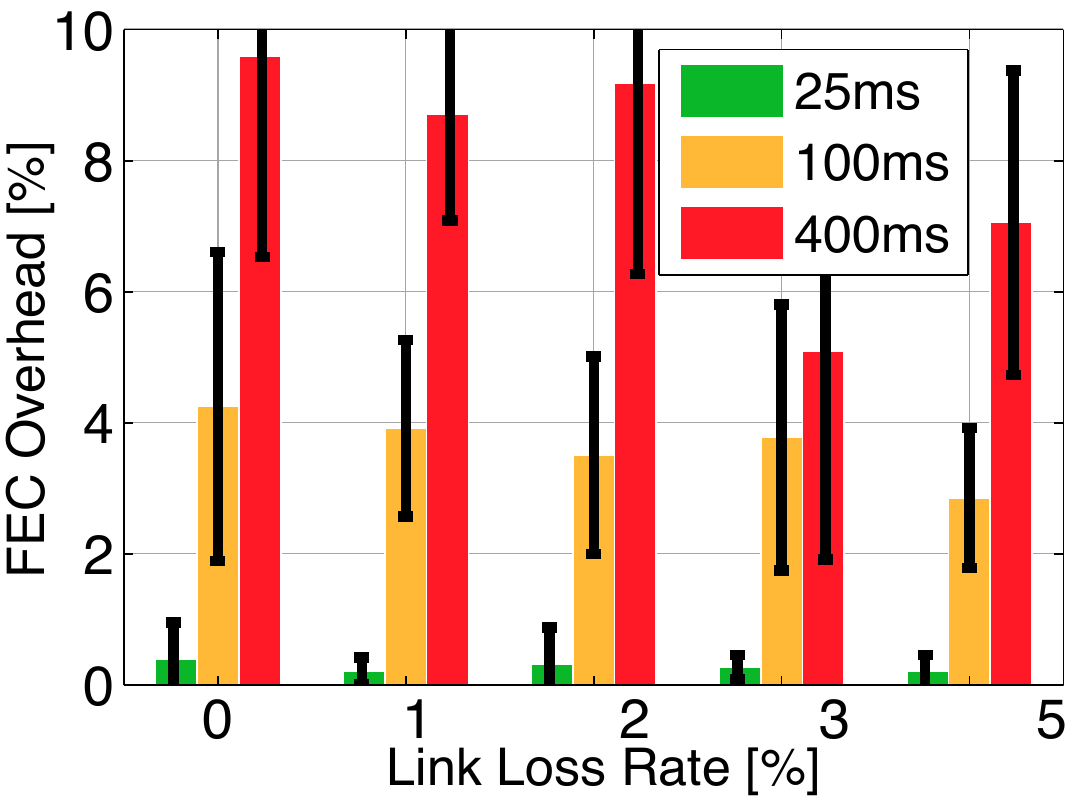}}
  \vspace{-3mm}
  \caption{\textbf{MPTCP-dFEC for bulk: }MPTCP Completion time, Subflow $B1$ and $B2$ utilisation and FEC overhead.}
  \label{fig:fec:dynamic:mptcp:bulk}
  \vspace{-1mm}
\end{figure*}


%

\begin{figure}
  \vspace{-2.5mm}
  \centering
  \includegraphics[width=.29\textwidth]{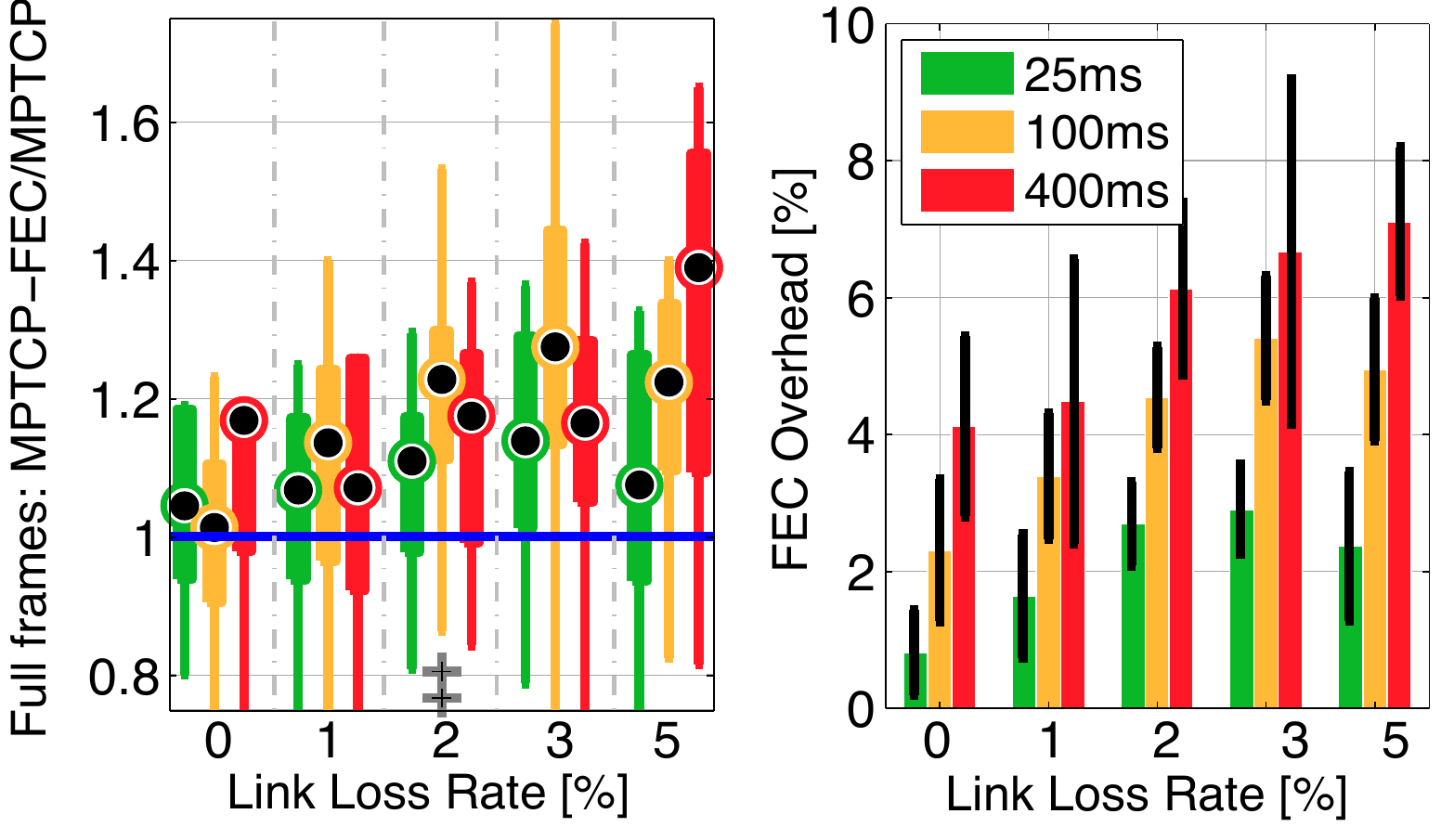}
    \vspace{-2mm}
    \caption{\textbf{MPTCP-dFEC for H.264: }Background traffic with losses 0, 1, 2, 3 or 5\% in $B1$ and RTT 25, 100 and 400 ms in $B2$, see Table~\ref{tab:emulation:parameters}. 1min. of \textit{Big Buck Bunny} encoded at 3,4 Mibps. Goodput (MPTCP-FEC/MPTCP) and FEC Overhead.}
    \label{fig:h264:mptcp:dynamic}
    \vspace{-2.5mm}
\end{figure}

\textbf{H.264: }Figures~\ref{fig:h264:mptcp:dynamic} illustrates MPTCP-dFEC's performance compared to regular MPTCP with H.264, where the left-hand side figure shows the ratio of number of full frames, i.e., I, P and B, and the right-hand side figure shows the FEC overhead. One can see that MPTCP-dFEC brings a smaller benefit compared to bulk, but constant of ca. 10 to 20\% for 25~ms and more than 20\% for 100~ms and 400~ms at times.

\begin{figure*}
  \vspace{-2.5mm}
  \centering
  \subfigure [\textbf{Google}\label{fig:http2:mptcp:dynamic:google}]{
   \includegraphics[width=.29\textwidth]{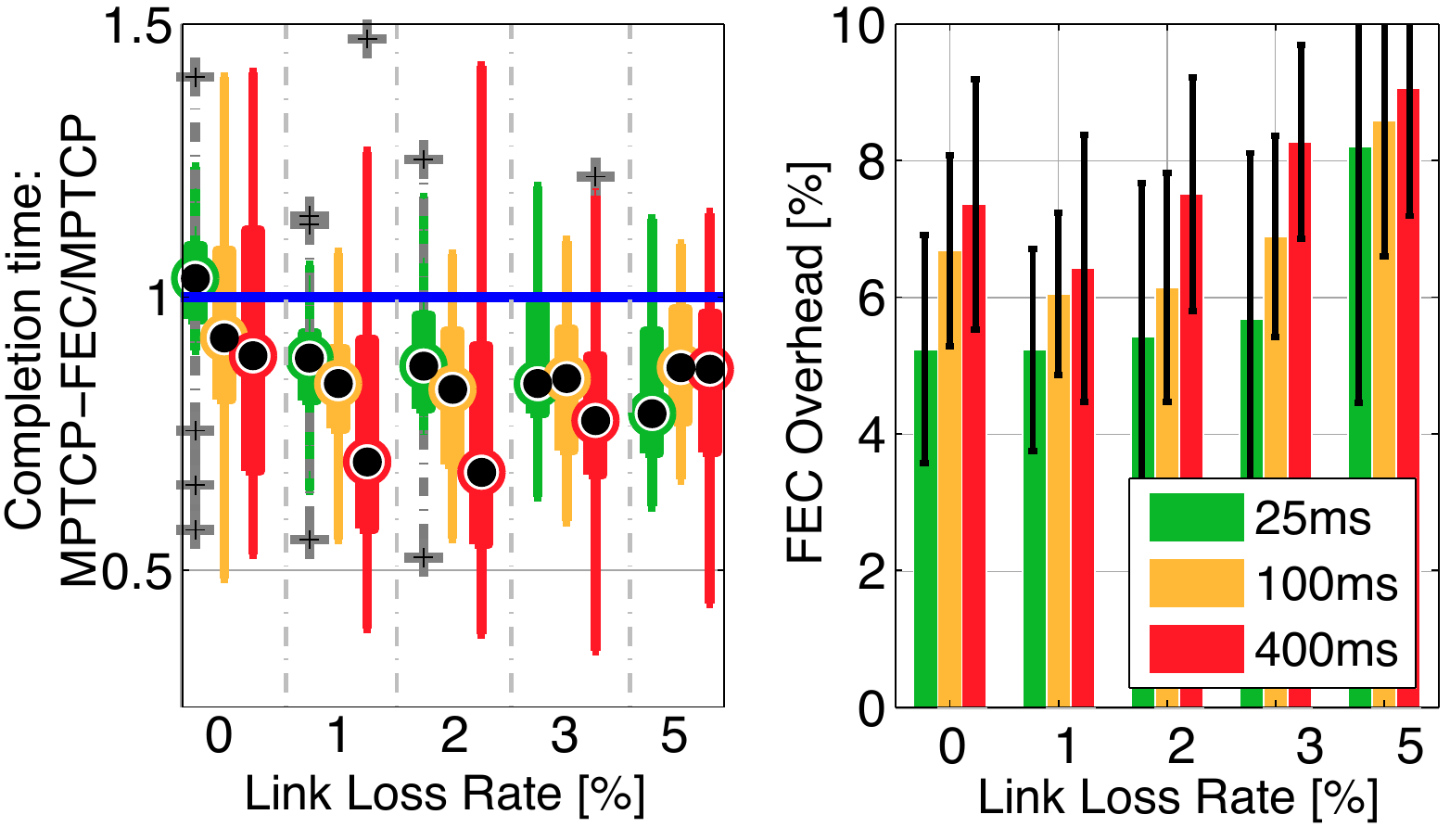}}
  \hspace{-1mm}
  \subfigure[\textbf{YouTube}\label{fig:http2:mptcp:dynamic:youtube}]
  {\includegraphics[width=.29\textwidth]{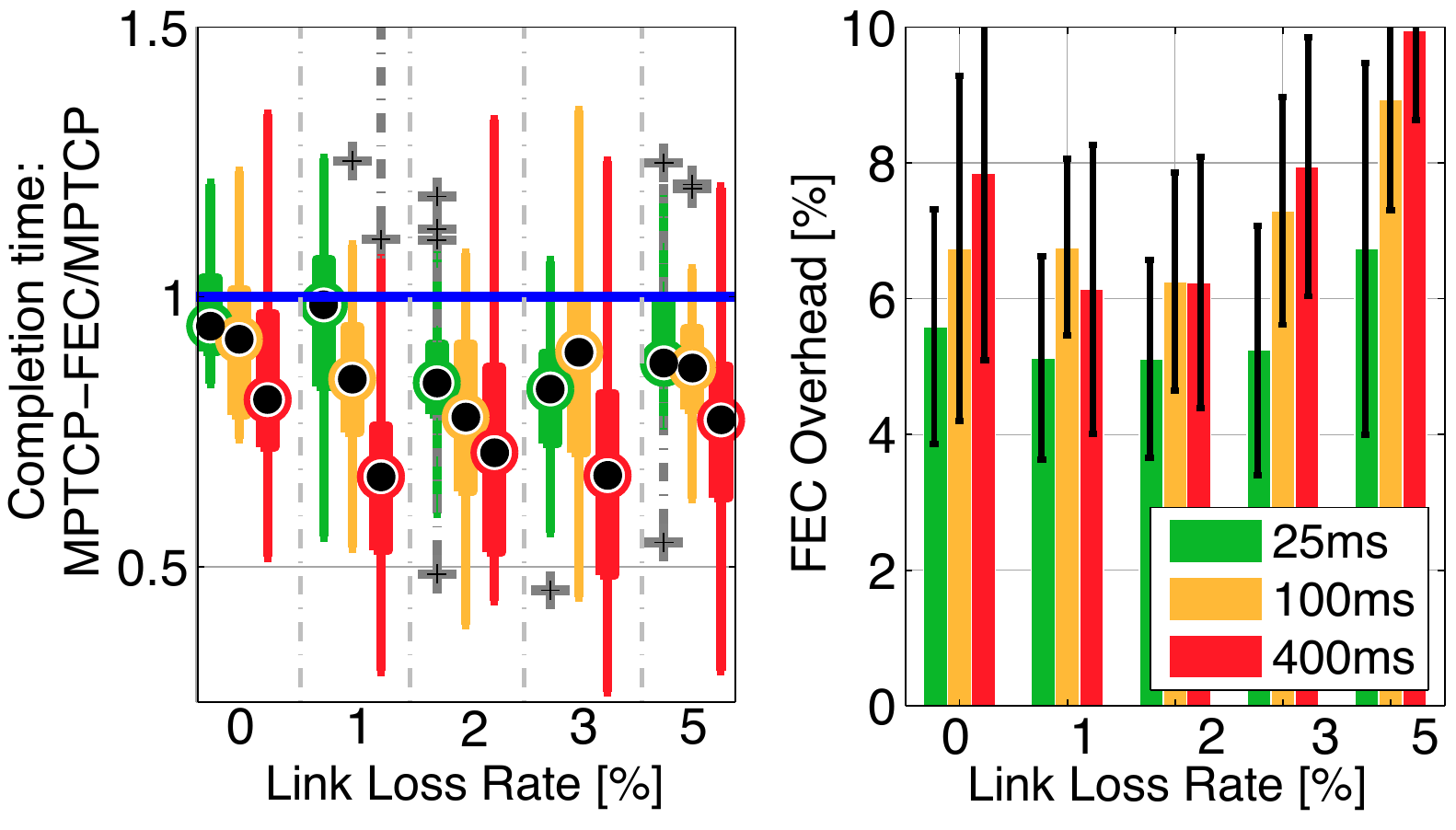}}
  \hspace{-1mm}
  \subfigure[\textbf{ESPN}\label{fig:http2:mptcp:dynamic:espn}]
  {\includegraphics[width=.29\textwidth]{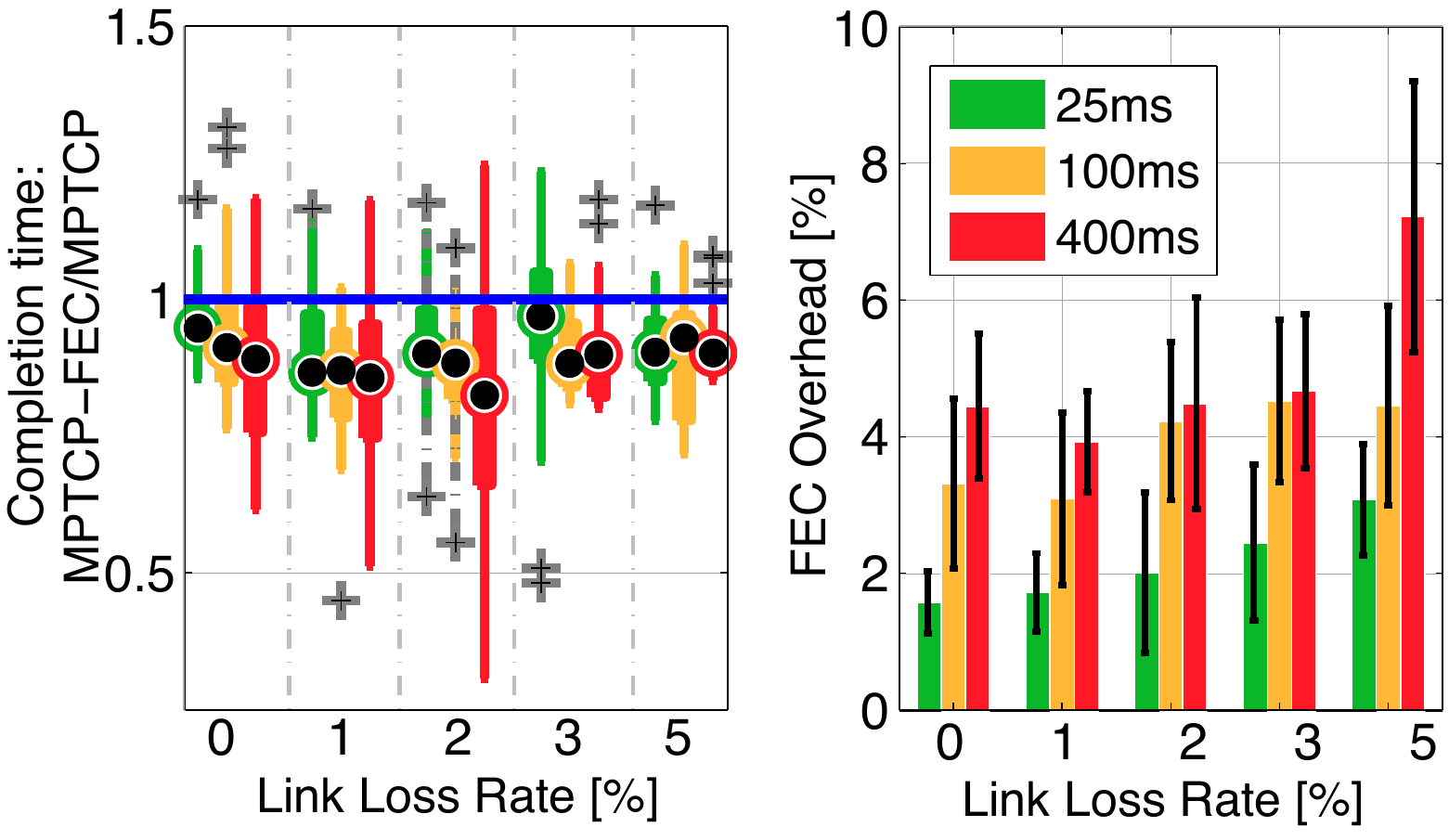}}
\vspace{-2mm}
\caption{\textbf{MPTCP-dFEC for HTTP/2: }Completion Time (MPTCP-FEC/MPTCP) and FEC Overhead with background traffic: losses between 0, 1, 2, 3 and 5\% in $B1$ and RTT between 25, 100 and 400 ms in $B2$, see Table~\ref{tab:emulation:parameters}.}
  \label{fig:http2:mptcp:dynamic}
  \vspace{-2.5mm}
\end{figure*}

\textbf{HTTP/2: }Figure~\ref{fig:http2:mptcp:dynamic} illustrates MPTCP-dFEC's performance compared to MPTCP with HTTP/2 for the websites from Table~\ref{tab:web-profile}. The the left-hand side figures show the completion time ratio and the right-hand side figures the FEC overhead. One can see that FEC brings benefit improving the completion time in all cases. The experiments with loss rates of 0\% show less benefit and a FEC overhead of up to 5\%. This is due to the FEC dynamic ratio starting a 10\% rate in TCP's IW. For Google, finishing within few RTTs, dFEC does not have the necessary time, i.e., RTTs, to reduce the overhead significantly. However, with all other RTTs for $B2$, link losses for $B1$ and website sizes, dFEC can bring a benefit of more than 40\% with YouTube and a constant 10 to 20\% benefit with ESPN.

\subsection{Real-Network Experiments}\label{subsec:real:network:experiments}
\begin{figure}[h!]
  \vspace{-3mm}
\centering
    \includegraphics[width=0.95\columnwidth]{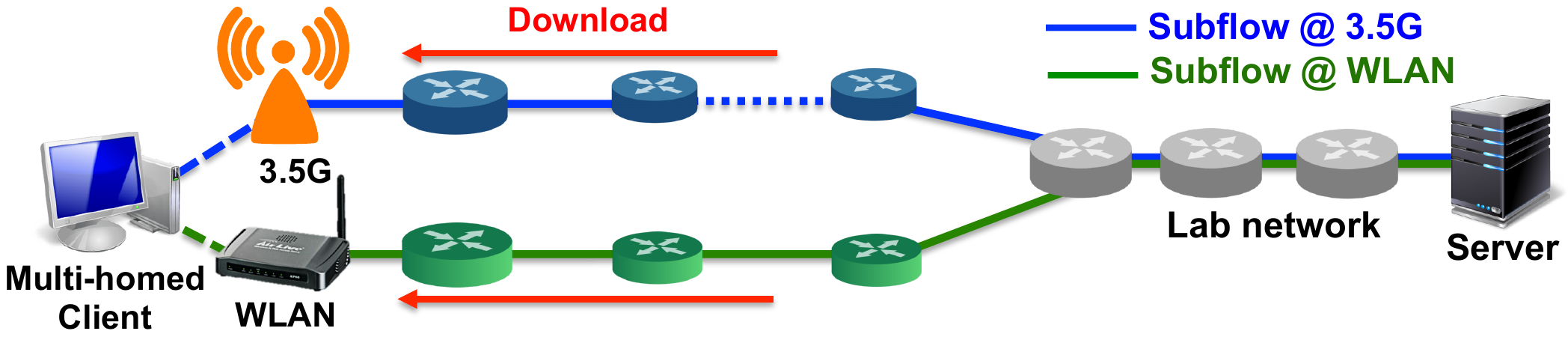}
\caption{Real-network experiment scenario.}
\label{fig:real:network:experiment}
  \vspace{-3mm}
\end{figure}

Finally, we validate the performance of the dFEC algorithm with real-network experiments within the topology as shown in Figure~\ref{fig:real:network:experiment}, i.e. non-shared bottleneck, now constructed over NorNet~\cite{ComNets2013-Core}.
We also use consumer hardware connected to a 20~Mbps DSL via WLAN and 10~Mbps via cellular 3.5G. We hold the same network capacities as shown in Table~\ref{tab:emulation:parameters}, however, for the delays, we select two different settings: 25~ms and 100~ms and 400~ms and 100~ms for WLAN and 3.5G. While the first scenario is very close to a smartphone, the second aims at experimenting with a path with a higher delay and loss rates, i.e. satellite terminal, with a cellular network. For the loss rates, we set 0.5\% and 1\% for 3.5G and WLAN networks, respectively. 

\begin{figure}
  \vspace{-3mm}
  \centering
  \includegraphics[width=.27\textwidth]{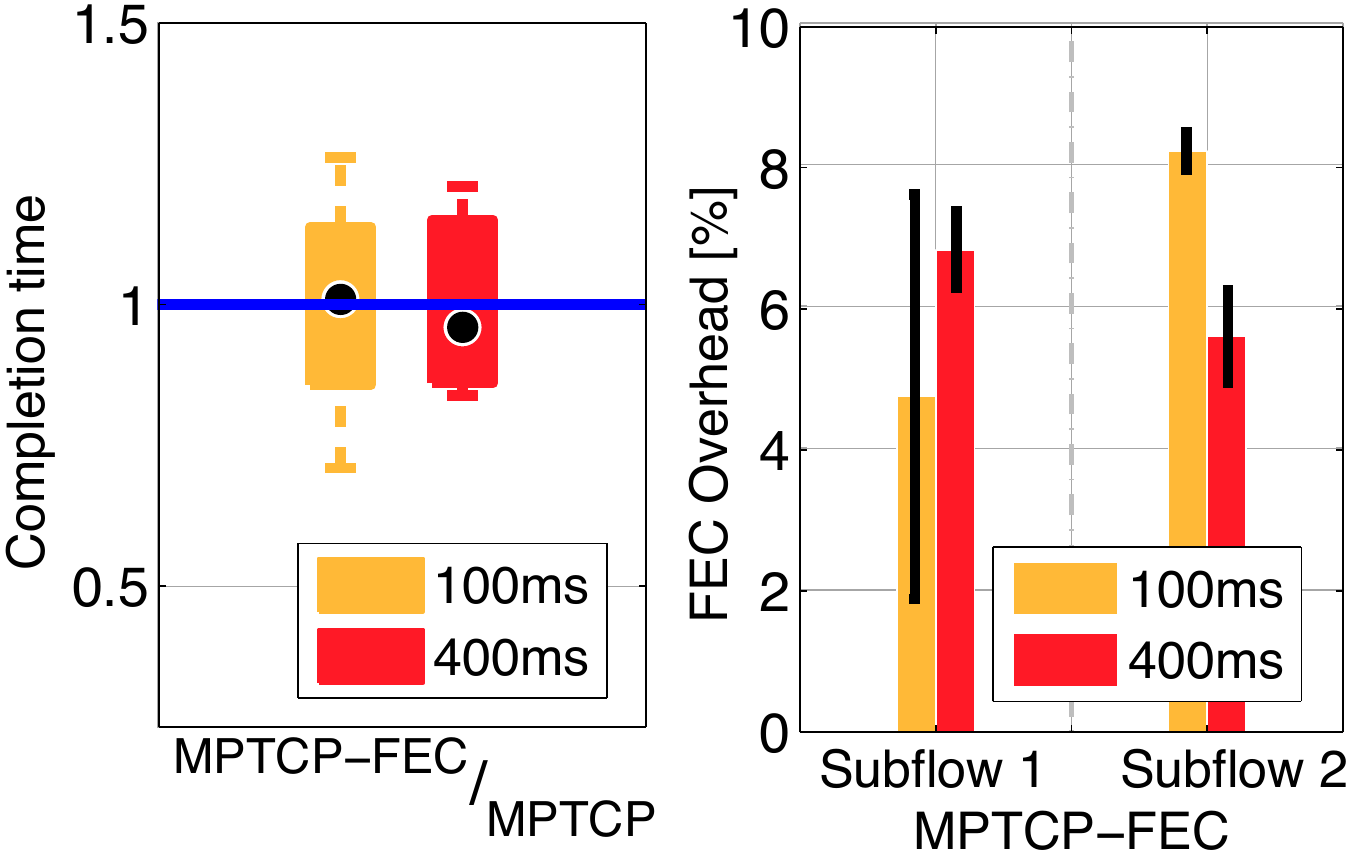}
    \vspace{-1mm}
    \caption{\textbf{MPTCP-dFEC/MPTCP with Bufferbloat: }Completion time and FEC overhead with real-network experiments.}\label{fig:bulk:nornet:bufferbloat}
        \vspace{-3mm}
\end{figure}

\begin{figure}
  \vspace{-1mm}
  \centering
  \includegraphics[width=.27\textwidth]{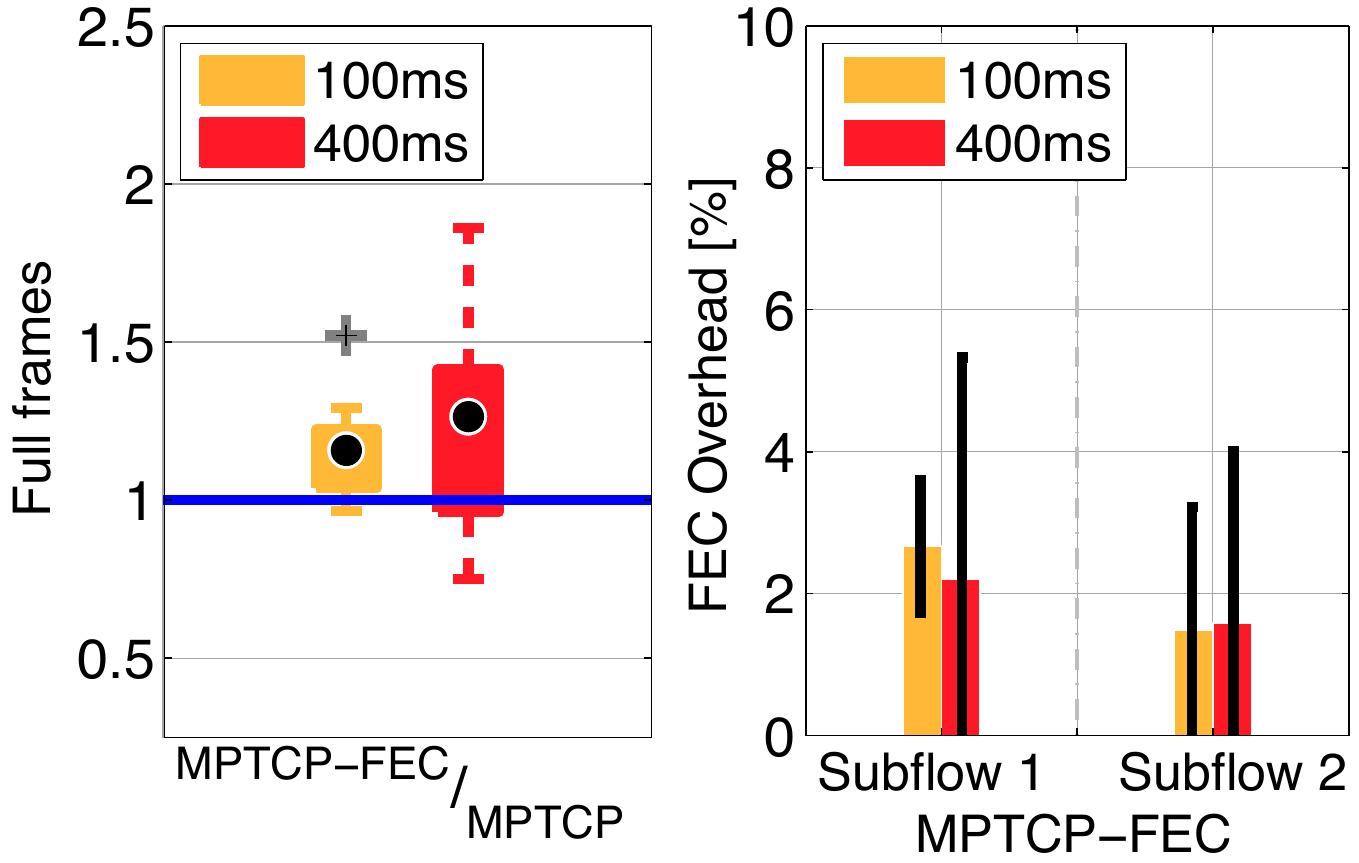}
    \vspace{-1mm}
    \caption{\textbf{MPTCP-dFEC/MPTCP for H.264: 1min. of \textit{Big Buck Bunny} encoded at 3,4 Mibps: } Full frame ratio with real-network experiments.}\label{fig:h264:nornet}
    \vspace{-3mm}
\end{figure}

\begin{figure*}
  \vspace{-3mm}
  \centering
  \subfigure [\textbf{Google}\label{fig:http2:mptcp:dynamic:google:nornet}]{
   \includegraphics[width=.27\textwidth]{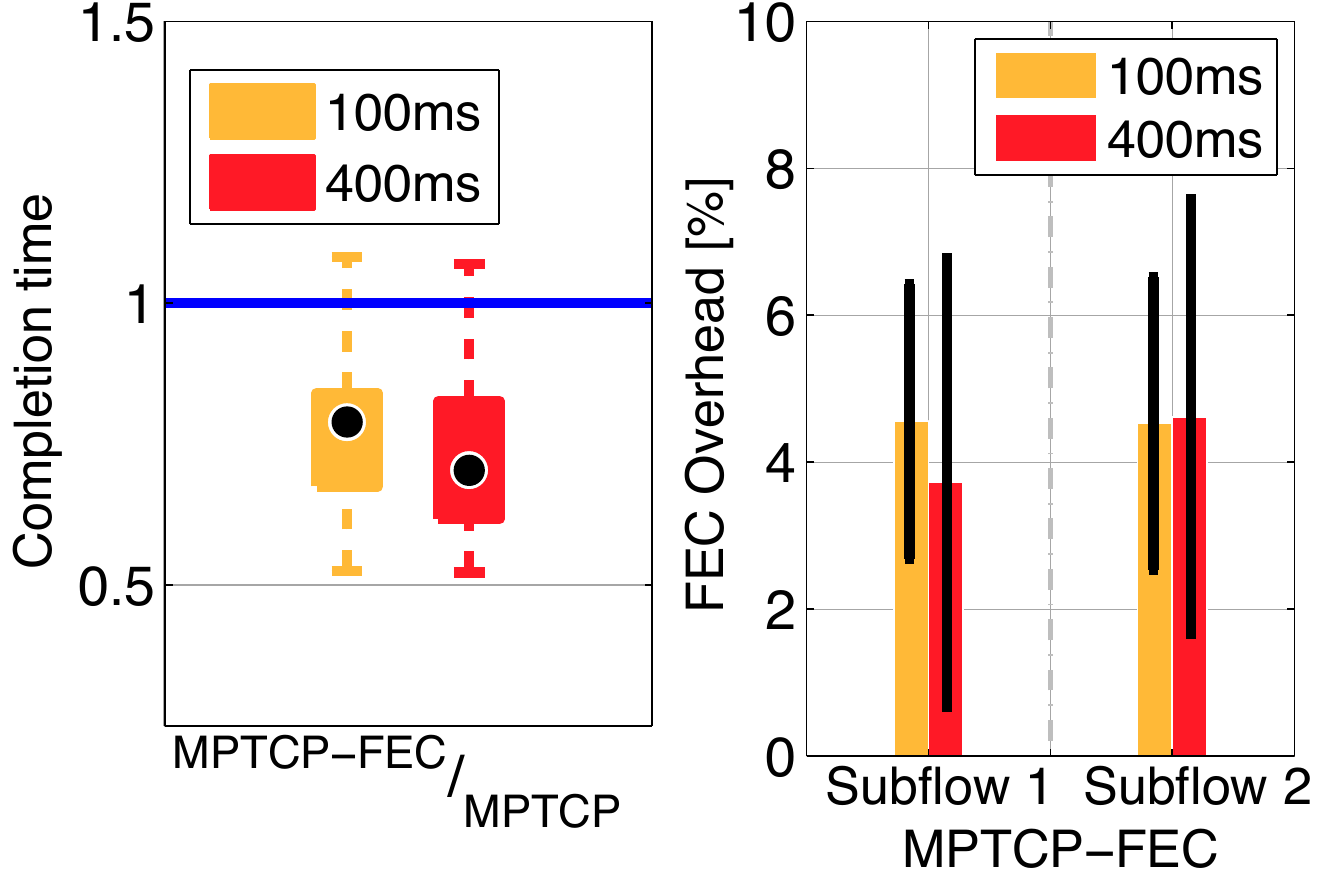}}
  \hspace{-0.5mm}
  \subfigure[\textbf{YouTube}\label{fig:http2:mptcp:dynamic:youtube:nornet}]
  {\includegraphics[width=.27\textwidth]{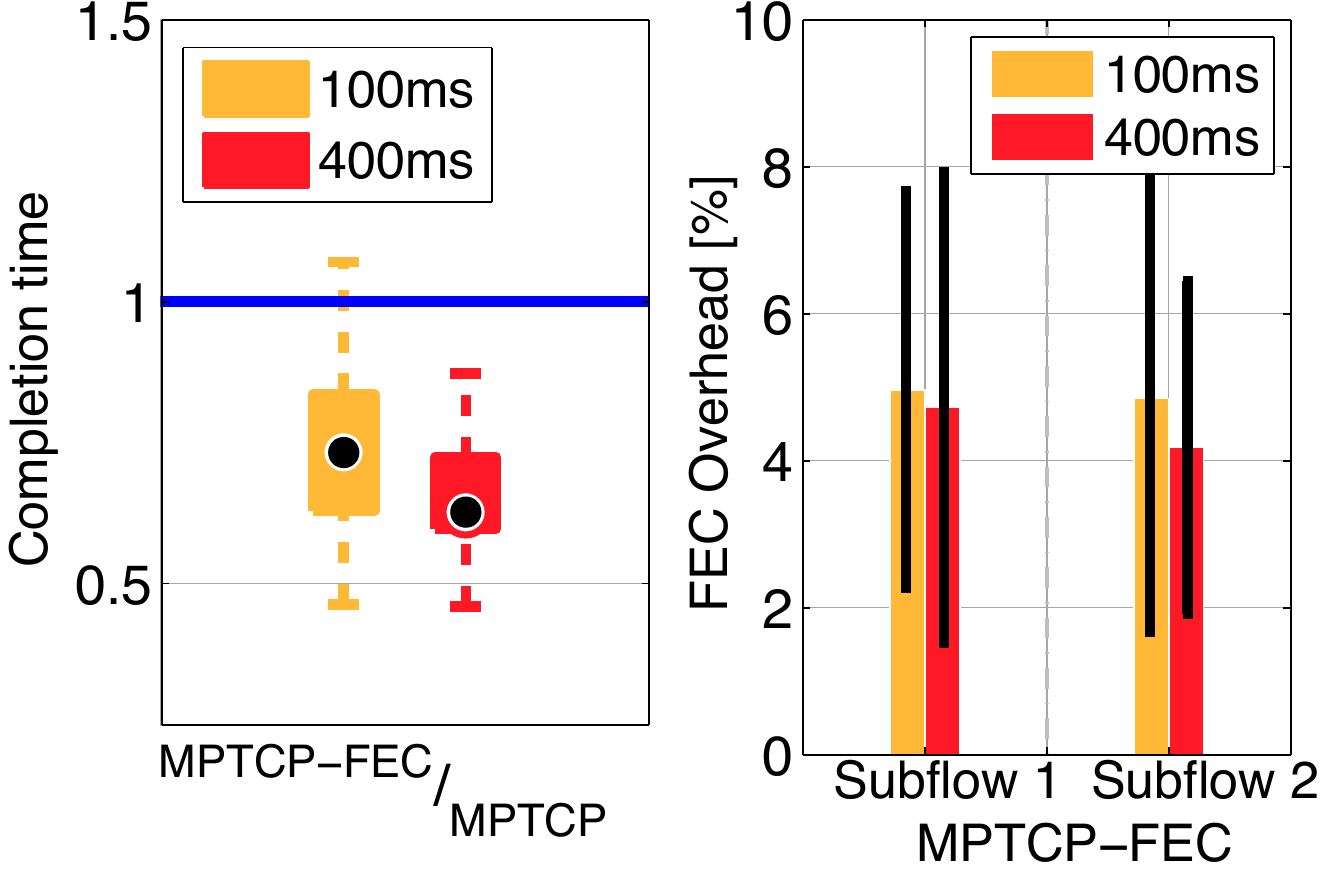}}
  \hspace{-0.5mm}
  \subfigure[\textbf{ESPN}\label{fig:http2:mptcp:dynamic:espn:nornet}]
  {\includegraphics[width=.27\textwidth]{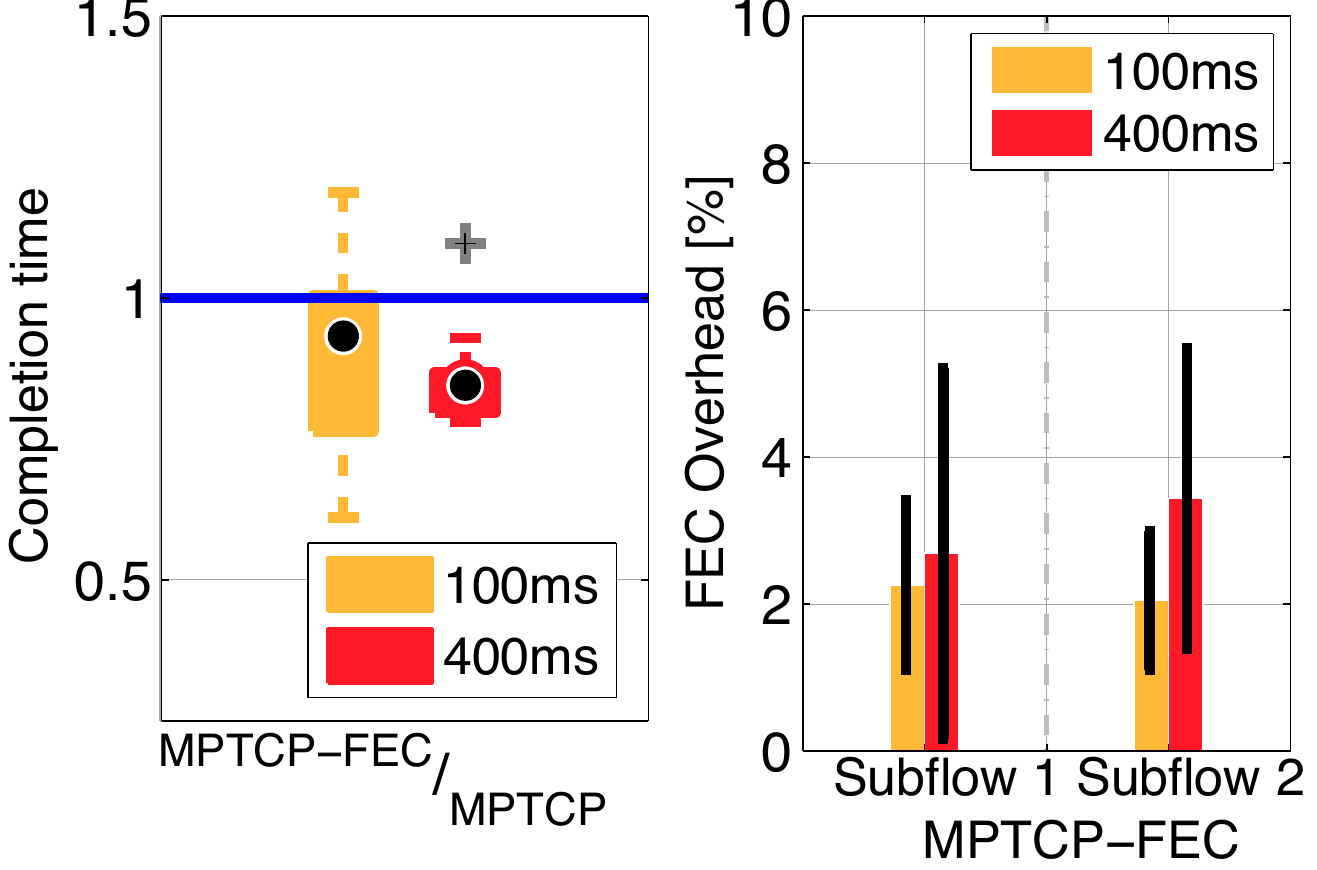}}
    \vspace{-3mm}
    \caption{\textbf{MPTCP-dFEC for HTTP: }Completion time for MPTCP-dFEC/MPTCP with real-network experiments.}\label{fig:http:nornet}
    \vspace{-3mm}
\end{figure*}

We introduced losses with $\texttt{netem}$ on the client side, in addition to non-influenceable concurrent traffic from other users to have some control over the experiment. The introduced losses have the goal to create a form of~\textit{ground truth}, checking how some of the average loss rate combinations affect dFEC compared to non-FEC experiments\footnote{The expected average loss rate, set with help of $\texttt{netem}$, were monitored during each experiment; counting the total amount of retransmissions (RTX) over the total traffic amount sent.}. In addition, we also evaluated the effect of~\textit{bufferbloat} on dFEC, when excessive network buffering reduce the ability of TCP loss-based congestion control to be responsive. Hence, we aim at showing dFEC's performance under more realistic conditions in a constructed non-shared bottleneck scenario. Figures~\ref{fig:bulk:nornet:bufferbloat},~\ref{fig:h264:nornet} and~\ref{fig:http:nornet} show bulk, H.264 and HTTP/2, respectively, for MPTCP and MPTCP-dFEC in the constructed testbed.

\textbf{Bufferbloat: }We run bulk transfers, removing the injected losses from both networks. The 3.5G network buffered up to 8~MiB data at times without any packet loss, we compared some of the results using the receive window and the congestion to compare. We observed no penalty with respect to completion time compared to MPTCP without FEC in any of the two scenarios. The traffic distribution is also similar in both scenarios compared to MPTCP without FEC. In Figure~\ref{fig:bulk:nornet:bufferbloat} we show that the dFEC with~\textit{bufferbloat} does not lose performance. We counted less than 0.005\% loss on both networks and the 3.5G path with up to 8~MiB without loss.

Figure~\ref{fig:h264:nornet} shows the full frames for H.264 with MPTCP with and without FEC. dFEC maintains its benefit close to 22\% compared to MPTCP, also in the~\textit{bufferbloat} scenario.

Figure~\ref{fig:http:nornet} shows the completion times for HTTP/2 with the websites from Table~\ref{tab:web-profile} for MPTCP with and without FEC in the~\textit{bufferbloat} scenario. dFEC reduces completion times of up to 30\% with a relatively low FEC overhead of less than 5\% on average in all cases.

\subsection{Dynamic FEC and System Performance}
\begin{figure}[h!]
  \vspace{-3mm}
  \centering
  \subfigure[\textbf{25 ms}\label{fig:}]
  {\includegraphics[width=.16\textwidth]{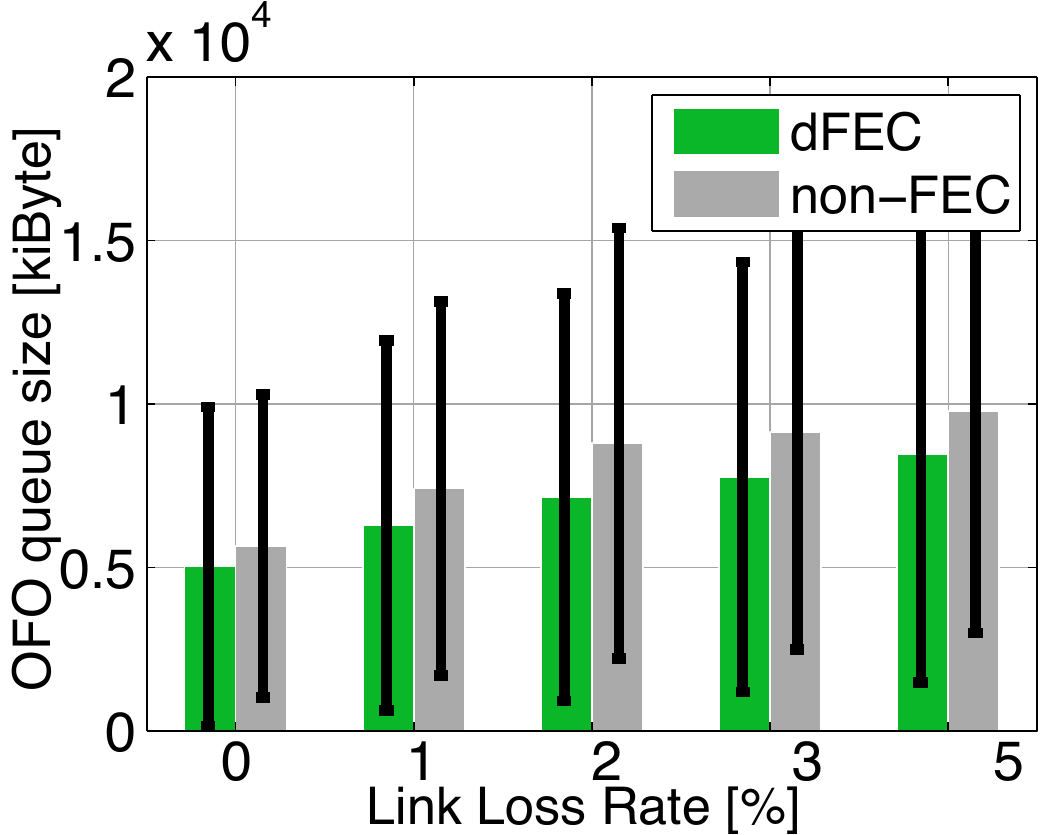}}
  \hspace{-2mm}
  \subfigure[\textbf{100 ms}\label{fig:}]
  {\includegraphics[width=.155\textwidth]{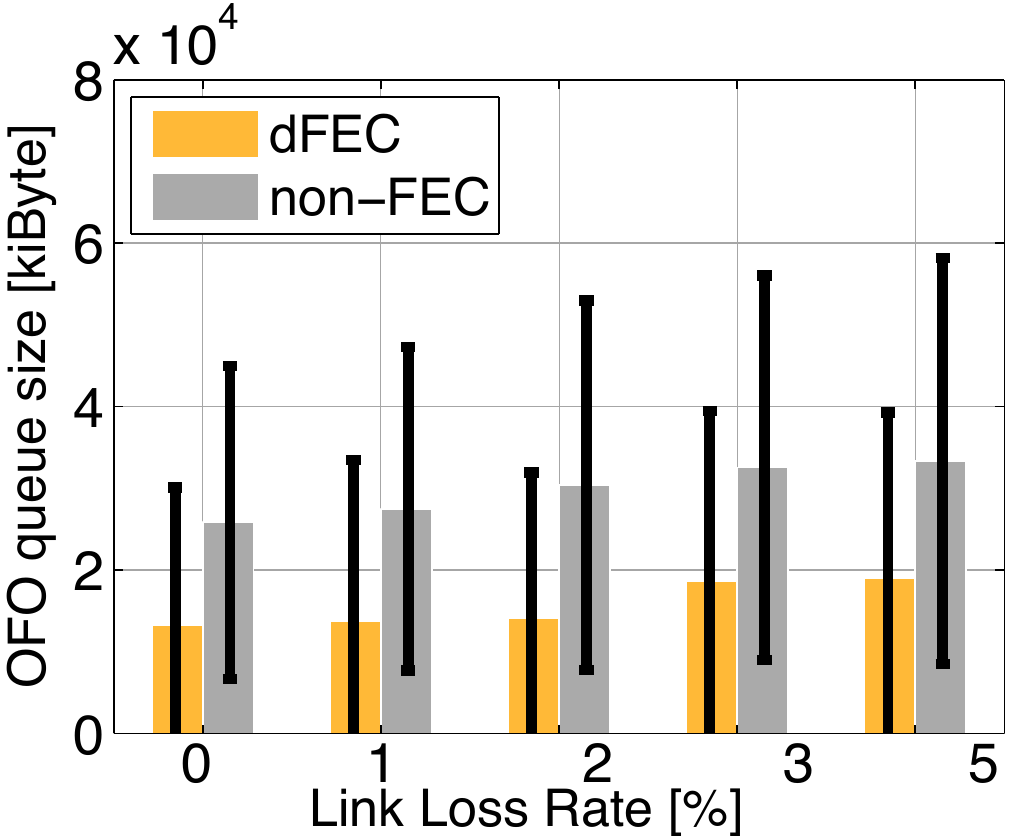}}
  \hspace{-2mm}
  \subfigure[\textbf{400 ms}\label{fig:}]
  {\includegraphics[width=.16\textwidth]{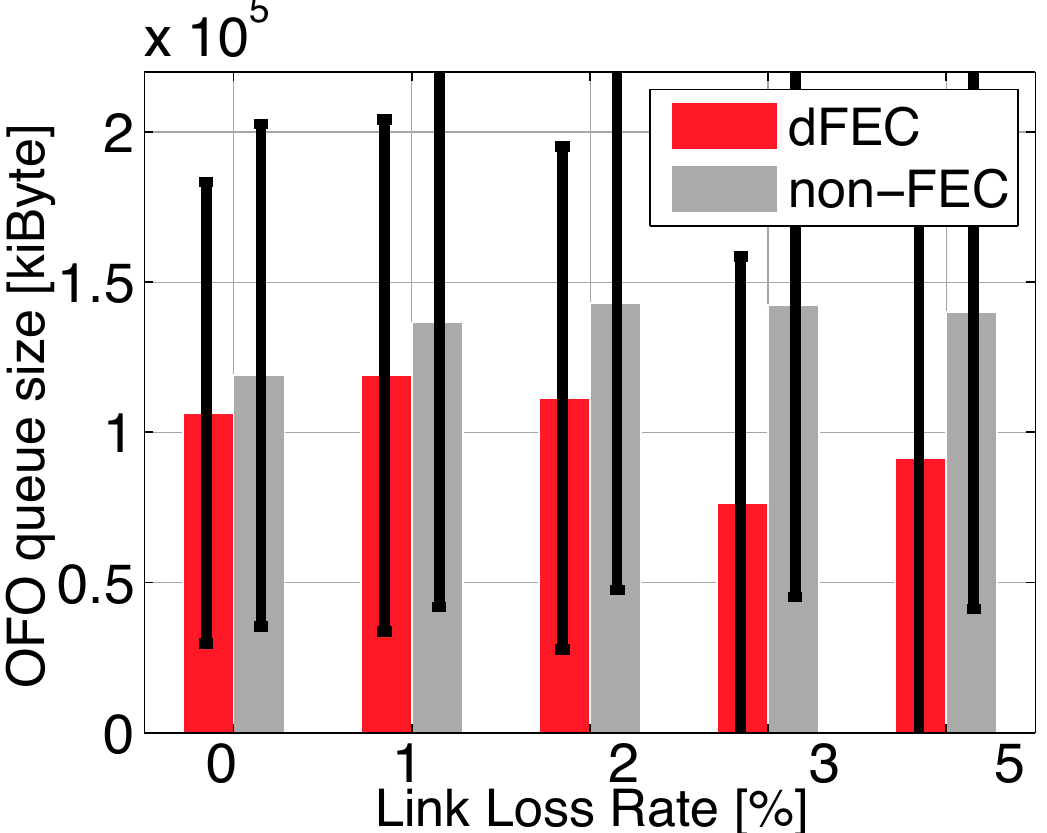}}
  \vspace{-3mm}
  \caption{\textbf{MPTCP-dFEC: }End-host average OFO queue size on the subflow level: Subflow 2}
  \label{fig:fec:ofo:subflow2}
  \vspace{-3mm}
\end{figure}

In this section, we analyse how dFEC affects the end-host in terms of memory usage. We take, as a measure, data being queued in the Out-Of-Order (OFO) queue, waiting for missing packets to be in-order delivered to either MPTCP-level or to the application. We performed tests with bulk and measured all changes in the Out-Of-Order (OFO) queue sizes on both subflows and on the MPTCP level.  
\begin{figure} [h!]
  \vspace{-3mm}
  \centering
  \subfigure [\textbf{25 ms}\label{fig:ofo:mptcp:25ms}]
  {\includegraphics[width=.16\textwidth]{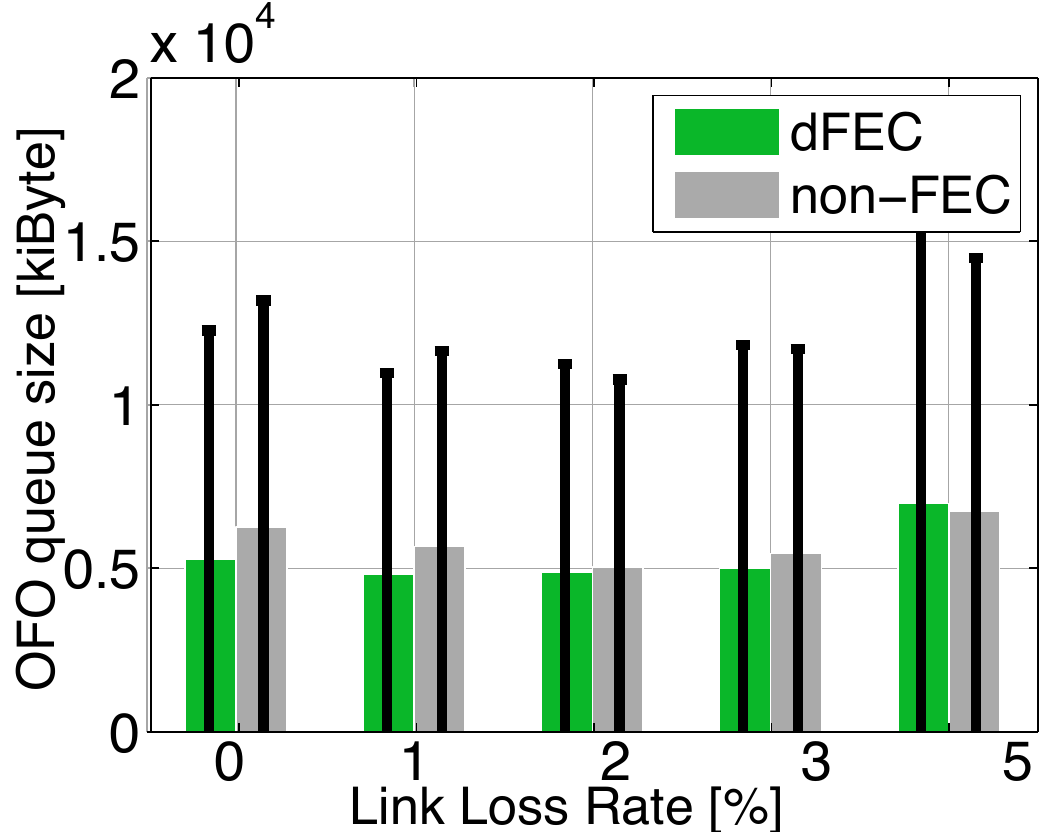}}
  \hspace{-2mm}
  \subfigure [\textbf{100 ms}\label{fig:ofo:mptcp:100ms}]
  {\includegraphics[width=.155\textwidth]{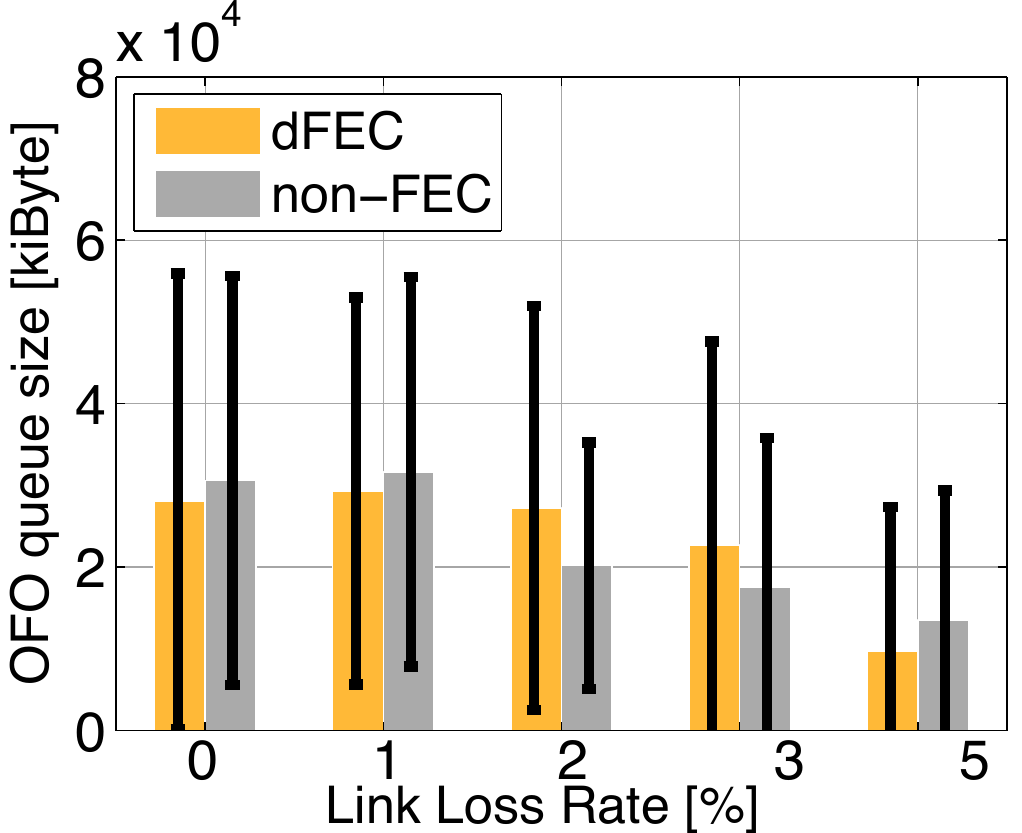}}
  \hspace{-2mm}
  \subfigure [\textbf{400 ms}\label{fig:ofo:mptcp:400ms}]
  {\includegraphics[width=.16\textwidth]{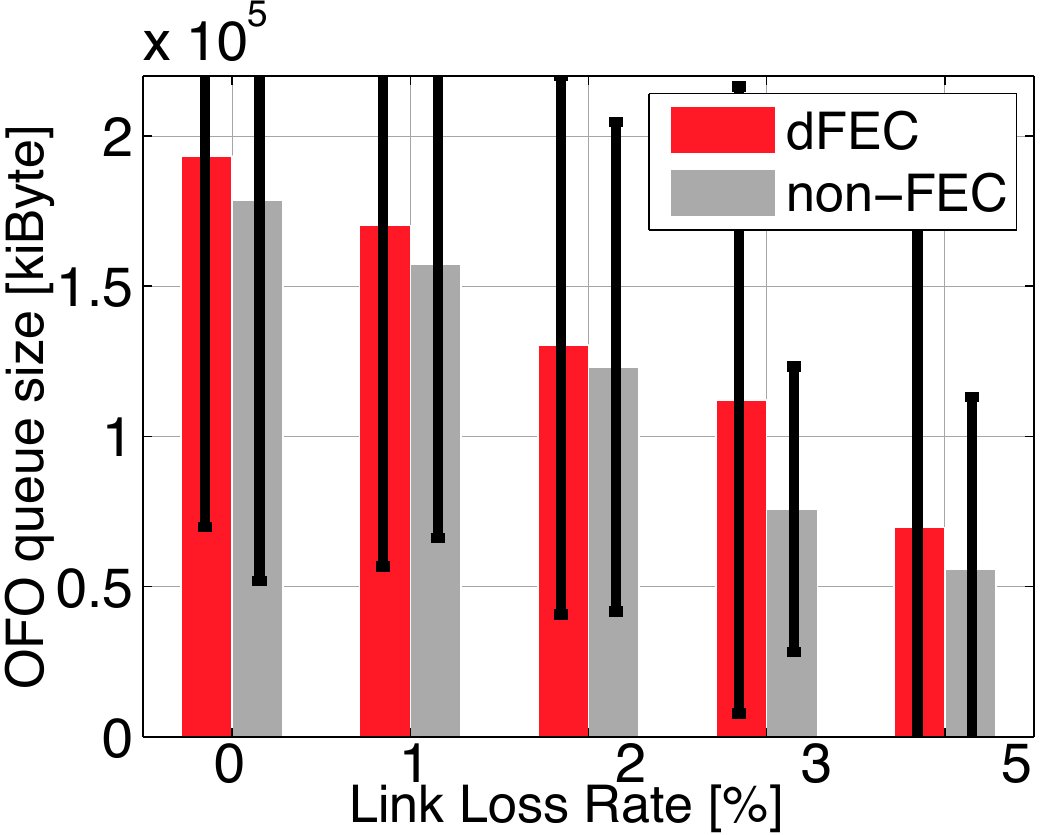}}
  \vspace{-3mm}
  \caption{\textbf{MPTCP-dFEC: }End-host average OFO queue size on the subflow level: MPTCP level}
  \label{fig:fec:ofo:mptcp}
  \vspace{-3mm}
\end{figure}
Figure~\ref{fig:fec:ofo:mptcp} shows the OFO queue sizes for 25, 100 and 400 ms RTTs. One can observe that dynamic FEC does not improve MPTCP's OFO queue occupancy when the subflows are homogeneous in terms of RTTs, see Figure~\ref{fig:ofo:mptcp:25ms}, and it even worsen the scenario with heterogeneous RTTs, see Figures~\ref{fig:ofo:mptcp:100ms} and~\ref{fig:ofo:mptcp:400ms}. To better understand this, we looked under MPTCP, into the subflows:
\begin{figure} [h!]
  \vspace{-3mm}
  \centering
  \subfigure[\textbf{25 ms}\label{fig:}]
  {\includegraphics[width=.16\textwidth]{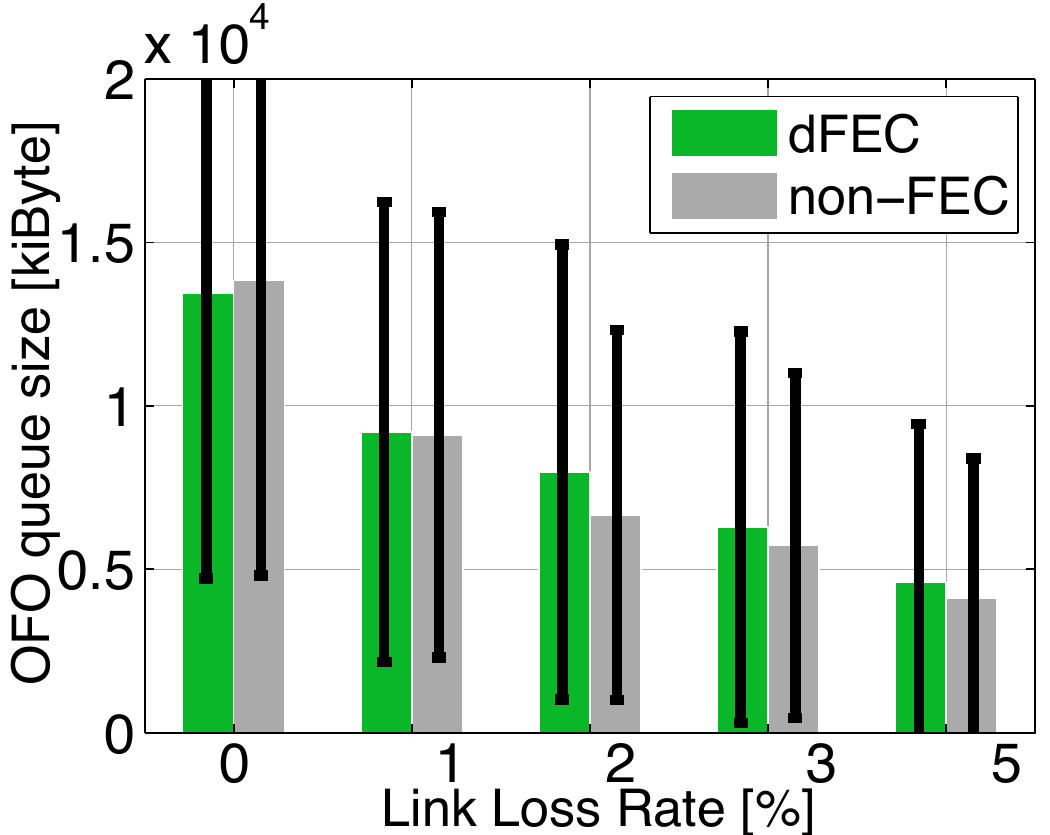}}
  \hspace{-2mm}
  \subfigure[\textbf{100 ms}\label{fig:}]
  {\includegraphics[width=.155\textwidth]{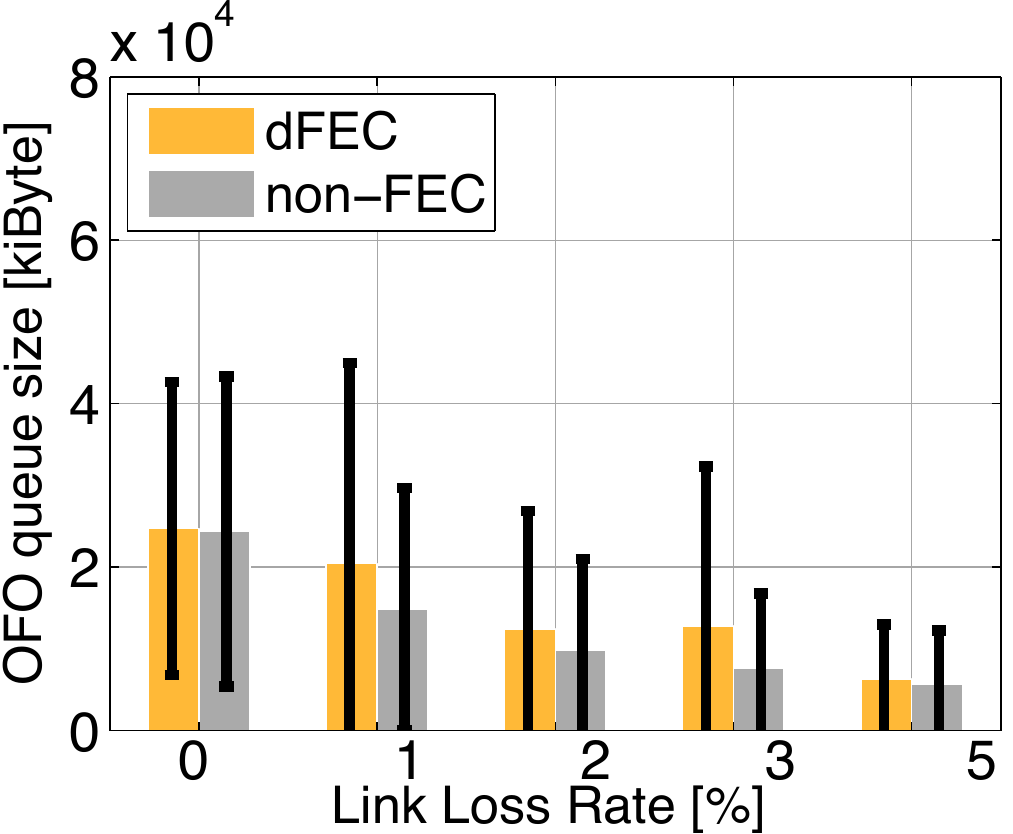}}
  \hspace{-2mm}
  \subfigure[\textbf{400 ms}\label{fig:}]
  {\includegraphics[width=.16\textwidth]{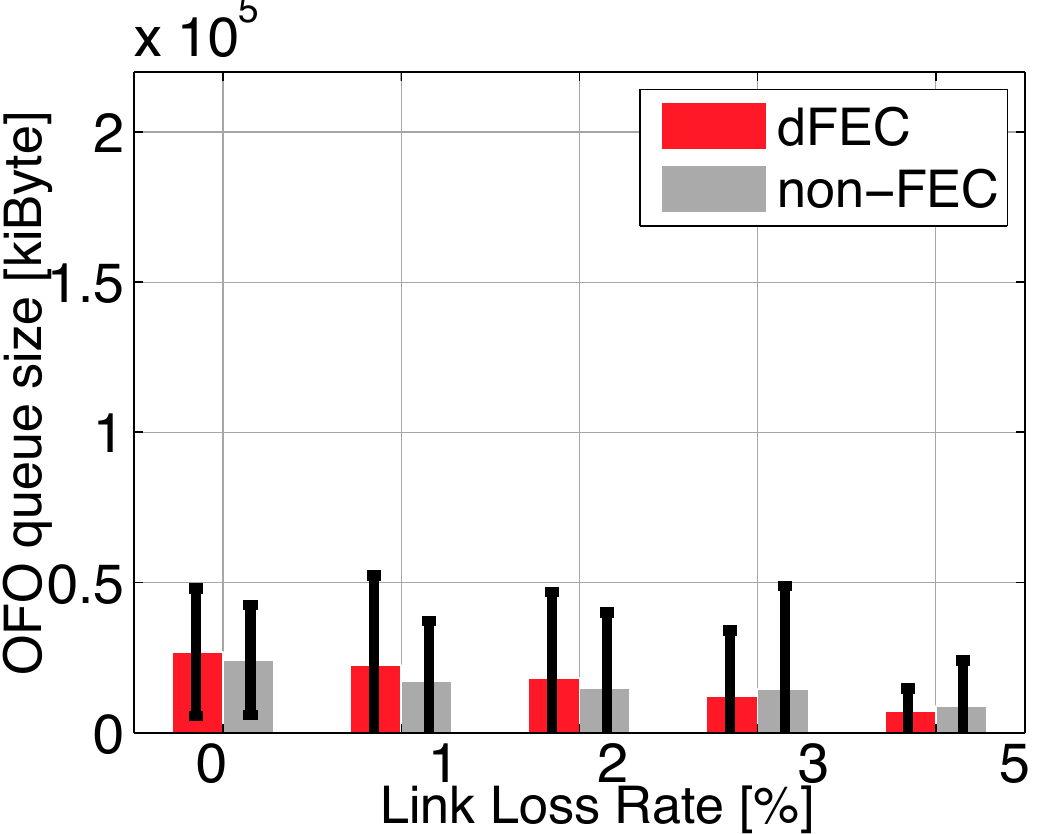}}
  \vspace{-3mm}
  \caption{\textbf{MPTCP-dFEC: }End-host average OFO queue size on the subflow level: Subflow 1}
  \label{fig:fec:ofo:subflow1}
  \vspace{-3mm}
\end{figure}
Figure~\ref{fig:fec:ofo:subflow1} shows the OFO queue size for the subflow on $B1$, see Table~\ref{tab:emulation:parameters}, hence the subflow with only the average loss rate changed. Here, one can hardly see a difference compared to regular MPTCP. In Figures~\ref{fig:fec:ofo:subflow2}, however, the OFO queue occupation of the subflow on $B2$ is considerably lower compared to default MPTCP.
We explain this by showing the effect of dynamic FEC into default MPTCP's congestion control and scheduler algorithms: The subflow on $B1$ is the lossier subflow compared to the subflow on $B2$, where we only vary its RTT between 25, 100 and 400~ms. dFEC improves the utilisation on subflow on $B1$ on average by 15\% meaning that the average multipath rate is increased on $B1$'s subflow. In such case, MPTCP's couple congestion control exposes a larger CWND for the min-RTT scheduler, which, in turn, prefers $B1$'s subflow over $B2$, due to its minimum RTT scheduling policy. 

Hence, while dynamic FEC increases the average utilisation on the lossy $B1$ subflow by shifting traffic from $B2$'s subflow and, consequently, increases the overall multipath throughput, it cannot guarantee improvements in the system resource's utilisation. At this stage, a tighter integration between FEC and MPTCP's scheduler and congestion control algorithms is required to minimise the impact on the receiver. This will be considered as part of our future research.

\section{Related Work}\label{sec:related_work}
This section is divided in two parts:~\ref{subsection:tcp:wireless} comments on TCP and MPTCP over wireless networks and~\ref{subsection:fec:transport} comments on FEC strategies within the transport layer. Finally, Table~\ref{tab:fec:approaches} summarises the key aspects of the systems in~\ref{subsection:fec:transport}.

\subsection{TCP and MPTCP over Wireless Networks}\label{subsection:tcp:wireless}
In its initial design, TCP was not meant to operate in wireless environments with links that often face random effects and, depending on the congestion control, make TCP drastically reduce its sending rate; having a long-term detrimental impact~\cite{Parsa2000,fu03,tian05,mirza07,winstein13,sivaraman15,zaki15}. For example, TCP's Additive Increase and Multiplicative Decrease (AIMD) mechanism reduces its sending rate by 50\% over one Round-Trip-Time (RTT)\footnote{With a Fast Retransmission (FR), the rate is halved, according to TCP's Proportional Rate Reduction (PRR).} and, increase, roughly, one packet per RTT.

Hence, TCP's performance over wireless networks, such as cellular or satellite, is suboptimal~\cite{Martin2002,Johansson16}, where one of the main limiting factors is the necessary recovery time. Besides many ratifications, TCP's legacy loss detection and recovery mechanisms remained mostly unchanged and tied to time: A Retransmission Timeout (RTO) is applied after a timer expires, or a Fast Retransmission (FR) is performed after three duplicated acknowledgements (\texttt{DupACK}) arrive from the receiver to detect a loss, and a retransmission requires at least one extra RTT to perform. Hence, in such scenarios, TCP is often combined with FEC, where there has been interest in the interplay of FEC and TCP with more recent initiatives being TCP-Tail Loss Probe (TLP)~\cite{dukkipati01}, TCP-Instant Recovery (IR)~\cite{flach00} or QUIC~\cite{hamilton16}. 

More recently, Multipath TCP (MPTCP) emerged to address the necessity of exploiting network resources with multi-connected devices, e.g., smartphones via WLAN and cellular networks~\cite{paasch14}, IPv4 and IPv6 addresses~\cite{Livadariu2015_GIS} and mobility resilience support~\cite{bagnulo10, paasch12, oh16}. However, having to remain deployable and operable in today's Internet, required MPTCP to be built on top of regular TCP~\cite{barre11,RFC6182}.

MPTCP started off by focusing on bandwidth aggregation in scenarios with homogeneous network paths~\cite{raiciu11, barre11}, e.g., data center networks. However, there was an increasing interest to explore MPTCP capabilities with applications that have tighter Quality Of Service (QoS) requirements, such as web traffic and video streaming, in mobile wireless Internet. These scenarios showed to be more critical, especially when the underlying network paths are heterogeneous in terms of delay and loss rate~\cite{Chen2013,ferlin16_IFIP,Kiran2016}.


\subsection{Forward Error Correction (FEC)}\label{subsection:fec:transport}
FEC is commonly used to improve reliability, with an active interest to integrate it into the transport layer~\cite{dukkipati01,flach00, hamilton16}. However, this has been prohibitively complex, with most of the proposals focusing on system simulations. Here we comment on the closest and most recent proposals:

MPLOT~\cite{Sharma08}, is a new transport protocol built in~\texttt{ns-2} to exploit multipath diversity with erasure codes. The paper, however, does not mention transport protocol-related details, e.g., congestion control and signalling. MPLOT focus solely on bandwidth aggregation aspects in both homogeneous and heterogeneous settings with delay and loss rates, comparing its approach against regular TCP. For homogeneous scenarios MPLOT shows gains of over 20\% and 37\% in heterogeneous scenarios the larger the number of paths. For heterogeneous RTTs, the gains are reduced the larger the delay heterogeneity. The evaluation was done with only bulk traffic.

Coded TCP (C-TCP)~\cite{kim12b} built in user-space, uses multiple homogeneous paths, i.e., two WLAN paths, with systematic block codes, transmitting data over UDP through a self-designed system based on delay and loss to realise congestion control and signalling. C-TCP shows its results against regular TCP using file transfer with FTP. C-TCP outperforms TCP with artificially injected losses up to 5\% by ca. 69\%, however, C-TCP implements an hybrid congestion control mechanism with both loss and delay signals, realised through tokens, whereas regular TCP is using CUBIC, a loss based congestion control. With two homogeneous paths, C-TCP outperforms regular TCP by 98\% with 5\% injected loss.

FMTCP~\cite{Cui12} proposes a fountain code-based system in~\texttt{ns-2} to help mitigate path heterogeneity in MPTCP. FMTCP focuses on evaluating heterogeneous settings by varying both RTTs between 25, 50, 100 and 150~ms and loss rates from 2 up to 15\% in one of the subflows, while the other subflow's characteristics were kept constant with 100~ms and no injected loss. They experiment with a non-shared bottleneck scenario, but it is not evident, which congestion control FMTCP adopts, since they claim to focus only on data distribution. They claim gains of more than 50\% in aggregation over MPTCP. 

SC-MPTCP~\cite{Li14} uses linear systematic encoding within~\texttt{ns-3} focusing on bandwidth aggregation focusing on heterogeneous settings, varying loss and RTTs between 1 and 5\% and 20 to 60~ms, respectively. They focus on demonstrating that SC-MPTCP aggregated bandwidth loses ca. 10\% compared to MPTCP with 5\% loss, and it uses needs much less buffering, e.g., 270~kiB against 8~MiB in MPTCP. 

ADMIT~\cite{Wu15} is a multipath system focusing on real-time high definition H.264 using a MPTCP-model within the \texttt{Exata} emulator. It applies systematic Reed-Solomon codes, describing also an adaptive FEC algorithm. It focuses on goodput, end-to-end delay and PSNR metrics, since it is designed for video. Goodput gains are ca. 20\% compared to regular MPTCP. 
Similarly, Bandwidth-Efficient Multipath Streaming (BEMA)~\cite{Wu16}, similar to ADMIT~\cite{Wu15}, is built for H.264 video over multiple paths, hence, it uses metrics such as goodput, end-to-end delay and PSNR. BEMA uses UDP and TCP with T-FRC~\cite{rfc3448} within the~\texttt{Exata} emulator, applying systematic Raptor codes with FEC adaptivity. 

Stochastic Earliest Delivery Path First (S-EDPF)~\cite{garcia15} is a user-space system built to stream video over multiple paths with MPTCP using low delay random linear codes. They reuse CTCP's framework~\cite{kim12a}, however, not considering FEC adaptivity. S-EDPF (no coding) performs as good as MPTCP's low-RTT, whereas S-EDPF-8 and 16 can improve goodput aggregation by  40\%, while halving the end-to-end delay.

Table~\ref{tab:fec:approaches} summarises the different FEC schemes, methodology and metrics of all proposals showed in~\ref{subsection:fec:transport}. One can notice that all systems design their own FEC inside the application, using MPTCP underneath only to distribute data over multiple paths, and they are evaluated with simulations or protocol models inside a network emulator. Although interesting, this cannot capture the interaction of such a design with a complex framework such as TCP and MPTCP. Hence, we depart from the proposed systems, proposing an entire XOR-based adaptive FEC implementation inside TCP and MPTCP.

In such case, the advantages of a XOR-based FEC approach are low computational overhead and simple implementation, where TCP's original segment structure can be maintained. However, the obvious disadvantage is that it can only recover one segment per block, e.g., if two or more packets are lost within a block, see Figure~\ref{fig:fec:wire}, the FEC packet is sent in vain, and the missing packets have to be retransmitted with Fast Retransmission (FR) of after a Retransmission Timeout (RTO).


\begin{table*}
 \vspace{-3mm}
  \centering
\setlength{\tabcolsep}{0.4em} 
  \caption{Some key characteristics of the systems described in Section~\ref{sec:related_work}.}
  \label{tab:fec:approaches}
    \resizebox{\textwidth}{!}{%
    \begin{tabular}{c|c|c|c|c|c|c}
      \toprule
      \multicolumn{1}{c|}{\textbf{Systems}} & \textbf{\shortstack{Implementation \\ Layer}} & \textbf{\shortstack{Transport \\ Protocol}} & \textbf{FEC Algorithm} & \textbf{\shortstack{FEC \\ Adaptivity}} &  \textbf{Evaluation} &  \textbf{\shortstack{Application(s) and \\ Evaluation Metrics}}\\
      \midrule

  MPLOT & Transport & MPLOT & Erasure & \ding{52} & \texttt{ns-2} & Goodput\\
  C-TCP & Application & UDP & Systematic block & \ding{52} & Emulation & Throughput of each path \\

  FMTCP & Transport & TCP & Rateless Fountain& \ding{52} & \texttt{ns-2} & Goodput, delivery delay and jitter\\

  SC-MPTCP & Transport & MPTCP  & Linear systematic & \ding{52} & \texttt{ns-3} & Goodput and buffer size and delay\\

  ADMIT & Transport & MPTCP-model & Syst. Reed-Solomon & \ding{52} & \texttt{Exata} emulator & PSNR, e2e delay and goodput \\

  S-EDPF & Application & MPTCP & Random linear & \ding{53} & Real-network & Goodput, e2e and reordering delay\\

  BEMA & Transport & UDP and TCP & Systematic raptor & \ding{52} & \texttt{Exata} emulator & PSNR, e2e delay and goodput \\
      \bottomrule
    \end{tabular}
    }
   \vspace{-3mm}
\end{table*}


\textbf{Summary: }Although there has been interest in adding FEC to the transport layer, in particular to TCP due to its loss detection and recovery mechanisms being tied to time, it has been prohibitively complex and it has been implemented with some simplifications. Most of the proposals suggest implementations in user-space, where applications have to be modified and FEC is, thus, application-specific. Also, in the application layer, the knowledge about network conditions is less granular, e.g., delay and loss rates. Therefore, in this paper we aim for a XOR-based FEC implementation within TCP, to aid multipath transport with heterogeneity with MPTCP.

\section{Conclusion}\label{sec:conclusion}
The performance of TCP over wireless high delay and lossy networks is known to be suboptimal~\cite{Martin2002,Johansson16}, with one of the main limiting factors being the loss recovery time. In such scenarios, it is often the option to replace TCP by UDP, even it compromises on benefits. However, another option is to tackle long loss recovery time of TCP by adding FEC to TCP, even though this has shown so far to be prohibitively complex.
Since, MPTCP is closely tied to regular TCP, this brings many benefits when it comes to deployability, but it also comes with challenges hindering MPTCP, in particular, when the underlying network paths are heterogeneous~\cite{Kiran2016,ferlin16_IFIP}. 

In this work, we designed and implemented a XOR-based dynamic FEC scheme for TCP and MPTCP. We showed that with the proposed framework, for links having low loss rates, the FEC overhead is relatively small and for lossy links, significant performance gains can be achieved for different applications, such as HTTP/2 with different website sizes, adaptive video with HTTP-DASH, non-adaptive video with H.264 and bulk transfers.

For future study we plan to investigate FEC in MPTCP across subflows where the interaction with the scheduler and congestion control need to be taken into consideration.  We will than be able to compare an intra-subflow FEC with an inter-subflow FEC approach, under different network settings. Also, real experiments with static and mobility scenarios will help to shed light of real network effects.





%

\bibliographystyle{IEEEtran}
\bibliography{ToN17_FINAL_MAIN}

%

\vskip +4\baselineskip plus -1fil

\begin{IEEEbiography}[{\includegraphics[width=1.05in,height=1.5in,clip,keepaspectratio]{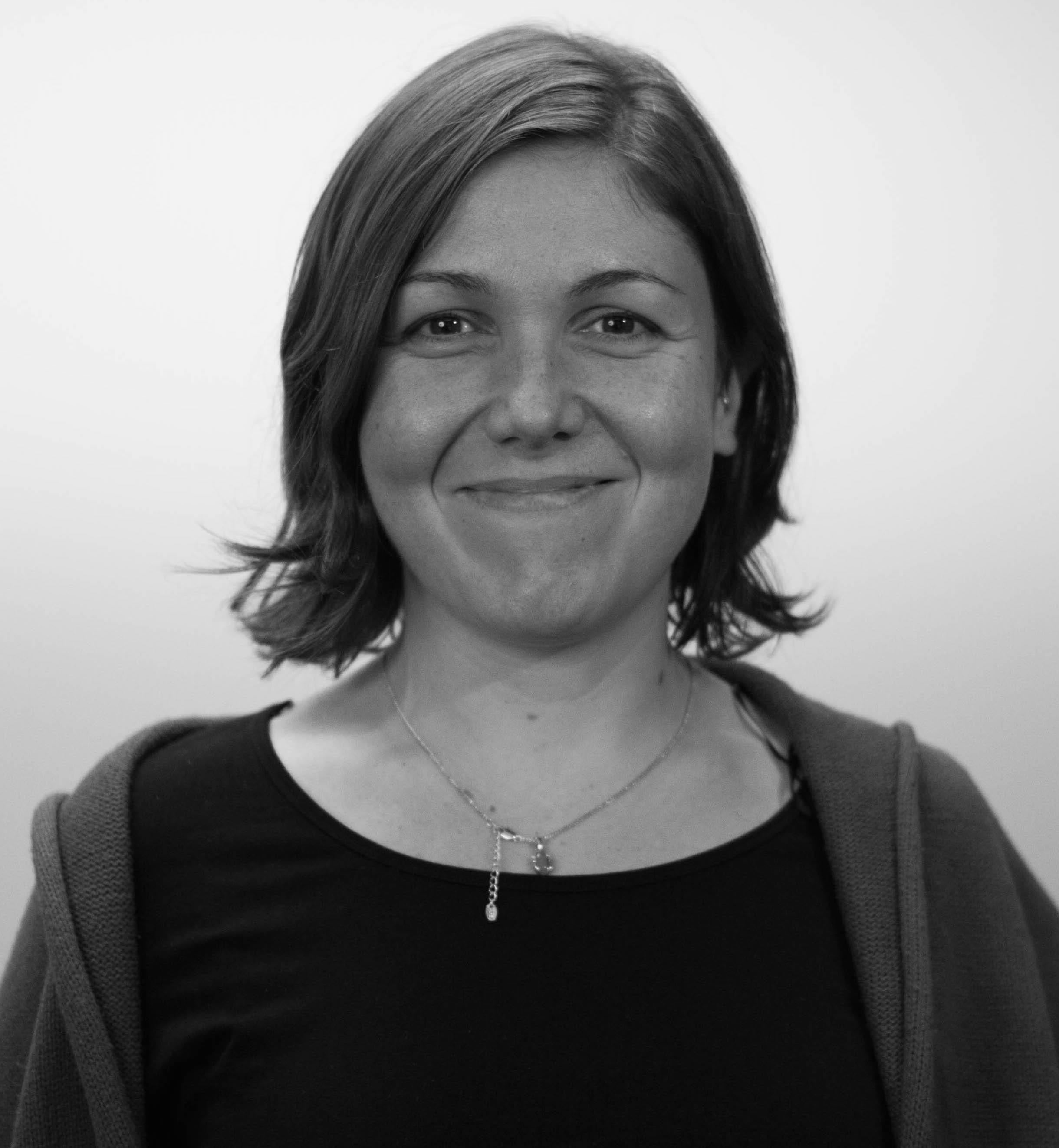}}]{Simone Ferlin}
received her Dipl.-Ing. degree in Information Technology with major in Telecommunications from Friedrich-Alexander Erlangen-Nuernberg University, Germany in 2010 and her PhD degree from the university of Oslo, Norway in 2017. Her research interests lie in the areas of computer networks, transport protocols, congestion control, network measurements and data analysis. Her dissertation focuses on improving robustness in multipath transport for heterogeneous networks.
\end{IEEEbiography}
\vskip -2\baselineskip plus -1fil

\begin{IEEEbiography}
[{\includegraphics[width=1.05in,height=1.5in,clip,keepaspectratio]{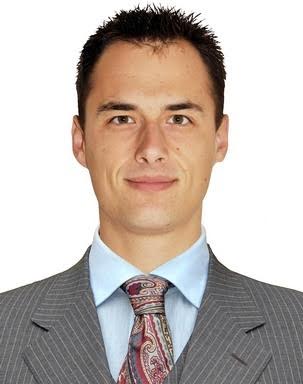}}]
{Stepan Kucera} (S'14 M'06) is a senior research scientist in the Small Cells Group of Nokia Bell Laboratories in Ireland. He is engaged as principal investigator and technical lead in projects aiming to create innovative technologies for next-generation mobile networking. His expertise lies mainly in the area of wireless and IP networking technology, including both 3GPP/IEEE/IETF standards and proprietary solutions. Dr. Kucera filed nearly 50 patents and published over 30 book chapters, transactions and conference papers in peer-reviewed IEEE venues. He is also the (co)-recipient of four professional awards (the Irish Laboratory of the Year Award and the Irish Commercial Laboratory of the Year Award in 2014, the IEEE Kansai Student Researcher Encouragement Award in 2007, the Best Student Paper Award at IEEE Vehicular Technology Conference Fall 2006) as well as was nominated for the Ericsson Young Scientist Award in 2010 and the IEICE Young Researcher Encouragement Award in 2009. Between 2008 and 2011, he was a research scientist at the New Generation Wireless Communications Research Center at the Keihanna Research Laboratories, NICT, Japan. He received his Ph.D. degree in Informatics from the Graduate School of Informatics at Kyoto University, Kyoto, Japan, in 2008, and his M.Sc. degree from Czech Technical University in Prague, Czech Republic, in 2003. He is a Senior Member of IEEE, and actively serves on technical boards of major IEEE journals and conferences.
\end{IEEEbiography}
\vskip -2\baselineskip plus -1fil

\begin{IEEEbiography}
[{\includegraphics[width=1.05in,height=1.5in,clip,keepaspectratio]{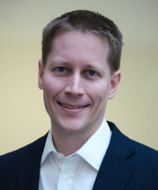}}]{Holger Claussen}
{Holger Claussen} Dr. Holger Claussen is leader of Small Cells Research Department at Bell Labs, Nokia with a team in Ireland and the US. In this role, he and his team are innovating in all areas related to future evolution, deployment, and operation of small cell networks to enable exponential growth in mobile data traffic. His research in this domain has been commercialized in Nokia's (formerly Alcatel-Lucent's) Small Cell product portfolio and continues to have significant impact. He received the 2014 World Technology Award in the individual category Communications Technologies for innovative work of "the greatest likely long-term significance". Prior to this, Holger was head of the Autonomous Networks and Systems Research Department at Bell Labs Ireland, where he directed research in the area of self-managing networks to enable the first large scale femtocell deployments from 2009 onwards. Holger joined Bell Labs in 2004, where he began his research in the areas of network optimization, cellular architectures, and improving energy efficiency of networks. Holger received his Ph.D. degree in signal processing for digital communications from the University of Edinburgh, United Kingdom in 2004. He is author of more than 90 publications and 110 filed patent applications. He is Fellow of the World Technology Network, senior member of the IEEE, and member of the IET.
\end{IEEEbiography}
\vskip -2\baselineskip plus -1fil

\begin{IEEEbiography}
[{\includegraphics[width=1.05in,height=1.5in,clip,keepaspectratio]{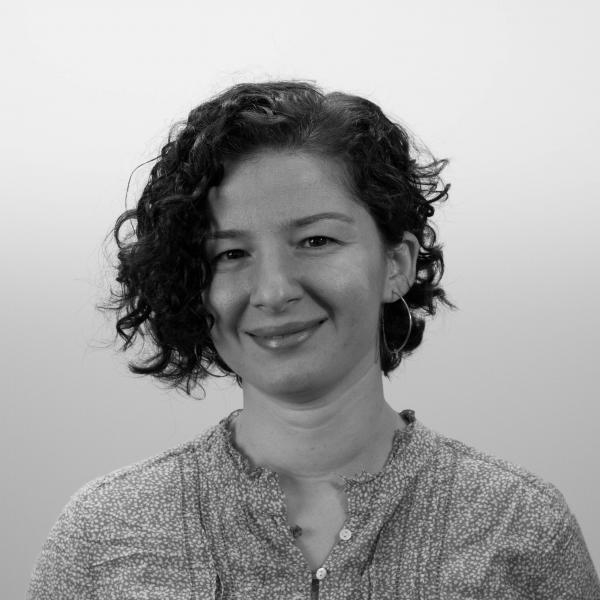}}]{Ozgu Alay}
Dr. Ozgu Alay received the B.S. and M.S. degrees in Electrical and Electronic Engineering from Middle East Technical University, Turkey, and Ph.D. degree in Electrical and Computer Engineering at Tandon School of Engineering at New York University.
Currently, she is a senior research scientist at Networks Department of Simula Research Laboratory, Norway.
Her research interests lie in the areas of mobile broadband networks, multi-path protocols and robust multimedia transmission over wireless networks. Her research is mostly experimental and her current focus is on data analytics for mobile networks.
\end{IEEEbiography}
\end{document}